\documentclass[twocolumn]{aastex631} 
\usepackage{apjfonts}
\usepackage{graphics,graphicx,subfigure,float,color}
\usepackage{xfrac}    
\usepackage{amsmath}  
\usepackage[]{units} 
\usepackage{hyperref}
\usepackage{enumitem}

\newcommand{\M}{{\ $\mid$}}
\newcommand{\PM}{{$\pm$}}
\newcommand{\ii}{{\sc ii}}
\newcommand{\iii}{{\sc iii}}
\newcommand{\iv}{{\sc iv}}
\newcommand{\W}{{$\lambda$}}
\newcommand{\CH}[1]{\colhead{#1}}
\setcitestyle{notesep={; },round,aysep={},yysep={;}}

\submitjournal{AASJournal ApJ}
\shortauthors{Berg et al.}
\shorttitle{EELG Properties}

\begin{document}

\title{
Characterizing Extreme Emission Line Galaxies I: \\
A Four-Zone Ionization Model for Very-High-Ionization Emission\footnote{
Based on observations made with the NASA/ESA Hubble Space Telescope,
obtained from the Data Archive at the Space Telescope Science Institute, which
is operated by the Association of Universities for Research in Astronomy, Inc.,
under NASA contract NAS 5-26555.}}

\author[0000-0002-4153-053X]{Danielle A. Berg}
\affiliation{Department of Astronomy, The University of Texas at Austin, 2515 Speedway, Stop C1400, Austin, TX 78712, USA}
\email{daberg@austin.utexas.edu}

\author[0000-0002-0302-2577]{John Chisholm}
\affiliation{Department of Astronomy, The University of Texas at Austin, 2515 Speedway, Stop C1400, Austin, TX 78712, USA}

\author[0000-0001-9714-2758]{Dawn K. Erb}
\affiliation{Center for Gravitation, Cosmology and Astrophysics, Department of Physics, University of Wisconsin Milwaukee, 3135 N Maryland Ave., Milwaukee, WI 53211, USA}

\author[0000-0003-0605-8732]{Evan D. Skillman}
\affiliation{Minnesota Institute for Astrophysics, University of Minnesota, 116 Church Street SE, Minneapolis, MN 55455, USA}

\author[0000-0003-1435-3053]{Richard W. Pogge}
\affiliation{Department of Astronomy, The Ohio State University, 140 W 18th Avenue, Columbus, OH 43210, USA}
\affiliation{Center for Cosmology \& AstroParticle Physics, The Ohio State University, 191 W Woodruff Avenue, Columbus, OH 43210}

\author[0000-0003-1435-3053]{Grace M. Olivier}
\affiliation{Department of Astronomy, The Ohio State University, 140 W 18th Avenue, Columbus, OH 43210, USA}


\begin{abstract}
Stellar population models produce radiation fields that
ionize oxygen up to O$^{+2}$, defining the
limit of standard \ion{H}{2} region models ($<54.9$ eV).
Yet, some {\it extreme emission line galaxies}, or EELGs, 
have surprisingly strong emission 
originating from much higher ionization potentials.
We present UV-{\it HST}/COS and optical-LBT/MODS
spectra of two nearby EELGs that have
very-high-ionization emission lines (e.g., \ion{He}{2}\W\W1640,4686 
\ion{C}{4}\W\W1548,1550, [\ion{Fe}{5}]\W4227, [\ion{Ar}{4}]\W\W4711,4740).
We define a 4-zone ionization model that is augmented by a very-high-ionization 
zone, as characterized by He$^{+2}$ ($>54.4$ eV). 

The 4-zone model has little to no effect on the measured total nebular abundances,
but does change the interpretation of other EELG properties:
we measure steeper central ionization gradients, higher volume-averaged 
ionization parameters, and higher central $T_e$, $n_e$, and log$U$ values.
Traditional 3-zone estimates of the ionization parameter can under-estimate the average 
log$U$ by up to 0.5 dex.
Additionally, we find a model-independent dichotomy in the abundance patterns,
where the $\alpha$/H-abundances are consistent 
but N/H, C/H, and Fe/H are relatively deficient, suggesting these 
EELGs are $\alpha$/Fe-enriched by $>3\times$.
However, there still is a {\it high-energy ionizing photon production problem} (HEIP$^3$). 
Even for such $\alpha$/Fe-enrichment and very-high log$U$s, 
photoionization models cannot reproduce the 
very-high-ionization emission lines observed in EELGs.

\end{abstract} 



\section{Introduction}
\subsection{Background}
The 21st century of astronomy has been marked by deep imaging surveys with the
{\it Hubble Space Telescope} ({\it HST}) that have opened new windows onto the high-redshift
universe, unveiling thousands of $z>6$ galaxies
\citep[e.g.,][]{bouwens15,finkelstein15,livermore17,atek18,oesch18}.
From these studies, and the numerous sources discovered, a general consensus
has emerged that low-mass galaxies host a substantial fraction of the star formation in 
the high-redshift universe and are likely the key contributors to reionization 
\citep[e.g.,][]{wise14,robertson15,madau15,stanway16}. 

Significant observational efforts have been invested to the study of these reionization era systems, 
revealing a population of compact, metal-poor, low-mass sources with blue UV continuum slopes 
that are rare at $z\sim0$ \citep[e.g.,][]{stark17,laporte17,mainali17,hutchison19}. 
Deep rest-frame UV spectra of $z>5$ galaxies have revealed prominent 
high-ionization nebular emission lines (i.e., \ion{O}{3}], \ion{C}{3}], \ion{C}{4}, \ion{He}{2}),
with especially large \ion{C}{3}] and \ion{C}{4} equivalent widths ($\sim20-40$ \AA),
indicating that extreme radiation fields characterize reionization-era galaxies 
\citep{sobral15,stark15,stark16,mainali17,mainali18}.
Further, in the spectral energy distributions of $z>6$ galaxies, 
{\it Spitzer}/IRAC 3.6$\mu-4.5\mu$ photometry has revealed strong, excess emission 
attributed to nebular H$\beta+$[\ion{O}{3}] \W\W4959,5007 emission
(rest-frame EW(H$\beta+$[OIII])$\sim600-800$\AA; e.g., \citealt{labbe13,smit15,debarros19,endsley20}). 
These large optical and UV nebular emission equivalent widths (EWs) require small continuum
fluxes relative to the emission lines, which can result from large bursts of star formation.
Despite these considerable advances in characterizing reionization era galaxies,
the spatial and spectral limitations of observing faint, distant galaxies have
left the physical processes regulating this dynamic evolutionary phase poorly constrained. 

\subsection{Extreme Emission-Line Galaxies}
In order to characterize the most distant galaxies that the next generation of telescopes
will observe, an expanded framework of local galaxies encompassing more extreme properties
is needed.
In particular, it is important to understand the conditions that produce
similarly large emission-line EWs in star-forming galaxies as seen at high redshifts,
so-called {\it extreme emission line galaxies} (EELGs).
In the past few years, progress has been made by observational campaigns focused on 
EELGs at lower redshifts with very large optical emission-line EWs.
At $z\sim0-2$, studies of large samples of galaxies with large [\ion{O}{3}]$+$H$\beta$ EWs 
find that the extreme nebular emission is associated with a recent burst of star formation
in low-mass galaxies that results in highly-ionized gas 
\citep[e.g.,][]{atek11, vanderwel11,maseda13,maseda14,chevallard18,tang19}. 

Other studies have focused on EELGs with large EWs of UV emission lines.
For instance, some studies of lensed galaxies at $z\sim2-3$ measure strong nebular 
\ion{C}{4} \W\W1548,1550 and \ion{He}{2} \W1640 emission 
\citep[e.g.,][]{christensen12,stark14,vanzella16,vanzella17,schmidt17,smit17,berg18,mcgreer18},
while studies of nearby dwarf galaxies have empirically demonstrated that strong 
\ion{C}{4} \W\W1547,1550, \ion{He}{2} \W1640, and \ion{C}{3}] \W\W1907,1909 emission requires 
low metallicities (Z $<0.1$ Z$_\odot$) and young, large bursts of star formation
(as indicated by large [\ion{O}{3}] \W5007 EWs; e.g., 
\citealt{rigby15,berg16,senchyna17,senchyna19,berg19a,berg19b,tang20}.)

However, even amongst these EELG studies, it is difficult to find galaxies with UV 
emission comparable to that seen in reionization era systems. 
Recently, \citet{tang20} observed the rest-frame UV emission lines in a sample of $1.3<z<3.7$ galaxies
with high specific star formation rates (sSFRs), finding that only metal-poor emitters with intense 
H$\beta+$[\ion{O}{3}] \W5007 EWs $>1500$ \AA\ had \ion{C}{3}] emission strengths comparable to those
seen at $z>6$.
While previous UV studies of local, metal-poor galaxies have reported a handful of 
\ion{C}{3}] \W\W1907,1909 EWs $>15$ \AA, these observations lacked the coverage and 
resolution necessary for detailed nebular studies
(e.g., \citealt{berg16}: J082555, J104457; 
\citealt{berg19a}: J223831, J141851, J121402, J171236, J095430, J094718). 
Here, we study high-quality UV and optical spectra of two nearby EELGs 
with the largest reported \ion{C}{3}] \W\W1907,1909 EWs at $z\sim0$ to date.

\subsection{Two Nearby EELGs: J104457 and J141851}
J104457 ($10^{\mathrm{h}}44^{\mathrm{m}}57\fs790+03\degr53\arcmin13.\arcsec10$)
and J141851 ($14^{\mathrm{h}}18^{\mathrm{m}}51\fs119+21\degr02\arcmin39.\arcsec84$])
were originally selected for UV spectroscopic study based on their 
properties as derived from their optical Sloan Digital Sky Survey observations.
Specifically, J104457 and J141851 are nearby, compact, low-stellar-mass, metal-poor,
UV-bright galaxies with high specific star formation rates and significant high-ionization 
emission (EW [\ion{O}{3}] \W5007 $> 1000$ \AA; see Table~\ref{tbl1}).
These properties place J104457 and J141851 in the class of blue compact dwarf (BCD) galaxies, 
with intense starburst episodes on spatial scales of $\lesssim1$ kpc \citep[see, e.g.,][]{papaderos08}.
We use the detailed observations of these metal-poor galaxies as unique laboratories to 
investigate the nebular and stellar properties in nearly pristine conditions that are 
analogous to the early universe.

Here we present part {\sc i} of a detailed analysis of the UV {\it HST}/COS
G160M and optical LBT/MODS spectra of J104457 and J141851, focused on
the emission lines and nebular properties.
Part {\sc ii} will expand on this analysis by simultaneously modeling the 
ionizing stellar population and will be presented in G. Olivier et al.\ (2021, in preparation).

This paper is organized as follows.
We describe the UV and optical spectroscopic observations in Section~\ref{sec:2}.
In Section~\ref{sec:3.1}, we introduce a 4-zone nebular ionization model and calculate the 
subsequent physical properties: direct temperature and density measurements are presented 
in \S~\ref{sec:3.2.1}, followed by a discussion of their structure,
while the ionization structure is analyzed in \S~\ref{sec:3.2.2}.
We then determine nebular abundances, presenting O/H in \S~\ref{sec:4.1},
new ionization correction factors in \S~\ref{sec:4.2},
N/O in \S~\ref{sec:4.3}, C/O in \S~\ref{sec:4.4}, $\alpha$-elements/O in 
\S~\ref{sec:4.5}, and Fe/O in \S~\ref{sec:4.6}.
We discuss the physical properties of EELGs in Section~\ref{sec:5},
where we focus on the the resulting differences from using a 3-zone versus 4-zone 
ionization model in interpreting individual abundances in \S~\ref{sec:5.1}.
We introduce the high-energy ionizing photon production problem in
\S~\ref{sec:5.2} and then consider the overall abundance and ionization profiles 
of EELGs in \S~\ref{sec:5.3} and \S~\ref{sec:5.4}, respectively.
Finally, we make recommendations for interpreting the spectra of EELGs in
\S~\ref{sec:5.5} and summarize our findings in Section~\ref{sec:6}.\footnote{
A note about notation: We adopt standard ion and spectroscopic notation to describe 
the ionization states that give rise to different emission lines. 
In this manner, a given element, $X$, with $i$ ionizations is denoted as an $X^{+i}$ ion that 
can produce an emission line via radiative decay given by $X$ followed by the Roman numeral {$i+1$}
or via recombination given by $X$ followed by the Roman numeral {$i$}.
For example, the numeral {\sc i} is used to represent neutral elements, 
{\sc ii} to represent the first ionization state, {\sc iii} to represent 
the second ionization state, and so on.
Additionally, square brackets are used to denote forbidden transitions, whereas
semi-forbidden transitions use only the closing bracket and allowed transitions
do not use brackets at all.
For example, recombination of the He$^{+2}$ ion produces allowed \ion{He}{2} emission
and radiative decay of the collisionally-excited O$^{+2}$ ion produces
forbidden [\ion{O}{3}] emission.
We will use this ion and spectroscopic notation interchangeably throughout this work.}


\section{High S/N Spectral Observations}\label{sec:2}


\subsection{{\it HST}/COS FUV Spectra}
\begin{figure}
\begin{center}
    \includegraphics[width=1.0\columnwidth,trim=0mm 0mm 0mm 0mm,clip]{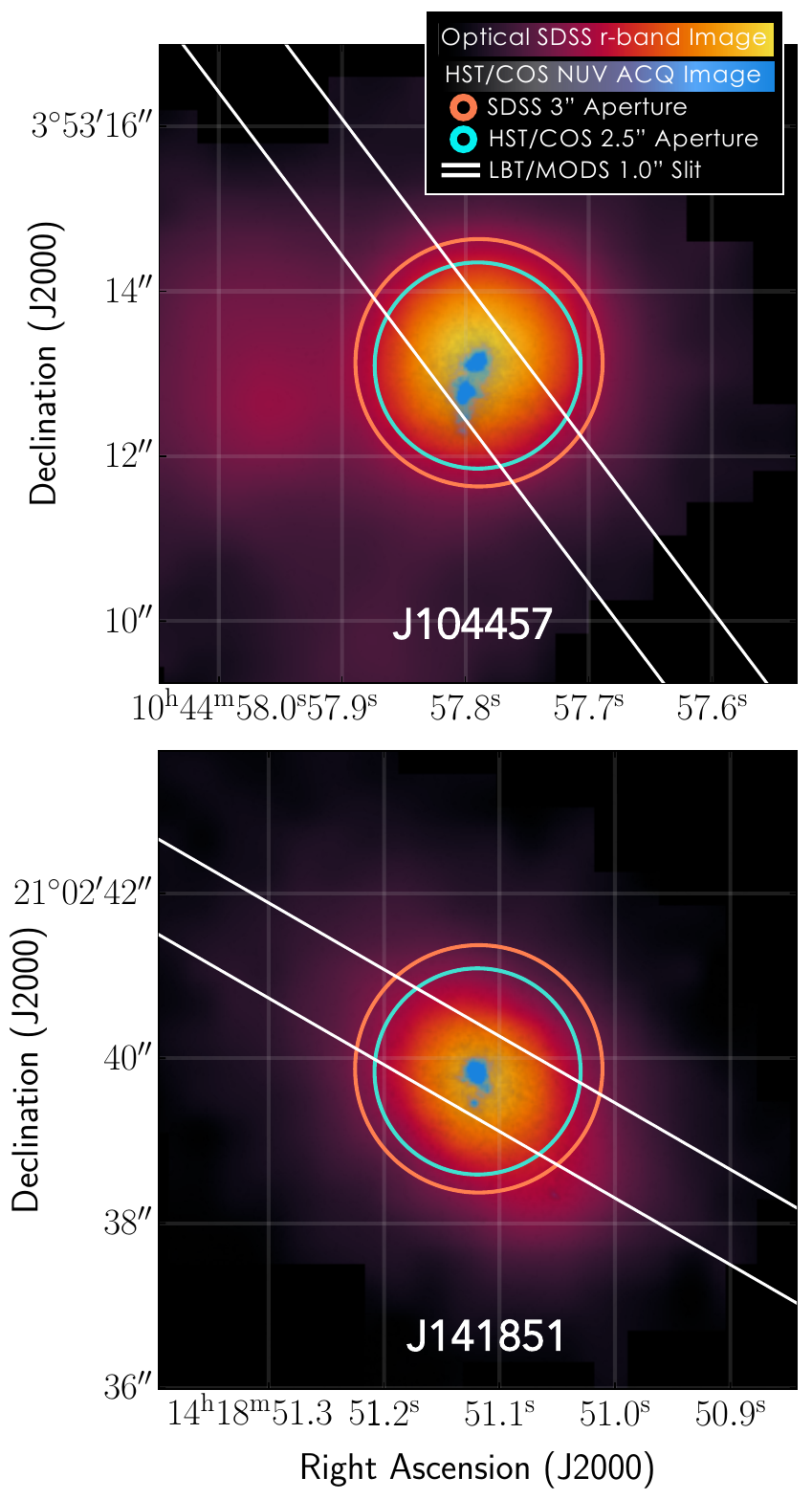}
\end{center}
\vspace{-2ex}
\caption{
The {\it HST}/COS NUV acquisition images of J104457 ({\it top}) and J141851 ({\it bottom}) are shown in blue, overlaid on top of the SDSS {\it r}-band image
(red-yellow color bar).
The 2.5\arcsec\ COS aperture used for the UV spectra is shown as a blue circle 
and is very similar to the SDSS 3\arcsec\ aperture in orange. 
In comparison, the 1\arcsec\ LBT/MODS slit (white lines) captures most of the NUV light, but
may miss a significant fraction of the extended nebular emission.
Note that J104457 has a fainter companion to the East (left) visible
in the optical, but this was not captured in the optical or UV spectra used here.}
\label{fig1}
\end{figure}

The high-resolution {\it HST}/COS G160M spectra for J104457 and J141851 were first presented in 
\citet{berg19b} to discuss their abnormally-strong C~\iv\ and He~\ii\ emission. 
We briefly summarize the observations here.
The {\it HST}/COS observations were observed by program HST-GO-15465 (PI: Berg). 
Utilizing the coordinates obtained through previous low-resolution COS G140L observations of these two targets,
target acquisitions were efficiently achieved using the IM/ACQ mode with the PSA
aperture and Mirror A.
The NUV acquisition images of J104457 and J141851 are shown in Figure~\ref{fig1},
demonstrating that they are high-surface-brightness, star-forming galaxies being 
dominated by just a few stellar clusters.

The COS FUV science observations were taken in the TIME-TAG mode using the 2.5\arcsec\ 
PSA aperture and the G160M grating at a central wavelength of 1589 \AA, for total 
exposures of 6439 and 12374s for J104457 and J141851, respectively. 
We used the FP-POS = ALL setting, which takes four images offset from one another in 
the dispersion direction, increasing the cumulative S/N and mitigating the effects 
of fixed pattern noise. 
Spectra were processed with CALCOS version 3.3.4.\footnote{\url{https://www.stsci.edu/hst/instrumentation/cos/documentation/calcos-release-notes}}

\begin{figure*}
\begin{center}
    \includegraphics[scale=0.25,trim=0mm 0mm 0mm 0mm,clip]{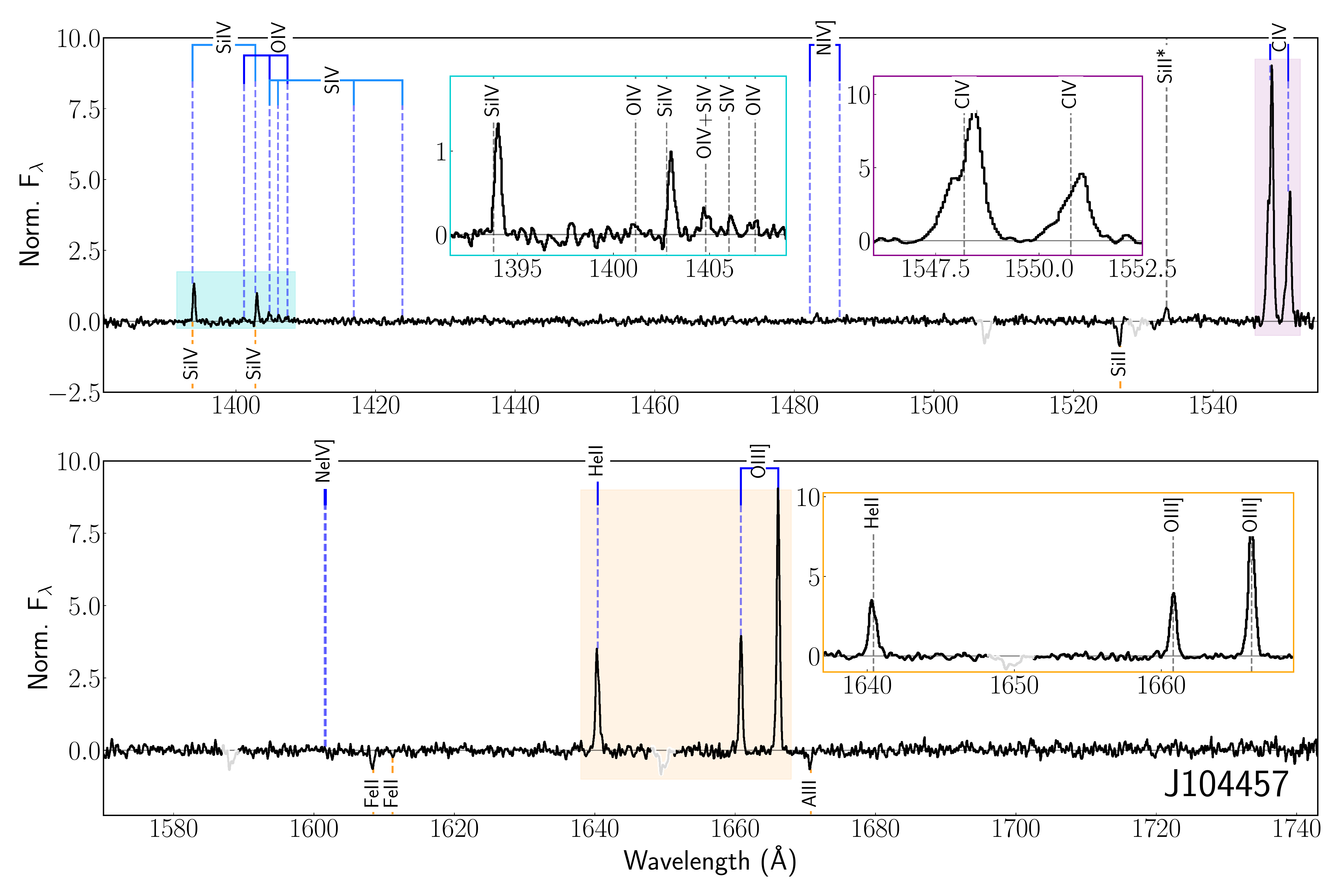} \\
    \includegraphics[scale=0.25,trim=0mm 0mm 0mm 0mm,clip]{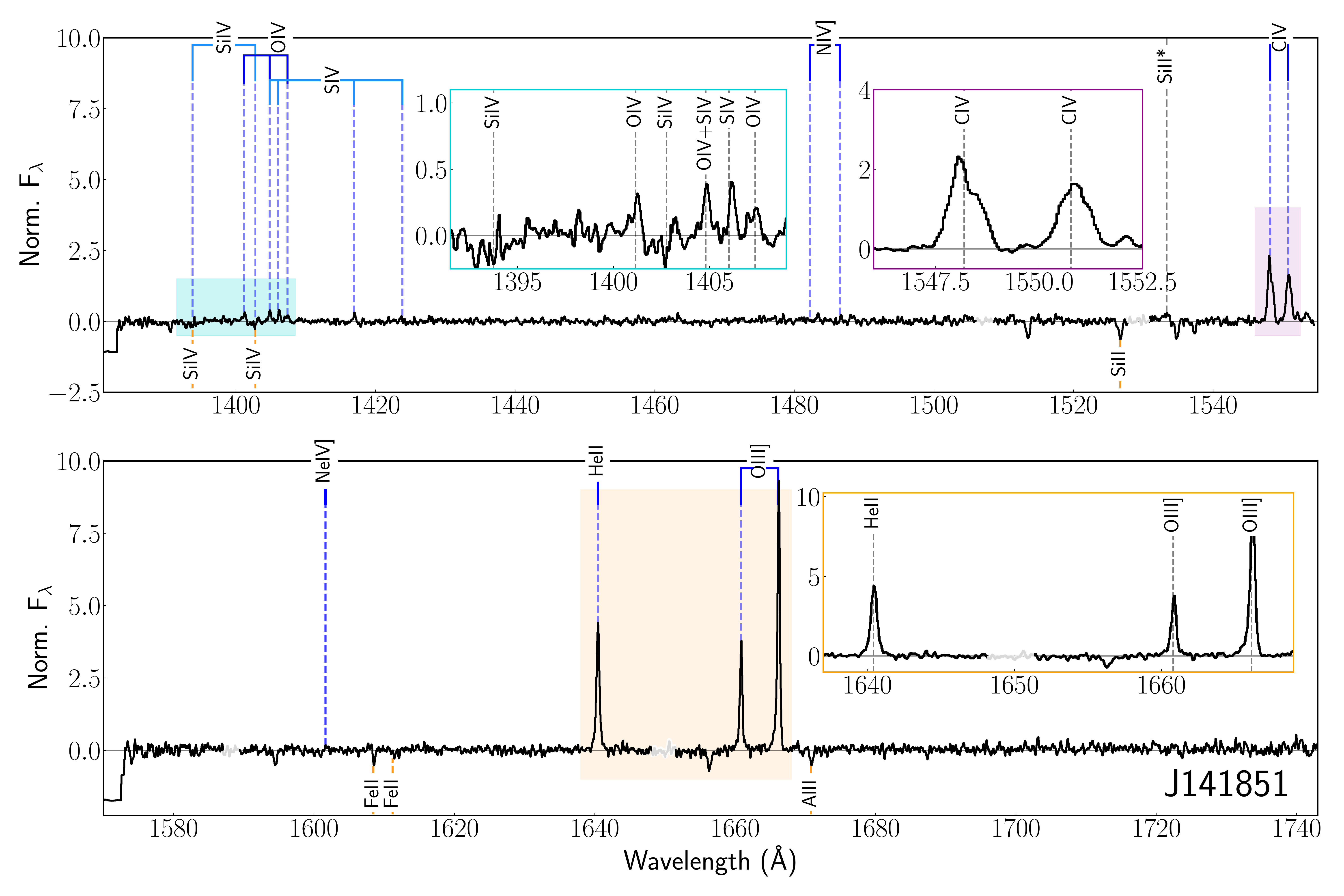}
\caption{
High-resolution {\it HST}/COS G160M UV spectra of J104457 ({\it top two panels}) and 
J141851 ({\it bottom two panels}).
Emission features are labeled at the top of each spectrum, whereas potential ISM absorption features
are labeled below and Milky Way lines are plotted in gray the strong, high-ionization emission 
lines that are characteristic of EELGs and are highlighted in the purple and orange
inset windows (i.e., \ion{C}{4}, \ion{He}{2}, and \ion{O}{3}]).
Plotting the full G160M spectra reveals additional high-ionization emission lines from
\ion{Si}{4}, \ion{O}{4}, and \ion{S}{4} as seen in the blue inset window.
These rare, extreme high-ionization emission lines support the hypothesis that these targets
are similar to reionization-era systems, producing copious amounts of very-high-energy ionizing photons.
}
\label{fig2}
\end{center}
\end{figure*}

In order to improve the signal-to-noise, we binned the spectra by 6 native COS pixels such 
that $\Delta v$ = 13.1 km s$^{-1}$, but the emission line FWHMs are still sampled by more 
than 4 pixels.
The resulting FUV spectra, shown in Figure~\ref{fig2}, have wavelength coverage that is 
rich in nebular features not found in the optical.


\begin{deluxetable}{lcc}
\tablecaption{Extreme Emission-Line Galaxy Properties}
\tablehead{
\multicolumn{1}{l}{Property}    & \CH{J104457} 		& \CH{J141851}  }
\startdata	
\multicolumn{3}{c}{\bf Adopted from Archival Sources:} \\
Reference			            & Berg+16           & Berg+19a          \\
R.A. (J2000)		            & \ \ 10:44:57.79	& \ 14:18:51.13	    \\
Decl. (J2000)		            & $+$03:53:13.15	& $+$21:02:39.74	\\
$z$					            & 0.013	            & 0.009         	\\
log $M_\star$ (M$_\odot$)       & 6.80		        & 6.63          	\\
log SFR	(M$_\odot$ yr$^{-1}$)   & $-0.85$		    & $-1.16$       	\\
log sSFR (yr$^{-1}$)		    & $-7.64$		    & $-7.79$         	\\
$E_(B-V)$ (mag)				    & 0.077				& 0.140		    	\\
$12+$log(O/H) (dex ($Z_\odot$)) & $7.45$ ($0.058$) 	& $7.54$ ($0.071$)	\\
log $U$						    & $-1.77$			& $-2.42$			\\
\\[-1ex]
\multicolumn{3}{c}{\bf Derived from the UV COS G160M Spectra:} \\
EW$_{\rm OIII]}$ (\AA)		    & $-2.89,-6.17$				& $-1.68,-4.78$		\\
EW$_{\rm CIV}$ (\AA)		    & $-6.71,-2.83$				& $-1.78,-1.43$		\\
EW$_{\rm HeII}$(\AA)		    & $-2.75$					& $-2.82$			\\
EW$_{\rm CIII]}$(\AA)		    & $-16.35$				    & $-18.41$			
\enddata	
\tablecomments{ 
Properties of the extreme emission-line galaxies presented here.
The top portion of the table lists properties previously reported by
\citet{berg16} for J104457 and \citet{berg19a} for J141851.
The R.A., Decl., redshift, total stellar masses, SFRs, and sSFRs were adopted from the 
SDSS MPA-JHU DR8 catalog$^a$,
whereas $E(B-V)$, $12+$log(O/H), and log $U$, were measured from the SDSS optical spectra.
The bottom portion of the table lists the properties derived from the UV
HST/COS G160M spectra.
Equivalent widths are listed for \ion{C}{4} \W\W1548,1550, \ion{O}{3}] \W\W1661,66, 
\ion{He}{2} \W1640, and \ion{C}{3}] \W\W1907,09. \\
$^a$ Data catalogues are available from \url{http://www.sdss3.org/dr10/spectro/galaxy_mpajhu.php}.
The Max Plank institute for Astrophysics/John Hopkins University(MPA/JHU) SDSS data base was produced by a 
collaboration of researchers(currently or formerly) from the MPA and the JHU. 
The team is made up of Stephane Charlot (IAP), Guinevere Kauffmann and Simon White (MPA),
Tim Heckman (JHU), Christy Tremonti (U. Wisconsin-Madison $-$ formerly JHU) and Jarle 
Brinchmann (Leiden University $-$ formerly MPA).
\\[-8ex]}
\label{tbl1}
\end{deluxetable}
   

\subsection{LBT/MODS Optical Spectra}
We obtained optical spectra of J104457 and J141851 using the Multi-Object 
Double Spectrographs \citep[MODS,][]{pogge10} on the Large Binocular Telescope \citep[LBT,][]{hill10}
on the UT dates of 2018 May 19 and 18, respectively.
The conditions were clear, with good seeing ($\sim0.7\arcsec$ for J104457 and
$\sim0.8\arcsec$ for J141851) and low variability ($<0.1\arcsec$ over the total
science integrations).
MODS is a moderate-resolution ($R\sim2000$) optical spectrograph with large wavelength
coverage ($3200$\AA$\lesssim \lambda \lesssim 10000$\AA). 
Simultaneous blue and red spectra were obtained using the 
G400L (400 lines mm$^{-1}$, R$\approx1850$) and 
G670L (250 lines mm$^{-1}$, R$\approx2300$) gratings, respectively.
J104457 and J141851 were observed using the 1\arcsec$\times$60\arcsec\ longslit for
3$\times$900s exposures, or 45 min of total exposure per object.
The slits were centered on the highest surface brightness knot of optical emission, as 
determined from the Sloan Digital Sky Survey (SDSS) {\it r}-band image.
and oriented to the parallactic angle at 
half the total integration time.
Both targets were observed at airmasses of less than 1.2, which served to 
minimize flux losses to differential atmospheric refraction \citep{filippenko82}.
The slit orientations of the MODS observations are shown relative to the {\it HST}/COS NUV acquisition 
images in Figure~\ref{fig1}, 
demonstrating that the peak of the optical and UV surface brightness profiles
are aligned and 
that most of the stellar light is captured within the slit.
However, in comparison to the extended nebular emission that is contained within the 2.5\arcsec\ 
COS aperture, the MODS observations may suffer from significant loses of extended emission. 

\begin{figure*}
\begin{center}
    \includegraphics[scale=0.21,trim=20mm 25mm 20mm 40mm,clip]{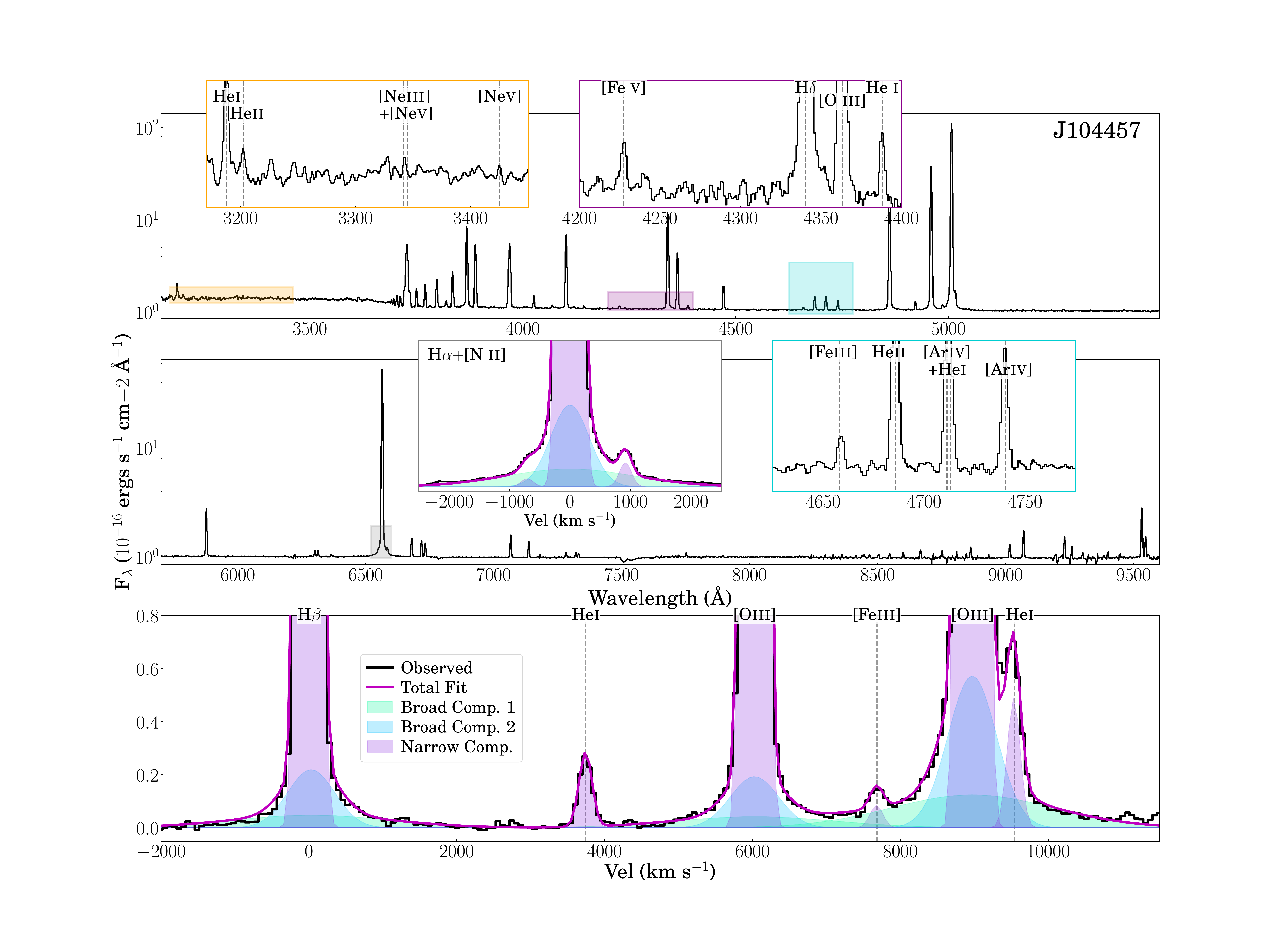}
    \includegraphics[scale=0.21,trim=20mm 30mm 20mm 30mm,clip]{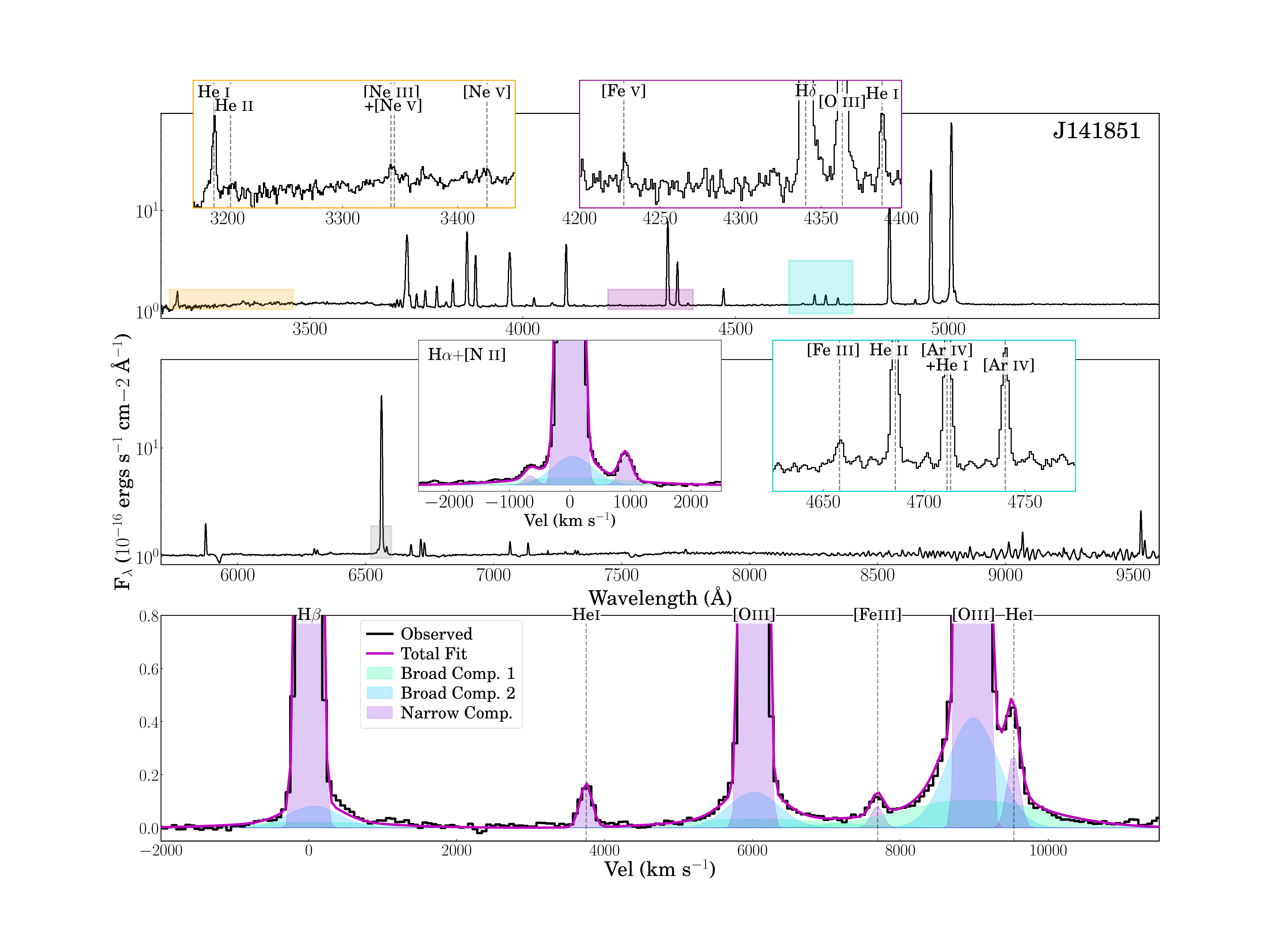}
\caption{
Optical MODS/LBT spectra of J104457 ({\it top three panels}) and J141851 ({\it bottom three panels}).
With expansive wavelength coverage from roughly 3,200--10,000 \AA, 
these spectra feature numerous $T_e$ and $n_e$ diagnostic emission lines that 
span a wide range of ionization zones.
Additionally, the colored inset windows highlight emission from very-high-ionization species:
the yellow and blue windows highlight the rare [\ion{Ne}{3}] \W3342 $T_e$--sensitive 
and [\ion{Ar}{4}] \W\W4711,4740 $n_e$--sensitive features that can be used to characterize the 
properties of the very-high-ionization zone, while the purple inset window highlights 
the narrow, rare [\ion{Fe}{5}] emission.
For the strongest emission lines, we also see weak broad emission.
The grey inset windows reveal broad emission at the base of the H$\alpha$ emission,
suggesting an additional source of energy such as from shocks.
The bottom panels show fits that are scaled from the multi-component H$\alpha$ fits.}
\label{fig3}
\end{center}
\end{figure*}

Spectra were reduced, extracted, analyzed using the beta-version of the MODS reduction 
pipeline\footnote{\url{http://www.astronomy.ohio-state.edu/MODS/Software/modsIDL/}} 
which runs within the XIDL \footnote{\url{http://www.ucolick.org/~xavier/IDL/}} 
reduction package.  
One-dimensional spectra were corrected for atmospheric extinction and flux calibrated 
based on observations of flux standard stars \citep{oke90}. 
The details of the MODS reduction pipeline are further described by 
\citet{berg15}; while that work analyzes multi-object multiplexed spectra,  
the major steps are identical to that of the present longslit reduction.
The resulting optical spectra are shown in Figure~\ref{fig3}.


\subsection{Emission Line Measurements}
For this work, we measured all of the emission line fluxes in a consistent manner when possible.
For the optical LBT/MODS spectra, we used the continuum modeling and line fitting
code developed as part of the CHAOS project \citep{berg15}.  
First, the underlying continuum of the optical spectra were fit by the 
{\sc starlight}\footnote{\url{www.starlight.ufsc.br}} spectral synthesis code 
\citep{fernandes05} using stellar models from \citet{bruzual03}. 
Next, emission lines were fit in the continuum-subtracted spectrum with Gaussian
profiles and allowing for an additional nebular continuum component.
The fit parameters (i.e., width, line center) of neighboring lines were constrained, 
allowing weak or blended features to be measured simultaneously. 
For both J104457 and J141581, we measure Gaussian full width half maximum (GFWHM)
values of roughly 3.5 \AA\ and 5.5 \AA\ in the blue and red spectra, respectively. 
With a linear spectral dispersion of 0.5 \AA/pix and 0.8 \AA/pix for the red and blue, respectively,
our narrow lines are well sampled.
However, our GFWHMs correspond to $\Delta v_{blue}\approx200$ km s$^{-1}$
at \W5007 and $\Delta v_{red}\approx250$ km s$^{-1}$ at \W6563, and, as a 
result, these spectra are not sensitive to broad components of comparable 
velocity width. 

Broad components can occur in the presence of stellar 
winds or shocks. 
In the case of the \ion{He}{2} \W1640 and \W4686 lines, we allow for additional broad components
but do not find evidence for any;
with velocity widths consistent with the other emission lines,
this suggests that the \ion{He}{2} emission is nebular in origin.
However, for the strongest optical emission lines (H$\beta$, [\ion{O}{3}] \W\W4959,5007, and H$\alpha$), 
we do measure multi-component fits, as shown in Figure~\ref{fig3}. 
These lines all have (1) strong, narrow nebular components, (2) moderate, broad components, and 
(3) weak, very broad components with similar velocity profiles, but where the scaled broad components 
of the H$\alpha$ and H$\beta$ features are stronger relative to those of the [\ion{O}{3}] emission lines.
This could suggest that the broad emission is more strongly produced in the low-ionization gas than the high-ionization gas.
Another possible explanation is that the broad emission is coming from
higher-density regions that suppress forbidden emission. 


\startlongtable

\begin{deluxetable}{lcc}
\tablewidth{\textwidth}
\setlength{\tabcolsep}{8pt}
\tabletypesize{\scriptsize}
\tablecaption{UV$+$Optical Emission-Lines from {\it HST}/COS and LBT/MODS Observations}
\tablehead{{Ion$+$Wavelength}       & {J104457}         & {J141851} \vspace{+0.05cm}}
\startdata
\multicolumn{3}{c}{\bf UV Lines: $I(\lambda)/I$(\ion{O}{3}])} \vspace{+0.05cm} \\
\hline \vspace{-0.15cm} \\
Si~\iv\ \W1393.76       & 10.4$\pm$2.5          & \nodata               \\
O~\iv\ \W1401.16        & 2.2$\pm$2.2           & 3.6$\pm$3.0           \\
Si~\iv\ \W1402.77       & 6.7$\pm$2.4           & \nodata               \\
O~\iv+S\iv\ \W1404.81   & 3.5$\pm$2.4           & 4.9$\pm$3.0           \\
S~\iv\ \W1406.02        & 2.9$\pm$2.4           & 3.3$\pm$3.0           \\
O~\iv\ \W1407.38        & 0.7$\pm$2.4           & 1.6$\pm$3.0           \\
S~\iv\ \W1416.89        & $<$2.7                & 1.8$\pm$3.0           \\
S~\iv\ \W1423.85        & $<$3.3                & 2.1$\pm$3.0           \\
N~\iv] \W1483.33        & 2.7$\pm$3.3           & 2.0$\pm$3.0           \\
N~\iv] \W1486.50        & 1.3$\pm$3.3           & 1.6$\pm$3.0           \\
Si\ii* \W1533.43        & 6.8$\pm$3.9           & 6.0$\pm$3.8           \\
C~\iv\ \W1548.19        & 149.0$\pm$9.6         & 39.5$\pm$4.3          \\
C~\iv\ \W1550.77        & 71.7$\pm$5.8          & 30.5$\pm$4.1          \\
He~\ii\ \W1640.42       & 42.8$\pm$6.3          & 58.8$\pm$5.8          \\
O~\iii] \W1660.81       & 45.7$\pm$6.4          & 40.2$\pm$5.3          \\
O~\iii] \W1666.15       & 100.0$\pm$8.2         & 100.0$\pm$7.1         \\
{\it Si~\iii] \W1883.00}& {\it 43.9$\pm$14.3}   & {\it 42.0$\pm$13.1}   \\
{\it Si~\iii] \W1892.03}& {\it 36.7$\pm$2.8}    & {\it 29.9$\pm$13.0}   \\
{\it C~\iii] \W1906.68} & {\it 135.8$\pm$18.4}  & {\it 162.2$\pm$9.7}   \\
{\it [C~\iii] \W1908.73}& {\it 106.7$\pm$17.7}  & {\it 110.0$\pm$6.6}
\vspace{+0.05cm} \\
\hline \vspace{-0.2cm} \\
E(B$-$V)$_{R16}$        & 0.086$\pm$0.042       & 0.036$\pm$0.050       \\
F$_{\rm O~III]}$        & 523.9$\pm$30.5        & 275.7$\pm$13.8        \\ 
\vspace{-0.1cm} \\
\hline\hline \vspace{-0.15cm} \\
\multicolumn{3}{c}{\bf Optical Lines: $I(\lambda)/I(\mbox{H}\beta)$} \vspace{+0.05cm} \\
\hline \vspace{-0.15cm} \\
He~{\sc i} \W3188.75     & 2.86$\pm$0.11    & 3.85$\pm$0.17     \\
He~\ii\ \W3203.00        & 0.60$\pm$0.10    & 0.47$\pm$0.16     \\
{[Ne~\iii] \W3342.18}    & 0.33$\pm$0.10    & 0.59$\pm$0.16     \\
{[Ne~{\sc v}] \W3345.82} & $<0.06$          & 0.64$\pm$0.16     \\
{[Ne~{\sc v}] \W3425.88} & 0.11$\pm$0.10    & 0.60$\pm$0.16     \\
{[O~\ii] \W3726.04}      & 9.60$\pm$0.15    & 14.43$\pm$0.27    \\
{[O~\ii] \W3728.80}      & 16.77$\pm$0.23   & 34.91$\pm$0.43    \\
He~{\sc i} \W3819.61     & 0.92$\pm$0.02    & 1.12$\pm$0.03     \\
H9 \W3835.39             & 7.19$\pm$0.10    & 7.24$\pm$0.11     \\
{[Ne~\iii] \W3868.76}    & 31.73$\pm$0.45   & 37.64$\pm$0.51    \\
He~{\sc i} \W3888.65     & 18.82$\pm$0.26   & 1.96$\pm$0.04     \\
H8 \W3889.06             & 0.82$\pm$0.13    & 17.10$\pm$0.20    \\
He~{\sc i} \W3964.73     & 6.89$\pm$0.10    & 3.15$\pm$0.08     \\
{[Ne\iii] \W3967.47}     & 13.41$\pm$0.18   & 10.62$\pm$0.17    \\
H7 \W3970.08             & 5.47$\pm$0.20    & 13.80$\pm$0.21    \\
He~{\sc i} \W4026.19     & 1.72$\pm$0.03    & 1.86$\pm$0.04     \\
{[S~\ii] \W4068.60}      & 0.38$\pm$0.01    & 0.71$\pm$0.05     \\
{[S~\ii] \W4076.35}      & 0.15$\pm$0.01    & 0.28$\pm$0.03     \\
H$\delta$ \W4101.71      & 25.59$\pm$0.42   & 25.76$\pm$0.39    \\
He~{\sc i} \W4120.81     & 0.26$\pm$0.02    & 0.30$\pm$0.02     \\
He~{\sc i} \W4143.15     & 0.25$\pm$0.09    & 0.30$\pm$0.16     \\
{[Fe~{\sc v}] \W4227.19} & 0.03$\pm$0.01    & 0.15$\pm$0.02     \\
H$\gamma$ \W4340.44      & 46.55$\pm$0.67   & 47.31$\pm$0.68    \\
{[O\iii] \W4363.21}      & 13.51$\pm$0.21   & 13.80$\pm$0.23    \\
He~{\sc i} \W4387.93     & 0.47$\pm$0.01    & 0.49$\pm$0.01     \\
He~{\sc i} \W4471.48     & 3.97$\pm$0.09    & 7.08$\pm$0.16     \\
{[Fe~\iii] \W4658.50}    & 0.34$\pm$0.03    & 0.43$\pm$0.02     \\
He~\ii\ \W4685.70        & 1.80$\pm$0.04    & 2.15$\pm$0.03     \\
{[Ar~\iv] \W4711.37$^a$} & 1.65$\pm$0.06    & 3.21$\pm$0.08     \\
He~{\sc i} \W4713.14$^b$ & 0.32$\pm$0.01    & 0.57$\pm$0.02     \\
{[Ar~\iv] \W4740.16}     & 1.16$\pm$0.02    & 2.51$\pm$0.06     \\
H$\beta$ \W4861.35$^c$   & 100.0$\pm$1.4  & 100.0$\pm$1.4   \\
{[Fe~\iv] \W4906.56}     & 0.08$\pm$0.06    & $< 0.03$          \\
He~{\sc i} \W4921.93     & 1.05$\pm$0.19    & 0.99$\pm$0.29     \\
{[O~\iii] \W4958.91}$^c$ & 143.0$\pm$1.5  & 162.5$\pm$1.7   \\
{[Fe~\iii] \W4985.87}$^c$& 0.04$\pm$0.09    & 0.85$\pm$0.15     \\
{[O~\iii] \W5006.84}$^c$ & 427.5$\pm$4.3  & 500.7$\pm$5.0   \\
He~{\sc i} \W5015.68$^c$ & 1.82$\pm$0.05    & 1.65$\pm$0.12     \\
{[Fe~\ii] \W5158.79}     & 0.08$\pm$0.09    & $< 0.03$          \\
{[Fe~\iii] \W5270.40}    & 0.15$\pm$0.09    & $< 0.03$          \\
{[N~\ii] \W5754.59}      & $< 0.09$         & $< 0.33$          \\
He~{\sc i} \W5875.62     & 10.17$\pm$0.15   & 9.96$\pm$0.18     \\
{[O~{\sc i}] \W6300.30}  & 0.86$\pm$0.05    & 1.50$\pm$0.08     \\
{[S~\iii] \W6312.06}     & 0.80$\pm$0.03    & 1.21$\pm$0.11     \\
{[O~{\sc i}] \W6363.78}  & 0.25$\pm$0.03    & 0.66$\pm$0.05     \\
{[N~\ii] \W6548.05}$^c$  & 0.27$\pm$0.05    & 0.56$\pm$0.05     \\
H$\alpha$ \W6562.79$^c$  & 296.7$\pm$4.2  & 275.8$\pm$3.9   \\
{[N~\ii] \W6583.45}$^c$  & 0.81$\pm$0.05    & 1.68$\pm$0.05     \\
He~{\sc i} \W6678.15     & 2.85$\pm$0.05    & 2.63$\pm$0.06     \\
{[S~\ii] \W6716.44}      & 2.50$\pm$0.04    & 4.04$\pm$0.13     \\
{[S~\ii] \W6730.82}      & 2.04$\pm$0.05    & 2.99$\pm$0.06     \\
He~{\sc i} \W7065.19     & 3.57$\pm$0.05    & 3.36$\pm$0.07     \\
{[Ar~\iii] \W7135.80}    & 2.42$\pm$0.09    & 2.98$\pm$0.11     \\
{[O~\ii] \W7319.92}      & 0.60$\pm$0.02    & 0.80$\pm$0.05     \\
{[O~\ii] \W7330.19}      & 0.48$\pm$0.07    & 0.71$\pm$0.06     \\
{[Ar~\iii] \W7751.06}    & 0.45$\pm$0.03    & 1.03$\pm$0.05     \\
P13 \W8665.02            & 1.05$\pm$0.03    & 2.18$\pm$0.07     \\
P12 \W8750.46            & 0.93$\pm$0.05    & 0.78$\pm$0.05     \\
P11 \W8862.89            & 1.44$\pm$0.07    & 2.34$\pm$0.07     \\
P10 \W9015.30            & 1.80$\pm$0.05    & 3.46$\pm$0.10     \\
{[S~\iii] \W9068.60}     & 4.24$\pm$0.07    & 6.18$\pm$0.10     \\
P9 \W9229.70             & 3.06$\pm$0.06    & 1.58$\pm$0.09     \\
{[S~\iii] \W9530.60}     & 10.10$\pm$0.14   & 15.39$\pm$0.23 
\vspace{+0.05cm}                                                \\
\hline \vspace{-0.2cm}                                          \\
E(B$-$V)                & 0.039$\pm$0.017   & 0.019$\pm$0.016   \\
F$_{H\beta}$            & 95.2$\pm$1.0    & 55.7$\pm$0.6 
\enddata
\tablecomments{
Reddening-corrected emission-line intensities from the high-resolution UV 
{\it HST}/COS G160M spectra and optical LBT/MODS spectra for J104457 and J141851.
The \ion{Si}{3}] and \ion{C}{3}] lines (italicized) 
are exceptions and are from the low-resolution {\it HST}/COS G140L spectra.
Fluxes for undetected lines are given as $<$ their 3$\sigma$ upper-limits.
The UV fluxes have been modified to a common scale and are given relative
to the \ion{O}{3}] \W1666 flux, multiplied $\times100$, from the G160M spectra.
The optical fluxes are given relative to H$\beta$$\times$100.
The last two rows below the UV lines list the dust extinction derived using the \citet{reddy16} 
reddening law and the raw, observed fluxes for \ion{O}{3}] \W1666, in units of 
$10^{-16}$ erg s$^{-1}$ cm$^{-2}$.
The last two rows below the optical lines list the dust extinction using the \citet{cardelli89} reddening law
and the raw, observed fluxes for H$\beta$, in units of $10^{-16}$ erg s$^{-1}$ cm$^{-2}$.
Details of the spectral reduction and line measurements are given in Section~\ref{sec:2}.  \\
$^a$ At the spectral resolution of the LBT/MODS spectra, we observe 
[Ar~\iv] \W4711$+$He~{\sc i} \W4713 as a blended line profile.
Therefore, the predicted He~{\sc i} \W4713 flux is subtracted to determine the 
residual [Ar~\iv] \W4711 flux. \\
$^b$ The He~{\sc i} \W4713 flux was predicted from the observed He~{\sc i} \W4471 flux 
and their relative emissivities, as determined by {\sc PyNeb}.
$^c$ These line fluxes were corrected for the additional broad emission components
seen in Figure~\ref{fig3}. Only the narrow components are listed here.}
\label{tbl2}
\end{deluxetable}
\clearpage


Interestingly, the two broad components of both galaxies 
have large velocity widths of roughly 2500 and 750 km/s.
These are especially large velocities compared to the small circular velocities of these galaxies
($v_{circ}=$ 16.6 and 12.1 km/s for J104457 and J141851, respectively, 
derived using the equation from \citealt{reyes11}) and the
lack of outflows measured from the UV absorption line spectra \citep[see Figure 3 in][]{berg19b}.
However, each broad component only accounts for 1--3\%\ of the total H$\alpha$ flux.
Broad component emission for H$\beta$, [\ion{O}{3}], and H$\alpha$ are often seen in the spectra of 
blue compact dwarf galaxies \citep[BCDs; e.g.,][]{izotov06,izotov07} with similar widths (1000--2000 km/s)
and fractional fluxes of 1--2\%.
While broad emission represents a small fraction of the H$\beta$ and H$\alpha$ fluxes, 
it can significantly affect the fit to the weak [\ion{N}{2}] \W\W6548,6584 lines. 
For this reason we adopt the narrow-line fluxes from our multi-component fits for our analysis, 
but do not investigate the the broad emission further. 

For the UV {\it HST}/COS spectra, no continuum model was used. 
Similar to the optical lines, we measured the nebular emission line strengths 
using constrained Gaussian profiles. 
In addition to the \ion{C}{4}, \ion{He}{2}, \ion{O}{3}], \ion{Si}{3}], and \ion{C}{3}] 
features that were previously detected in the low-resolution G140L spectra, 
we identify and measure emission from \ion{Si}{4}, \ion{O}{4}, \ion{S}{4},
and \ion{Si}{2}* features in the G160M spectra (see Figure~\ref{fig2}).

Flux measurements for both the UV and optical lines were corrected for Galactic 
extinction using the {\sc python dustmaps} interface \citep{green18} to query 
the \citet{green15} extinction map, with a \citet{cardelli89} reddening law.
Then, the relative intensities of the four strongest Balmer lines 
(H$\alpha$/H$\beta$, H$\gamma$/H$\beta$, H$\delta$//H$\beta$) were used to determine 
the dust reddening values, $E(B-V)$, for both the \citet{cardelli89} and \citet{reddy16} laws.
Finally, these $E(B-V)$ values were used to reddening-correct the other emission lines, assuming a 
\citet{cardelli89} extinction law in the optical and a \citet{reddy16} law in the UV.
The uncertainty measured for each line is a combination of the 
spectral variance, flux calibration uncertainty, Poisson noise, read noise, sky noise, 
flat fielding uncertainty, and uncertainty in the reddening determination.

We note that the significant detections of the \ion{O}{3}] \W1661,1666 emission doublet in both the 
G140L and G160M allowed us to place our relative emission line measurements on a common scale.
We, therefore, scaled the \ion{Si}{3}] and \ion{C}{3}] emission line fluxes from the low-resolution
G140L spectra by the \ion{O}{3}] $I_{\lambda1666,{\rm G160M}}$/$I_{\lambda1666,{\rm G140L}}$ flux ratio
and included them in our subsequent analysis of the G160M emission lines.
The reddening-corrected, scaled emission line intensities measured for the UV {\it HST}/COS and
optical LBT/MODS spectra of J104457 and J141851 are reported in Table~\ref{tbl2}, respectively.


\subsection{The Effect of Aperture on Relative Flux}
In order to utilize our high S/N {\it HST}/COS UV spectra and {\it LBT}/MODS optical
spectra together, we must consider the effect of aperture losses in the 1\arcsec\
MODS longslit versus the COS 2.\arcsec5 aperture. 
Such a comparison can be done using the {\it LBT}/MODS optical spectra for J104457 and J141851 
and their corresponding SDSS optical spectra, which were observed with an aperture (3\arcsec)
that is similar to that of COS. 
To do so, we normalized each spectrum by its average continuum flux in the relatively featureless
wavelength regime of 4500--4600 \AA\ in order to compare the relative line strengths of interest.
We find that the percent differences in emission line fluxes 
are typically $<$ 5\%, suggesting that an aperture correction is not required.
Further, these differences are not systematic, precluding an accurate aperture correction 
unless the exact 2D ionization structure can be determined. 

Interestingly, the differences between low-ionization and high-ionization species are similarly small 
for J104457, however, the differences are larger for the low-ionization species than the high-ionization
species in the J141851 spectra. 
This situation would naturally result from the aperture differences given a simple \ion{H}{2} 
region structure with the high-ionization region concentrated in the center and the low-ionization 
region being more extended.
Additionally, we compared the same temperature and density measurements we describe in 
\S~\ref{sec:3.2.1} for the {\it LBT}/MODS spectra and SDSS spectra, but find that the 
results agree within the uncertainties.
We, therefore, do not apply any aperture corrections to the MODS optical spectra
and do not find any strong evidence that this will affect comparisons between the UV and optical data. 
However, it is important to note that this result is only true for the relative flux comparisons;
the absolute flux correction of SDSS/MODS is roughly a factor of 51 for J104457 and
47 for J141851. 


\begin{figure*}
\begin{center}
    \includegraphics[scale=0.475,trim=0mm 0mm 0mm 0mm,clip]{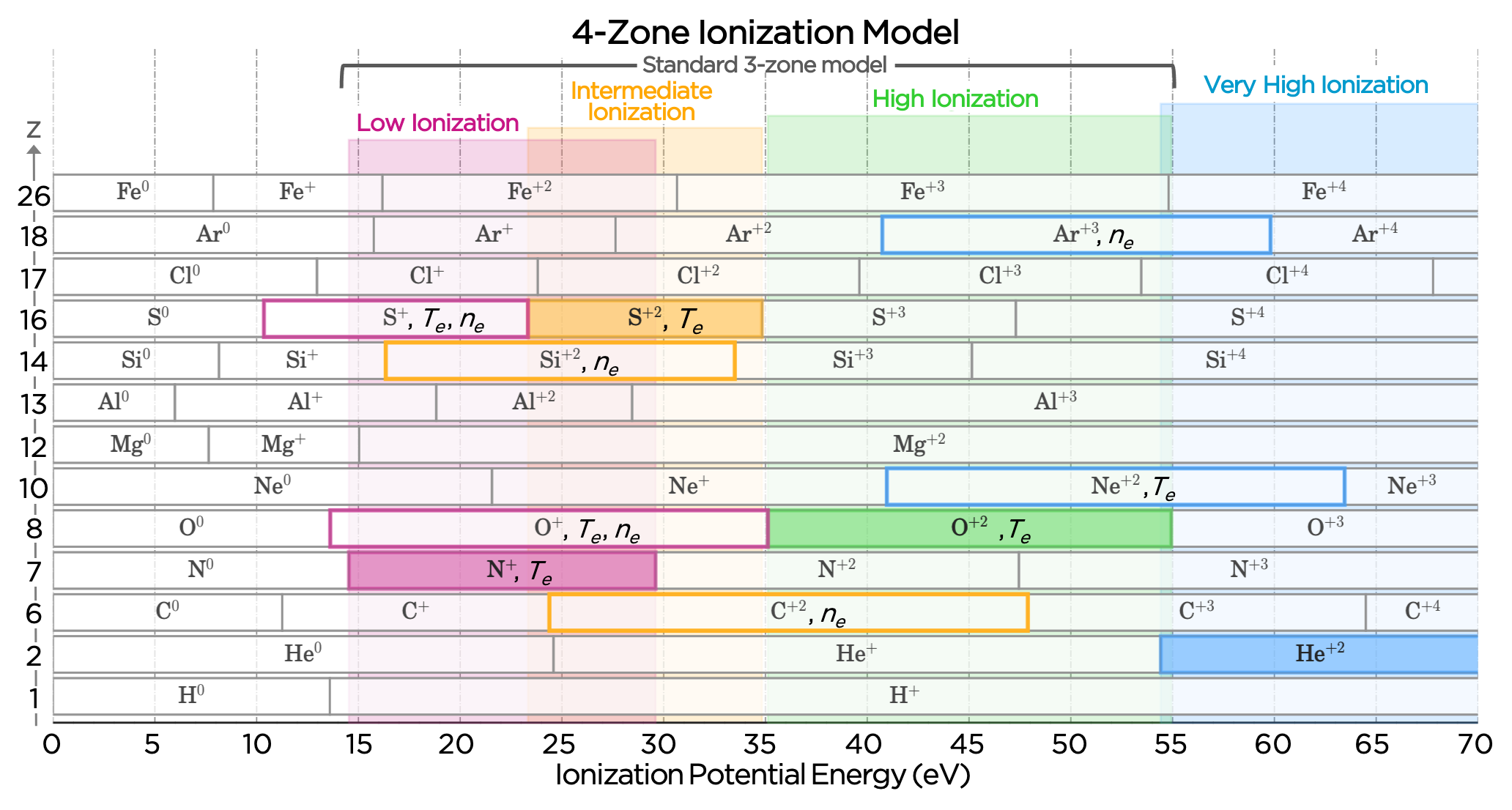} 
    \includegraphics[scale=0.475,trim=0mm 0mm 0mm 0mm,clip]{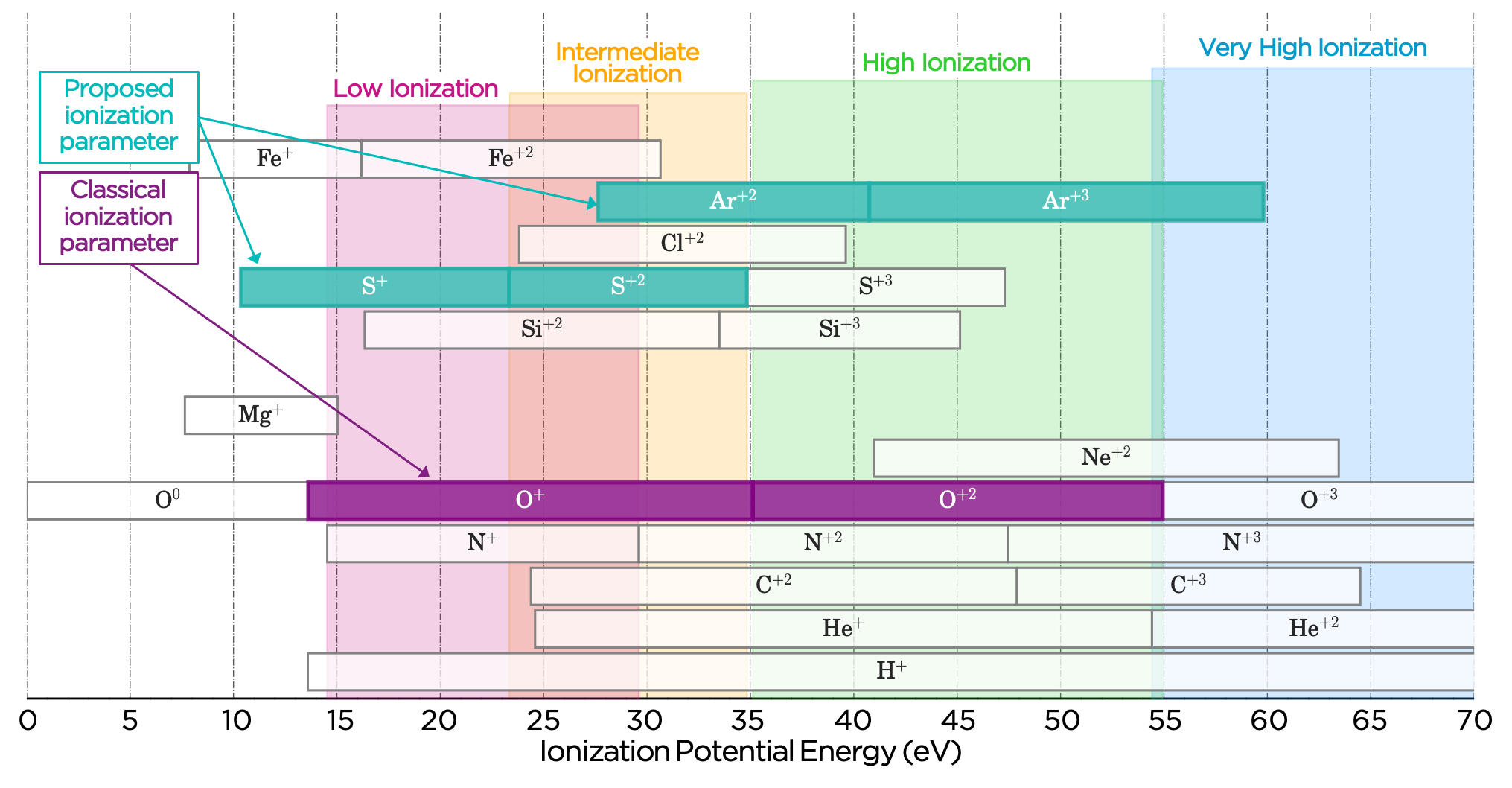}
\caption{
{\it Top:} 
The ionization potential energies (eV) of important ISM species,
ordered vertically by their atomic numbers (Z).
The figure is divided horizontally into different gas ionization zones, 
where the classical 3-zone model for star-forming regions defines
low-ionization by the N$^+$ species (pink shaded box), 
intermediate-ionization by the S$^{+2}$ species (yellow shaded box), and 
high-ionization by the O$^{+2}$ species (green shaded box).
However, observations of J104457, J141851, and other EELGs probe an extended range 
of gas ionization phase that are capable of, for instance, emitting nebular 
O$^{+3}$ (\ion{O}{4} \W\W1401,1404,1407), 
C$^{+3}$ (\ion{C}{4} \W\W1548,1550), 
He$^{+2}$ (\ion{He}{2} \W1640), 
N$^{+3}$ (\ion{N}{4}] \W\W1483,1487, and 
Ar$^{+3}$ ([\ion{Ar}{4}] \W\W4711,4740).
We, therefore, attempt to better characterize the very-high-ionization nebulae 
of EELGs by establishing a 4-zone ionization model with the addition of a very 
high-ionization zone defined by the He$^{+2}$ species (blue shaded box).
Additional $T_e$ and $n_e$ diagnostic species are denoted by a color-coded
outline for each ionization zone.
{\it Bottom:} 
Ionization potential energies, same as the top plot, but limited to species 
that are commonly observed in UV and optical spectra of \ion{H}{2} regions in EELGs.
The purple shaded boxes highlight the components of the O$^{+2}$/O$^{+}$ ratio that
is commonly used as diagnostic of the ionization parameter for the 3-zone nebula model.
In comparison, for EELGs, we recommend using the S$^{+2}$/S$^{+}$ ratio to probe the 
ionization parameter of the low- to intermediate-ionization zones and the 
Ar$^{+3}$/Ar$^{+2}$ ratio to probe the high- to very-high-ionization zones.
\label{fig4}}
\end{center}
\end{figure*}


\section{Improved Nebular Properties: Harnessing the UV$+$Optical} \label{sec:3}


\subsection{An Updated Ionization Model of EELGs} \label{sec:3.1}

Previous nebular abundance determinations for J104457 and J141851 were reported in
\citet{berg16} and \citet{berg19a}, respectively.
Those works followed the standard best-practice methodology of determining total and relative 
abundances using the {\it direct method} (i.e., measuring the electron temperature and density) 
and assuming a classic 3-zone ionization model.
In the top of Figure~\ref{fig4} we plot the ionization potential energies of several important 
interstellar medium (ISM) ions relative to the 3-zone ionization model, where the
ionization potential energy ranges of N$^+$, S$^{+2}$, and O$^{+2}$ define the low-, 
intermediate-, and high-ionization zones, respectively. 
Together these three zones are able to adequately characterize the \ion{H}{2} regions of typical 
star-forming galaxies.
For example, O$^{+}$ and O$^{+2}$ nicely span the entire ionization energy range of the 
3-zone nebula model and so are commonly used as a ratio that is diagnostic of the ionization parameter.
However, several of the important emission lines in our EELGs lie outside the high-ionization
zone at even greater energies (i.e., C$^{+3}$, He$^{+2}$, O$^{+3}$, Ne$^{+2}$, and Ar$^{+3}$)
such that their contributions to the ionization structure and abundances of these galaxies are missing.

The presence of \ion{He}{2} \W1640, \ion{C}{4} \W\W1548,1550, and \ion{O}{4} \W\W1401,1404,1407
emission lines in the {\it HST}/COS UV spectra in Figure~\ref{fig2}, as well as 
[\ion{Ne}{3}] \W3869, \ion{Fe}{5} \W4227, \ion{He}{2} \W4686, and [\ion{Ar}{4}] \W\W4711,4740 
emission lines in the LBT/MODS optical spectra in Figure~\ref{fig3}, reveal the interesting 
detection of a very-high-ionization zone region within these EELG nebulae.
We, therefore, attempt to better characterize the very-high-ionization nebulae of EELGs by defining 
a {\it 4-zone} ionization model. 
The 4-zone model simply extends the classical 3-zone model with the addition 
of a very-high-ionization zone that is designated by the He$^{+2}$ species 
(needed to produce the observed \ion{He}{2} emission via recombination).
In the bottom panel of Figure~\ref{fig4} we see that the [\ion{O}{3}] \W5007/[\ion{O}{2}] \W3727 
ratio, which is commonly used as a proxy for ionization parameter in a 3-zone nebula model, does 
not adequately characterize the full nebula of these EELGs, missing the very-high-ionization zone 
in particular. 

Alternatively, for EELGs, we recommend defining two additional ionization parameters 
to characterize the low-ionization and high-ionization volumes separately.
Expanding on the works of \citet{berg16} and \citet{berg19a}, 
we re-computed the temperature and density structure of J104457 and J141851 (Section~\ref{sec:3.2.1}),
as well as the ionization structure (Section~\ref{sec:3.2.2}) and chemical abundances 
(Section~\ref{sec:4}), incorporating the new UV and optical emission lines measured 
in this work and considering the four-zone ionization model proposed here.
To perform these calculations, we used the {\sc PyNeb} package in {\sc python} 
\citep{luridiana12, luridiana15} with the atomic data adopted in \citet{berg19a}. 

\subsubsection{Photoionization Models} \label{sec:3.1.1}
To aid in our interpretation of the four-zone ionization model, we employed a spherical 
nebula model composed of nested spheres of decreasing ionization, which is supported by the 
visual compactness and structural simplicity of these galaxies (see Figure~\ref{fig1}).
Additionally, we ran new photoionization models, which were especially useful for testing
ionization correction factors and understanding the ionization structure of J104457 and J141851.

Our photoionization models consist of a {\sc cloudy} 17.00 \citep{ferland13} grid
assuming a simple, spherical geometry and a full covering factor of 1.0.
For our central input ionizing radiation field, we use the
``Binary Population and Spectral Synthesis'' \citep[BPASSv2.14;][]{eldridge16, stanway16} 
single-burst models.
Appropriate for EELGs, our grid covers a range of 
ages: $10^{6.0}-10^{7.0}$ yrs for our young bursts, 
ionization parameters: $-3.0 <$ log$U <-1.0$,
matching stellar and nebular metallicities: 
Z$_{\star}$ = Z$_{neb} =0.001,0.002,0.004,0.006 = 0.05,0.10,0.20,0.30$ Z$_{\odot}$
(or $7.4<$ 12+log(O/H) $< 8.2$), and
densities: $n_e = 10^1-10^4$ cm$^{-3}$. 
The \citet{grevesse10} solar abundance ratios and Orion grain set were used to initialize the 
relative gas-phase and dust abundances. 
These abundances were then scaled to cover the desired range in nebular metallicity, 
and relative C, N, and Si abundances ($0.25 <$ (X/O)/(X/O)$_{\odot} < 0.75$).
The ranges in relative N/O, C/O, and Si/O abundances were motivated by the observed 
values for nearby metal-poor dwarf galaxies 
\citep[e.g.,][]{garnett95b, vanzee06a, berg11, berg12, berg16, berg19a}. 

These {\sc cloudy} models have a large number of zones (typically 200--300)
that represent the number of shells or radial steps outward considered in the calculations. 
It is important to note that these zones or shells are physically different than the 3- and 4-zone 
ionization models discussed throughout this work.
Specifically, the 3- and 4-zone ionization models are defined by the ionization
potential energies of the representative ions, where each ionization zone
is composed of many {\sc cloudy} shells.



\subsection{Measuring the Structure of Nebular Properties} \label{sec:3.2}
Detailed abundance determinations from collisionally-excited lines require knowledge
of the electron temperature ($T_e$) and density ($n_e$) structure in a galaxy such 
that the nebular physical conditions are known for each ionic species.
In the standard 3-zone model, the most common method uses the $T_e$-sensitive 
[\ion{O}{3}] \W4363/\W5007 ratio to directly calculate the electron temperature 
of the high-ionization gas.
The temperatures of the low- and intermediate-ionization zones are then inferred 
from photoionization model-based relationships.
In contrast, the density structure across ionization zones is more difficult to determine.
Therefore, the 3-zone model usually assumes a single, uniform density derived from the 
$n_e$-sensitive [\ion{S}{2}]\W6717/\W6731 ratio of the low-ionization zone.
\ion{H}{2} regions commonly have [\ion{S}{2}] ratios that are consistent with the 
low density upper limit, where even large fluctuations on the order of 100\% would
have negligible impact on abundance calculations, and thus motivates the assumption 
of a homogeneous density distribution of $n_e = 100$ cm$^{-3}$ throughout nebulae. 


\begin{figure*}
\begin{center}
    \includegraphics[scale=0.45,trim=0mm 0mm 0mm 0mm,clip]{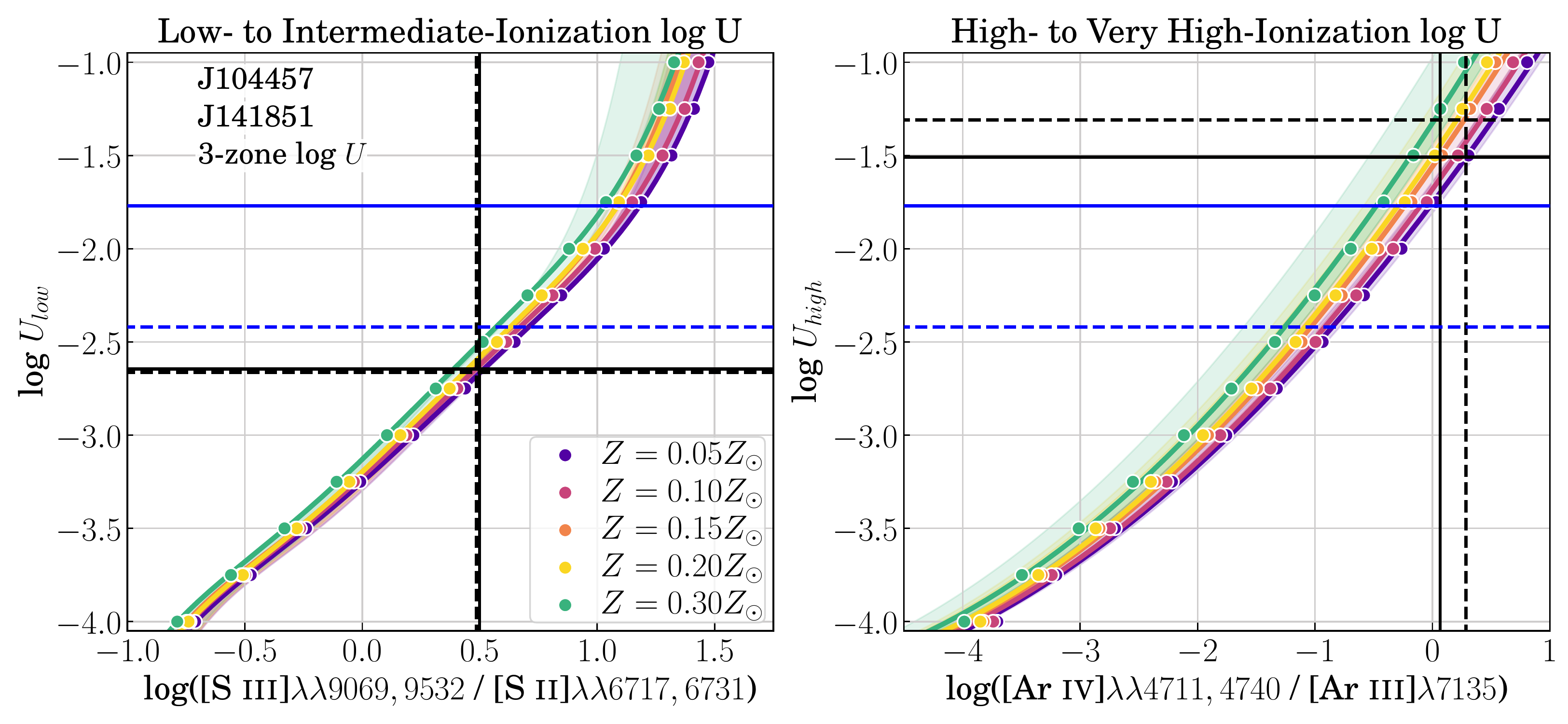} 
\caption{
Photoionization models characterizing ionization parameter in different parts of a nebula.
Models of a given metallicity (indicated by color) are shown for a burst age of $10^{6.7}$ yrs
(solid line) and spanning $10^6 - 10^7$ yrs (shaded regions).
{\it Left:} The low- and intermediate-ionization zones can be characterized by the 
[\ion{S}{2}] \W\W6717,31 and [\ion{S}{3}] \W\W9069,32 emission lines, whose ionization
energies span these zones. 
{\it Right:} The [\ion{Ar}{3}] \W7135 and [\ion{Ar}{4}] \W\W4711,40 emission lines can be used
to characterize the high- to very-high-ionization zone, as the ionization energy range of [\ion{Ar}{4}] 
extends beyond the classical 3-zone model and into the very-high-ionization zone. 
Using the [\ion{S}{3}]/[\ion{S}{2}] ratio, the log$U_{low}$ values inferred for J104457 and
J141851 are lower than their corresponding 3-zone log$U_{int}$ values determined from [\ion{O}{3}]/[\ion{O}{2}].
On the other hand, log$U_{high}$ values inferred from the [\ion{Ar}{4}]/[\ion{Ar}{3}] ratios are
considerably higher than the 3-zone log$U_{int}$ values.
\label{fig5}}
\end{center}
\end{figure*}


\subsubsection{Temperature and Density Structure}\label{sec:3.2.1}
Given the extreme nature of our EELGs, the simple 3-zone model structure cannot be assumed. 
Fortunately, owing to the improved resolution and S/N of the LBT/MODS spectra over existing 
optical spectra, we were able to directly probe the physical conditions across the entire
ionization energy range of J104457 and J141851.
Specifically, we use different electron temperature and density measurements for each of the
4 ionization zones: 

\begin{itemize}
\item[]{\it Low-ionization zone:}
We measure temperatures from the [\ion{O}{2}] 
\W\W7320,7330/\W\W3727,3729 ratio and density from [\ion{S}{2}] \W6717/\W6731.
The [\ion{N}{2}] \W5755/\W\W6548,6584 line ratio has been demonstrated to be a 
more robust measure of the electron temperature in the low-ionization zone \citep[e.g.,][]{berg15}, 
but the low N$^+$ abundances of our EELGs precluded detection of the $T_e$-sensitive 
[\ion{N}{2}] \W5755 auroral line.

\item[]{\it Intermediate-ionization zone:} 
We use the [\ion{S}{3}] \W6312/\W9069,9532 ratio, after checking for atmospheric  
contamination of the red [S\iii] lines\footnote{
Using {\sc PyNeb}, the theoretical ratio of emissivities of [\ion{S}{3}] \W9532/\W9069 
is $\epsilon_{\lambda9532}/\epsilon_{\lambda9069}= 2.47$, and remains consistent over a 
wide range of nebular temperatures ($0.5\times10^4 \leq T_e \leq 2.0\times10^4$) and 
densities ($10^2 \leq n_e \leq 10^4$).
For J141851, [\ion{S}{3}] \W9532/\W9069 = 2.48, consistent with the theoretical ratio.
However, for J104457, [\ion{S}{3}] \W9532/\W9069 = 2.37, and so [\ion{S}{3}] \W9532
is corrected to the theoretical ratio relative to \W9069 prior to determining $T_e$[\ion{S}{3}].}
to determine the intermediate-ionization zone temperature. 
Unfortunately, we do not have a robust probe of the density in this zone, but
we are able to use the \ion{Si}{3}] \W1883/\W1892 ratio from the archival low-resolution 
{\it HST}/COS spectra to measure an upper limit on the density in the 
low- to intermediate-ionization zone. 
We note that the optical [\ion{Cl}{3}] \W5517/\W5537 line ratio is an excellent
probe of the intermediate-ionization zone, however these lines are too faint
and lie too close to the dichroic to get adequate measurements from the LBT/MODS spectra. 

\item[]{\it High-ionization zone:} 
We use the standard [\ion{O}{3}] \W4363/\W\W4959,5007 ratio for the high-ionization zone temperature.
While we also lack a strong probe of the density in the high-ionization zone,
we estimate the \ion{C}{3}] density using the \W1907/\W1909 ratio from the archival low-resolution 
{\it HST}/COS spectra, where C$^{+2}$ spans intermediate- to high-ionization energies. 

\item[]{\it Very-high-ionization zone:}
With the very-high-ionization zone defined by He$^{+2}$ ($>54.42$ eV) in
Figure~\ref{fig4}, the only {\it pure} very-high-ionization emission lines
we observe are \ion{He}{2}, \ion{O}{4}, and [\ion{Fe}{5}].
Therefore, in order to characterize the very-high-ionization zone,
we also consider {\it bridge} ions, or ions
that partially span both the high- and
very-high-ionization zones, such as Ne$^{+2}$ (40.96--63.45 eV) and 
Ar$^{+3}$ (40.74--59.81 eV).
Specifically, we use the temperature-sensitive [\ion{Ne}{3}] \W3342/\W3868 ratio 
and the density-sensitive [\ion{Ar}{4}] \W4711/\W4740 ratio.
Note that at the resolution of the LBT/MODS spectra, 
the [\ion{Ar}{4}] \W4711 line is blended with \ion{He}{1} \W4713.
To correct for the \ion{He}{1} flux contribution, we first continuum subtract the spectra
to account for \ion{He}{1} absorption and then estimate the \ion{He}{1} \W4713 flux from
the measured \ion{He}{1} \W4471 flux and the theoretical \ion{He}{1} \W4713/\W4471 ratio. 
\end{itemize}

The temperatures and densities determined for each of the four ionization zones are listed in Table~\ref{tbl3}.
Assuming a simple high-to-low ionization gradient from center-to-edge of the nebula,
all of the measurements together describe an H\ii\ region with higher temperatures 
and densities in the center that decrease with distance outward. 
Comparing the extremes of the temperatures and densities across the different zones, 
J104457 has gradients spanning $\Delta T_e = 1,500$ K and $\Delta n_e = 1,350$ cm$^{-3}$, 
while the gradients of J141851 are somewhat steeper with 
$\Delta T_e = 9,000$ K and $\Delta n_e = 2,040$ cm$^{-3}$.
Note, however, that several of the temperature and density measurements have
significant uncertainties.

Within these measured temperature ranges, the high- and very-high-ionization zones of the 
J104457 nebula have temperatures that are consistent with one another and are roughly a thousand 
K hotter than the outter region of the  low- to intermediate-ionization zones (weighted average
$T_e$ = 17,840 K).
We note that the [\ion{O}{2}] temperature is consistent with the higher, central temperatures,
but [\ion{O}{2}] measurements are often systematically
biased to hotter temperatures \citep[e.g.,][]{esteban09,pilyugin09,berg20}.
For J141851, the intermediate- and high-ionization zones within the nebula have 
temperatures that are consistent with one another within the errors and are a few thousand 
K hotter than the outer low-ionization region.
In comparison to the very-high-ionization zones, however, the intermediate-ionization 
zones of both nebula are $\sim$1,500--5,600 K cooler. 
These temperature and density structures, paired with our assumed spherical ionization model,
suggest extreme radiation sources are at the center of these EELGs. 


\begin{deluxetable}{lccc}
\setlength{\tabcolsep}{4pt}
\tablecaption{3-Zone and 4-Zone Nebular Conditions for EELGs}
\tablehead{
\CH{}                        & \CH{Ion. Zone}    & \CH{J104457}                       & \CH{J141851}                       \\ 
\multicolumn{1}{l}{Property} &\CH{3 Zone\M\ 4 Zone}& \CH{ 3 Zone\M\ 4 Zone}           & \CH{ 3 Zone\M\ 4 Zone}               }
\startdata
$T_{e}$ [Ne~\iii] (K)        & VH               & 19,200\PM2,300                      & 23,600\PM3,200                      \\
$T_{e}$ [O~\iii] (K)         & H                & 19,200\PM200                        & 17,800\PM200                        \\
$T_{e}$ [S~\iii] (K)         & I                & 17,700\PM500                        & 18,000\PM1,200                      \\
$T_{e}$ [O~\ii] (cm$^{-3}$)  & L                & 19,100\PM1,500                      & 14,600\PM600                        \\
$\Delta T_e$ (K)             &                  & 1,500                               & 9,000                               \\
\\[-1.5ex] 
$n_{e}$ [Ar~\iv] (cm$^{-3}$) & VH               & 1,550\PM1,100                       & 2,110\PM1,300                       \\
$n_{e}$ C~\iii] (cm$^{-3}$)  & {\it H--L}       & +$<$ {\it 8,870}                    & +$<$ {\it 1,680}                    \\
$n_{e}$ Si~\iii] (cm$^{-3}$) & {\it I--L}       & +$<$ {\it 9,450}                    & +$<$ {\it 3,610}                    \\
$n_{e}$ [S~\ii] (cm$^{-3}$)  & L                & 200\PM40                            & 70\PM40                             \\
$\Delta n_e$  (cm$^{-3}$)    &                  & 1,350                               & 2,040                               \\
\\[-1.5ex] 
log $U_{low}$ (\nicefrac{[S\iii]}{[S\ii]})
                             & I/L               & \hspace{1.5ex}N/A\M\ $-2.65$        & \hspace{1.5ex}N/A\M\ $-2.66$       \\
log $U_{int}$ (\nicefrac{[O\iii]}{[O\ii]})
                            & All\M\ L/H        & $-1.77$\M\ $-1.77$                  & $-2.42$\M\ $-2.42$                  \\
log $U_{high}$ (\nicefrac{[Ar\iv]}{[Ar\iii]})
                            & \ N/A\M\ VH/H     & \hspace{1.5ex}N/A\M\ $-1.51$        & \hspace{1.5ex}N/A\M\ $-1.31$        \\
$\Delta$log $U$             & All               & $-1.14$                             & $-1.35$                             \\
log $U_{ave}$               & All               & $-1.77$\M\ $-1.66$                  & $-2.42$\M\ $-1.93$                  \\
\enddata
\tablecomments{
Nebular temperatures, densities, and ionization parameters for J104457 and J141851 using both the 3-zone and 4-zone models.
Column 2 specifies the ionization zone(s) of each property, where L = low, I =
intermediate, H = high, VH = very high, and All = all ionization zones.  
Temperatures and densities from ions spanning different ionization zones are given first,
followed by ionization parameters derived from three different line ratios. }
\label{tbl3}
\end{deluxetable}


\subsubsection{Characterizing the Ionization Parameter}\label{sec:3.2.2}
An important parameter for characterizing the physical nature of an \ion{H}{2} region 
is the ionization parameter, $q$, or the flux of ionizing photons (cm$^{-2}$ s$^{-1}$)
per volume density of H, $n_{\rm H}$ (cm$^{-3}$).
More commonly, we use the dimensionless log$U$ ionization parameter, defined as $U = q/c$.
While $U$ varies as a function of radius throughout a nebula, decreasing as the number 
of ionizing photons is geometrically diluted further from the central source, we can also
characterize the average ionization parameter, $U_{ave}$, of the entire nebula in a 
3-zone model as the degree of ionization of oxygen.
It has therefore become common to use photoionization models to determine the relationship 
of log$U_{ave}$ as a function of the optical [\ion{O}{3}] \W5007/[\ion{O}{2}] \W3727 ratio.
This is a reasonable quantity for typical star-forming \ion{H}{2} regions, where the 
variation in log$U$ across the nebulae declines gradually as a function of radius 
(see further discussion in Section~\ref{sec:5.4})
and has an average value of $-3.2 <$ log $U < -2.9$ \citep[e.g.,][]{dopita00,moustakas10}.
For J104457 and J141851, we use the equations from 
\citet[see Table 3]{berg19a}
with the observed [\ion{O}{3}] \W5007/[\ion{O}{2}] \W3727 ratios to determine ionization 
parameters for the 3-zone model of log $U_{ave} = -1.77, -2.42$, respectively.
Note, these ionization parameters are not only atypical compared to local populations of galaxies,
but are also likely underestimated, as the O$^{+}$ and O$^{+2}$ ions do not characterize
the full extent of the very-high-ionization zone in EELGs (see Figure~\ref{fig4}).

To better characterize the extreme, extended ionization parameter space of the nebular 
environments of EELGs, we recommend examining how the ionization parameter changes
across ionization zones.
In this context, the standard 3-zone log $U_{ave}$ is best equated with the 
intermediate-ionization zone, log $U_{int}$, but two additional ionization 
parameters are needed to represent the low-ionization and high-ionization 
volumes separately.
We use the photoionization models described in \S~\ref{sec:3.1.1} to estimate new ionization 
parameters to characterize the low- to intermediate-ionization volume, log$U_{low}$, as a 
function of the [\ion{S}{3}] \W\W9069,9532/[\ion{S}{2}] \W\W6717,6731 emission-line ratio and 
the high- to very-high-ionization volume, log$U_{high}$, as a function of the 
[\ion{Ar}{4}] \W\W4711,4740/[\ion{Ar}{3}] \W7135 emission-line ratios.
To summarize, we determine the ionization structure with the following relations:
\begin{itemize}
\setlength\itemsep{0.1em}
    \item log$U_{low} \propto$ [\ion{S}{3}] \W\W9069,9532/[\ion{S}{2}] \W\W6717,6731
    \item log$U_{int} \propto$ [\ion{O}{3}] \W5007/[\ion{O}{2}] \W3727
    \item log$U_{high} \propto$ [\ion{Ar}{4}] \W\W4711,4740/[\ion{Ar}{3}] \W7135
\end{itemize}
Note that all three of these diagnostic ratios utilize the same element 
and so are not vulnerable to variations in relative abundances.

Our log $U_{low}$ versus [\ion{S}{3}]/[\ion{S}{2}] and 
log $U_{high}$ versus [\ion{Ar}{4}]/[\ion{Ar}{3}] models are plotted in Figure~\ref{fig5}.
The light color shading depicts the minimal variation in the models with burst age,
centered on models with an age of $t=10^{6.7}$ yrs (colored lines) and extending from $t=10^{6.0}-10^{7.0}$ yrs.
We fit each metallicity model with a polynomial of the shape:
$y = c_3\cdot{x^2} + c_2\cdot{x} + c_1$,
where $y =$ log$U$, $x$ is the log of the observed line ratio,
and the $c$ coefficients are listed in Table~\ref{tbl4}.

The observed [\ion{S}{3}]/[\ion{S}{2}] line ratios of J104457 (black solid line) and J141851 
(black dashed line) are nearly the same, resulting in measured ionization parameter values of 
log $U_{low} = -2.65, -2.66$ that are lower than the standard 3-zone [\ion{O}{3}]/[\ion{O}{2}]-derived 
volume-averaged values (blue lines; log $U_{ave} = -1.77, -2.42$).
For the [\ion{Ar}{4}]/[\ion{Ar}{3}] line ratios, we measured log $U_{high}$ values of $-1.51, -1.31$
for J104457 and J141851, respectively, that are higher than the 3-zone log$U_{ave}$ values.


\begin{deluxetable}{lrrrrrr}[h!]
\setlength{\tabcolsep}{2pt}
\tablecaption{Coefficients for Ionization Parameter Model Fits}
\tablehead{
\multicolumn{1}{c}{} 	& \multicolumn{6}{c}{$Z(Z_{\odot})$} \\ 
\cline{2-7}
\CH{$y = f(x)$} 			& \CH{0.005} & \CH{0.05} & \CH{0.10} & \CH{0.20} & \CH{0.30} & \CH{0.40}}
\startdata	
\multicolumn{1}{l}{\bf{y = log $U_{low}$:}} & & & & \\
\multicolumn{1}{l}{$x =$ log([\ion{S}{3}]/[\ion{S}{2}])} & & & & \\
{\ \ \ $c_1$ ......................} & $-3.2705$ & $-3.2506$ & $-3.2204$ & $-3.1963$ & $-3.1295$ & $-3.1147$\\
{\ \ \ $c_2$ ......................} & $1.1163$	 & $1.1145$	 & $1.1397$	 & $1.1811$	 & $1.2104$  & $1.2280$ \\
{\ \ \ $c_3$ ......................} & $0.1692$	 & $0.2060$	 & $0.2154$	 & $0.2236$  & $0.1844$  & $0.1816$ \\
\vspace{-1ex} \\
\multicolumn{1}{l}{\bf{y = log $U_{high}$:}} & & & & \\
\multicolumn{1}{l}{$x =$ log([\ion{Ar}{4}]/[\ion{Ar}{3}])} & & & & \\
{\ \ \ $c_1$ ......................} & $-1.9370$ & $-1.6396$ & $-1.4934$ & $-1.2273$ & $-0.9737$ & $-0.7817$\\
{\ \ \ $c_2$ ......................} & $0.7662$	 & $0.8589$	 & $0.9093$	 & $0.9695$	 & $1.0210$  & $1.0545$ \\
{\ \ \ $c_3$ ......................} & $0.0554$	 & $0.0658$	 & $0.0711$	 & $0.0741$  & $0.0760$  & $0.0769$ \\
\vspace{-2ex} 
\enddata
\tablecomments{{\sc cloudy} photoionization model fits of the form 
$f(x)=c_3\cdot{x^2}+c_2\cdot{x}+c_1$ for the ionization parameters characterizing the
ionization parameter.
For the low- to intermediate-ionization region, log$U_{low}$ is determined from
$x =$ log([\ion{S}{3}] \W\W9069,9532/[\ion{S}{2}] \W\W6717,6731) and for the 
high- to very-high-ionization region, log$U_{high}$ is determined from
$x =$ log([\ion{Ar}{4}] \W\W4711,4740/[\ion{Ar}{3}] \W7135).
The best fits are for a burst of star formation with an age of $t=10^{6.7}$ yrs.
The model grids and polynomial fits are shown in Figure~\ref{fig5}.}
\label{tbl4}
\end{deluxetable}


\section{Abundance Determinations}\label{sec:4}


We compute absolute and relative abundances for J104457 and J141851 with 
both the standard 3-zone ionization model and the expanded 4-zone ionization model.
For all calculations, we use the {\sc PyNeb} package in {\sc python} with the atomic data
adopted in \citet{berg19b} for a 5-level atom model, plus a six-level atom model for oxygen 
in order to utilize the UV \ion{O}{3}] \W\W1661,1666 lines for C/O abundance determinations.
Ionic abundances were calculated from the optical spectra for 
O$^{0}$/H$^+$, O$^{+}$/H$^+$, O$^{+2}$/H$^+$, N$^{+}$/H$^+$, 
S$^{+}$/H$^+$, S$^{+2}$/H$^+$, Ar$^{+2}$/H$^+$, Ar$^{+3}$/H$^+$, Ne$^{+2}$/H$^+$,
Fe$^{+2}$/H$^+$, Fe$^{+3}$/H$^+$, and Fe$^{+4}$/H$^+$, whereas the
C$^{+2}$/O$^{+2}$, O$^{+3}$/O$^{+2}$, and S$^{+3}$/O$^{+2}$ relative abundances
were determined from the UV spectra.

To determine accurate ionic abundances, we adopt the characteristic temperature and 
density of each ionization species when available (see \S~\ref{sec:3.2.1}).
Specifically, we adopt the $T_e$[\ion{O}{2}] temperature and $n_e$[\ion{S}{2}] density 
for the low-ionization zone ions:
O$^0$, O$^+$, N$^+$, S$^+$,  N$^+$, and Fe$^+2$.
For the intermediate-ionization zone ions, 
 S$^{+2}$ and Ar$^{+2}$, 
we adopt the $T_e$[\ion{S}{3}] temperature.
However, owing to their large uncertainties, we do not use either of the intermediate-ionization 
zone densities ($n_e$ \ion{C}{3}], $n_e$ \ion{Si}{3}]), but rather adopt the $n_e$[\ion{S}{2}] density.
For the high-ionization O$^{+2}$, C$^{+2}$, S$^{+3}$, and Fe$^{+3}$ ions, we use the $T_e$[\ion{O}{3}]
temperature and $n_e$[\ion{Ar}{4}] very-high-ionization density.
Finally, we use the $T_e$[\ion{Ne}{3}] temperature and $n_e$[\ion{Ar}{4}] density 
to characterize the very-high-ionization zone and calculate O$^{+3}$ and Fe$^{+4}$. 
Note, Ne$^{+2}$ and Ar$^{+3}$ partially span both the high- and very-high-ionization
zones, and so an average of the temperatures and densities characterizing these zones
is used. 

In general, the total abundance of an element relative to hydrogen in an \ion{H}{2} region is 
calculated by summing the abundances of the individual ionic species together relative to hydrogen as:
\begin{equation}
	{\frac{N(X)}{N(H)}\ } = \sum_i {\frac{N(X^{i})}{N(H^{+})}\ } = \sum_i {\frac{I_{\lambda(i)}}{I_{H\beta}}\ } {\frac{j_{H\beta}}{j_{\lambda(i)}}\ },
	\label{eq:Nfrac}
\end{equation}
where the emissivity coefficients, $j_{\lambda(i)}$, are determined for the appropriate
ionization zone temperature and density.
Details of elemental abundance determinations are given below.


\subsection{Ionic And Total O Abundances} \label{sec:4.1}

The most common method of calculating O/H abundances involves adding together the 
dominant ionic abundances, O$^+$/H$^+$ and O$^{+2}$/H$^+$, determined from the 
[\ion{O}{2}] \W3727 and [\ion{O}{3}] \W\W4959,5007 emission lines.
Because the ionization energy ranges of O$^{+}$ and O$^{+2}$ span the full range of
a standard 3-zone \ion{H}{2} region, contributions from O$^0$ and O$^{+3}$ 
(and higher ionization species) can be ignored.
In our 4-zone ionization model this is not necessarily the case.
For J104457 and J141851 we detect weak \ion{O}{4} \W\W1401,1405,1407\footnote{
Note that the emission line at 1405 \AA\ is a blend of \ion{O}{4} \W1404.806 and \ion{S}{4} \W1404.808, 
and so is not used here.}  emission in their 
{\it HST}/COS spectra and so can directly estimate the impact of O$^{+3}$ on the total O abundance.
To do so, we calculated
O$^{+3}$/H$^+$ = [O$^{+3}$/O$^{+2}$]$_{UV}$/[O$^{+2}$/H$^+$]$_{opt.}$,
where the O$^{+3}$/O$^{+2}$ abundance was determined from the UV 
\ion{O}{4} \W\W1401,1407/\ion{O}{3}] \W1666 ratio.
We also detect \ion{O}{1} \W\W6300,6363 emission in the LBT/MODS spectra,
allowing a measure of the O$^{0}$/H$^{+}$ abundance.
Therefore, the total oxygen abundances (O/H) were calculated from the 
sum of four ionization species: 
\begin{equation}
    \frac{\rm O}{\rm H} = \frac{{\rm O}^0}{{\rm H}^{+}}    + \frac{{\rm O}^{+}}{{\rm H}^{+}} + 
                          \frac{{\rm O}^{+2}}{{\rm H}^{+}} + \frac{{\rm O}^{+3}}{{\rm H}^{+}}.
\end{equation}

Ionic and total O/H abundances determined for both the classical 3-zone
and expanded 4-zone ionization models are reported in Table~\ref{tbl5}.
The main differences are the inclusion of the O$^0$ and O$^{+3}$ species and the use of the 
very-high-ionization zone density for species in the high-ionization zone in the 4-zone model.
In general, we find that the O$^{+3}$/H$^+$ abundances are very small, with
the O$^{+3}$/O$_{tot.}$ fractions of only 1–2\%.
Additionally, the effects of the different density assumptions in the 3-zone versus 4-zone model, 
where the density was increased from $\sim10^2$ to $\sim10^3$ cm$^{-3}$, are negligible.
In fact, the O$^{+2}$/H$^+$ abundances differ by much less than 1\%. 


\subsection{Ionization Correction Factors} \label{sec:4.2}

If all of the species of an element present in the \ion{H}{2} region are not observed, 
then an ionization correction factor (ICF) must be used to account for the missing abundance.
We showed in the previous section that, even for EELGs, the rarely observed higher 
ionization species of O (above O$^{+2}$) represent a small fraction of the total O 
abundance, and so, total O abundances can still be accurately determined by the simple 
sum of the O$^+$/H$^+$ and O$^{+2}$/H$^+$ ratios. 
The other elements discussed here, namely N, C, S, Ar, Ne, and Fe, can have significant 
fractions of their species in unobserved ionic states and so require an ICF to infer 
their total abundances. 

For each element, we determined appropriate ICFs based on photoionization modeling as a 
function of ionization parameter. 
For the 3-zone model, we adopted the [\ion{O}{3}]/[\ion{O}{2}]-based log$U_{ave}$ 
to determine the appropriate ICFs.
We then determined a comparable volume-averaged ionization parameter to characterize the
4-zone model from the set of ionization parameters determined in \S~\ref{sec:3.2.2}.
To do so, we calculated the average ionization parameter, log $U_{ave}$, by weighting 
the log$U_{low}$, log$U_{int}$, and log$U_{high}$ values by their corresponding ionization 
fractions of oxygen, as determined in \S~\ref{sec:4.1}.
For J104457 and J141851, we measure log $U_{ave} = -1.70, -1.91$, respectively,
for the 4-zone model, and use these values to determine the appropriate ICFs. 


\subsection{N/O Abundances} \label{sec:4.3}

Relative N/O abundances are often determined by employing the simple assumption that N/O = N$^+$/O$^+$.
This method benefits from the similar ionization and excitation energies of N$^+$ and O$^+$ (see Figure~\ref{fig4}),
and is particularly useful for low- to moderate-ionization dominated nebula.
For high-ionization nebulae, a N ICF is needed to correct N abundances for higher ionization
species of N, where N/H = ICF(N$^+$)$\times$N$^+$/H$^+$.
Several N ICFs in the literature have been derived as a function of O$^+$/(O$^+$ $+$ O$^{+2}$) = 
O$^+$/O$_{tot.}$ \citep[e.g.,][]{peimbert69,izotov06,nava06,esteban20}; here we consider two of them. 
First, we calculate the simple N ICF from \citet{peimbert69}: ICF(N$+$) = O$_{tot.}$/O$^+$, 
and then consider the recent empirical fit from \citet{esteban20} to Milky Way data: 
ICF(N$^+$) = $0.39+1.19\times$(O$_{tot.}$/O$^+$).
Note that these methods require a calculation of an ionic and total O abundance,
where O$_{tot.}$ is just the sum of O$^+$ and O$^{+2}$ ions in the 3-zone model.
Therefore, we also use the photoionization models described in Section~\ref{sec:3.1.1} to 
investigate a N ICF that can be inferred from observations of strong emission lines alone. 

\begin{figure}
\begin{center}
    \includegraphics[scale=0.4,trim=0mm 0mm 0mm 0mm,clip]{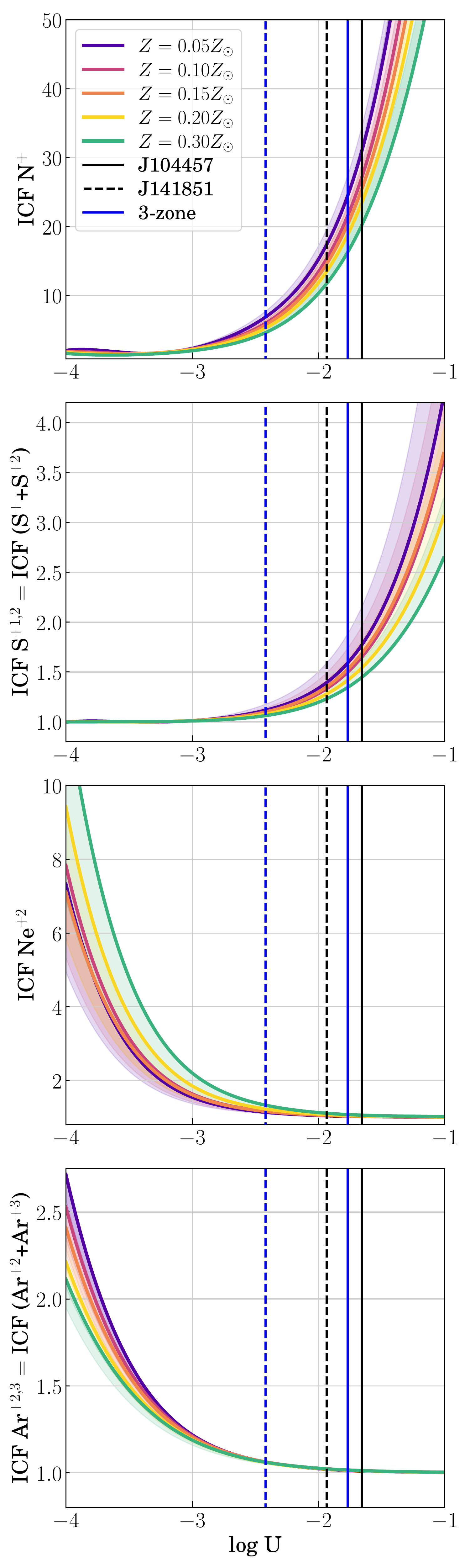}
\caption{
Photoionization models of N, S, Ne, and Ar ionization correction factors versus ionization parameter.
Lines are color-coded by the gas-phase oxygen abundance and are centered on models with an age of $t=10^{6.7}$ yrs.
The light color shading demonstrates that little variation is seen in the ICFs for bursts aging from
$t=10^{6.0}-10^{7.0}$ yrs.
For reference, the log$U_{ave}$ values of J104457 and J141851 are shown for
the 3-zone (blue) and 4-zone (black) models.}
\label{fig6}
\end{center}
\end{figure}

In the top panel of Figure~\ref{fig6} we plot our model N ICF versus log $U$ for a range of metallicities.
Here, the N ICF is the ionization fraction of N$^{+}$, or 
ICF(N$^+$) = N$_{tot.}$/N$^+$ = $[X({\rm N}^+)]^{-1}$,
and is used to correct N/H abundances as 
\begin{equation}
	{\frac{\mbox{N}}{\mbox{H} }} = {\frac{\mbox{N}^{+}}{\mbox{H}^{+}}}\times[X(\mbox{N}^{+})]^{-1}
	                             = {\frac{\mbox{N}^{+}}{\mbox{H}^{+}}}\times{\mbox{ICF(N}^{+})}. \nonumber
\end{equation}
We use the average ionization parameter values of J104457 (solid line) and J141851 
(dashed line) to determine their N ICFs in the 3-zone (blue lines) and 
4-zone (black lines) models.
We find that our N ICFs are generally smaller than those of \citet{peimbert69} and 
\citet{esteban20} for the 3-zone model and larger or equivalent than these literature
ICFs for the 4-zone model. 
These differences are not unexpected given the 2-zone model basis 
of the O$^+$/O$_{tot.}$ variable and dissimilar calibration samples used in 
\citet{peimbert69} and \citet{esteban20}.
Further, our models have an explicit benefit over the past models mentioned here:
since a number of different emission line ratio calibrations have been recommended 
to infer ionization parameter \citep[see, e.g.,][]{levesque14, berg19a}, 
our ICF models allow significantly greater applicability, particularly
in the absence of a direct oxygen abundance determination. 
Nitrogen ICFs and N/O abundances are reported in Table~\ref{tbl5}.

Despite the very-high ionization of J104457 and J141851, we determine smaller N ICFs than
the relationships of \citet{peimbert69} and \citet{esteban20} for the 3-zone model, 
but find ICFs more similar to \citet{esteban20} for the 4-zone model.
The resulting N/O values are low, but not at odds with the standards of metal-poor galaxies,
spanning just 27--32\%\ and 16--40\%\ solar for J104457 and J141851, respectively.

Given the very-high-ionization emission lines observed from other elements in our spectra,
we might expect there to also be emission from high-ionization species of N,
but this is also dependent on the abundance of N ions.
Unfortunately, the UV \ion{N}{3}] emission line quintuplet around \W1750 lies just 
outside the wavelength coverage of our high-resolution COS spectra, and we only 
weakly detect the \ion{N}{4}] emission lines at \W\W1483,1487.
However, we can use our optical [\ion{N}{2}] \W6584 and UV \ion{N}{4}] \W\W1483,1487 line 
measurements as a guide to compare to expectations from photoionization models. 

At the 4-zone average ionization parameters characterizing J104457 and J141851, 
the photoionization models predict relative N$^{+}$, N$^{+2}$, and N$^{+3}$ ionization 
fractions of 3.2\%, 79.2\%, and 17.4\%, respectively, for J104457, and 
5.4\%, 84.4\%, and 9.6\%, respectively, for J141851. 
N$^{+2}$ is the clearly dominant species of N in these EELGs.
However, at the temperatures and densities measured for J104475 and J141851, the summed 
emissivity of the two strongest lines of the \ion{N}{3}] quintuplet, \W\W1749,1752, is 
$\sim$20\% of that of the [\ion{N}{2}] \W6584 line. 
For the \ion{N}{4}] \W\W1483,1486 lines the emissivity is a bit stronger, at 37\%--65\%
of the [\ion{N}{2}] \W6584 line.
Therefore, the weak detections of the \ion{N}{4}] \W\W1483,1486 lines align with the 
expectations for the low-ionization fraction of N$^{+3}$ and low N abundance of our EELGs. 


\subsection{C/O Abundances} \label{sec:4.4}


In a simple 3-zone ionization model, C/O can be determined from the C$^{+2}$/O$^{+2}$ ratio alone,
where this assumption is most appropriate for moderate ionization nebulae.
For high-ionization nebulae resulting from a hard ionizing spectrum, 
we must also consider carbon contributions from the C$^{+3}$ species
to avoid underestimating the true C/O abundance. 
However, even if the \ion{C}{4} \W\W1548,1550 doublet is observed in emission, as is the case with
the EELGs studied here, these lines are resonant, and so, determining their intrinsic fluxes 
and subsequent C$^{+3}$/H$^+$ abundances is problematic. 
Instead, we use the photoionization-model-derived C ICF of \citet{berg19a}: 
\begin{equation}
	{\frac{\mbox{C}}{\mbox{O} }} = {\frac{\mbox{C}^{+2}}{\mbox{O}^{+2}}\ }\times \Bigg[{\frac{X(\mbox{C}^{+2})}{X(\mbox{O}^{+2})}}\Bigg]^{-1}
			     			     = {\frac{\mbox{C}^{+2}}{\mbox{O}^{+2}}\ }\times{\mbox{ICF(C}^{+2})}, \nonumber
\end{equation}
where $X(\rm{C}^{+2}$) and $X(\rm{O}^{+2}$) are the C$^{+2}$ and O$^{+2}$ ionization fractions, respectively.

Carbon ICFs and C/O abundances are reported in Table~\ref{tbl5}, where the average ionization
parameters, log $U_{ave}$, were used to determine the ICFs for the 3- and 4-zone models.
Note that the C and O abundances presented here have not been corrected for the fraction of 
atoms embedded in dust. 
However, the depletion onto dust grains is expected to be small for the low abundances and small extinctions
of J104457 and J141851, and so the relative dust depletions between C and O should be negligible. 


\subsection{$\alpha$-element/O Abundances}\label{sec:4.5}


Strong collisionally-excited emission lines for the $\alpha$-elements S, Ne, and Ar 
are observed in the optical LBT/MODS spectra of J104457 and J141851.
In particular, we observe significant [\ion{S}{2}] \W\W6717,6731, [\ion{S}{3}] \W\W9069,9532,
[\ion{Ar}{3}] \W7135, [\ion{Ar}{4}] \W\W4711,4740, and [\ion{Ne}{3}] \W3869 emission lines
that, with the application of appropriate ICFs, allow us to determine the relative 
abundances of these elements. 

For sulfur abundance determinations in a 3-zone nebula, 
contributions from S$^{+}$, S$^{+2}$, and S$^{+3}$ are relevant. 
Unfortunately, we only observe S emission in the optical spectra from the S$^{+}$ (10.36--22.34 eV) 
and S$^{+2}$ (22.34--34.79 eV) ions that probe the low- to intermediate-ionization zones. 
Note that while the ionization energy of S$^+$ is lower than that of H$^0$ (13.59 eV)
and [\ion{S}{2}] emission may therefore originate from outside the \ion{H}{2} region,
we showed in Section~\ref{sec:4.1} that ionic contributions from the neutral zone are 
negligible in very-high-ionization EELGs. 
Because there are no strong S$^{+3}$ or S$^{+4}$ emission lines in the optical, an ICF is 
typically required to account for the unseen S species whose ionization energies are concurrent 
with the O$^{+2}$ zone (35.12--54.94 eV).

Similar to sulfur, two species of Ar are observed, but originating from intermediate-
to very-high-ionization zones: Ar$^{+2}$ and Ar$^{+3}$.
While the Ar$^+$ volume (15.76--27.63 eV) will also be present within an \ion{H}{2} region,
its contribution should be very small for the very-high-ionizations characterizing EELGs.
For neon, however, only the high to very-high Ne$^{+2}$ ionization state (40.96--63.45 eV) 
is strongly observed in the optical or FUV spectra of EELGs.
While this is the dominant ionization zone of these nebulae, 
we must still correct for possible contributions from other ionization states.

Again, using the photoionization models described in Section~\ref{sec:3.1.1}, 
we plot S, Ne, and Ar ICFs as a function of log $U$ in the bottom three panels of Figure~\ref{fig6}.
For S we see that the ICFs are close to one at low ionization (low log $U$ values) and steeply 
increase for log $U > -2.5$ as the unobserved S$^{+3}$ and S$^{+4}$ ionization states become more prominent.
On the other hand, the opposite trend is seen for the Ne and Ar ICFs, as the observed high 
ionization species come to dominate the nebula for log $U > -2.5$. 
As expected, for the average ionization parameter values characterizing the 4-zone 
nebula model of J104457 (solid blue line) and J141851 (dashed blue line), which are significantly 
greater than $-2.5$, we measure Ar and Ne ICFs that are consistent with 1.0.
For sulfur we measure small, but important ICFs that serve to correct for the weak \ion{S}{4} 
\W\W1405,1406,1417 features observed in the FUV spectra of J104457 and J141851.


For reference, we also calculate S and Ar ICFs from \citet{thuan95} as
\footnotesize{
\begin{align}
        \rm{ICF(S}&^{+}+\rm{S}^{+2}) =  \frac{\rm{S}}{\mbox{S}^{+} + \rm{S}^{+2}} \nonumber \\
        	      & = \big[0.013 + x\{5.10 + x[-12.78 + x(14.77 - 6.11x)]\}\big]^{-1}, \rm{ and} \nonumber
\end{align}
\begin{align}
        \rm{ICF(Ar}^{+}+\rm{Ar}^{+2})& =  \frac{\rm{Ar}}{\mbox{Ar}^{+2} + \rm{Ar}^{+3}} \nonumber \\
        			   & = \big[0.99+ x\{0.091 + x[-1.14 + 0.077x]\}\big]^{-1}, \nonumber
\end{align}} \normalsize
\noindent where $x=$O$^+$/(O$^{+} +$ O$^{+2}$), and Ne ICFs from \citet{peimbert69} as
ICF(Ne$^{+2}$) = (O$^{+} +$ O$^{+2}$)/O$^{+2}$.
For Ar and Ne, where we observe the very-high-ionization species directly, the 3-zone ICFs
adopted from the literature agree within the uncertainties of our results.
However, we infer significantly lower S ICFs from our models than from 
\citet{thuan95}, resulting in smaller S/O abundances.
This difference may be due to significant changes in atomic data for S$^{+2}$ over
the past few decades \citep[see, e.g., Fig. 4 in][]{berg15}.
All $\alpha$-element ICFs and abundances are reported in Table~\ref{tbl5}. 


\subsection{Fe/O Abundances}\label{sec:4.6}


\begin{figure}
\begin{center}
    \includegraphics[width=0.85\columnwidth,trim=0mm 0mm 0mm 10mm,clip]{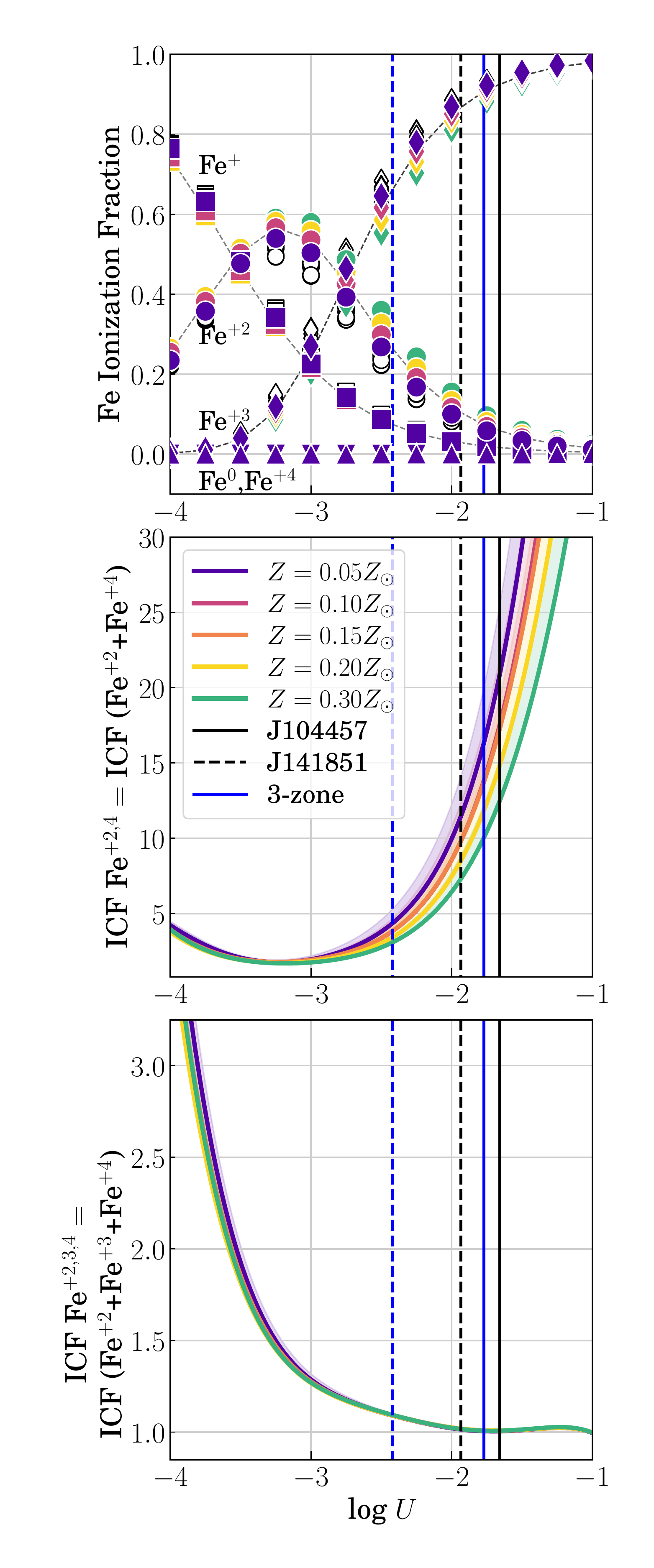}
    \vspace{-2ex}
\caption{
Photoionization models of Fe ionization fraction versus ionization parameter.
{\it Top:} The changing fractions of Fe species are shown as a function of ionization parameter,
where downward-pointing triangles, squares, circles, diamonds, and upward-pointing triangles 
represent the Fe$^0$, Fe$^+$, Fe$^{+2}$, Fe$^{+3}$, and Fe$^{+4}$ ions, respectively.
Open symbols designate deviations due to burst age, spanning $t=10^6$ to $t=10^7$ years. 
Dashed gray lines trace the $Z = 0.1 Z_\odot$, or 12 + log(O/H) = 7.7, models.
Note that the Fe$^{0}$ and Fe$^{+4}$ fractions are negligible regardless of ionization parameter.
{\it Middle:} The model Fe ICF to be used when only Fe$^{+2}$ or Fe$^{+2} +$Fe$^{+4}$ are observed.
{\it Bottom:} The model Fe ICF to be used when Fe$^{+2} +$Fe$^{+3}$ or Fe$^{+2} +$Fe$^{+3} +$Fe$^{+4}$ 
are observed.
For reference, the log$U_{ave}$ values of J104457 and J141851 are shown for
the 3-zone (blue) and 4-zone (black) models.}
\label{fig7}
\end{center}
\end{figure}


While collisionally-excited emission lines are commonly observed for one or two 
species of Fe in \ion{H}{2} regions, Fe abundance determinations are often avoided 
due to the importance of dust depletion, accurate ICFs, and fluorescence.
However, several recent studies have revived the interest in Fe abundances by 
suggesting that enhanced $\alpha$/Fe abundance ratios are responsible for the 
extremely hard radiation fields inferred from the stellar continua and 
emission line ratios in chemically-young, high-redshift galaxies 
\citep[e.g.,][]{steidel18,shapley19,topping20}.
Given the importance of $\alpha$/Fe (e.g., O/Fe) abundances to interpreting the ionizing 
continua of early galaxies, we were motivated to investigate the Fe/O abundances of our EELGs.

In the top panel of Figure~\ref{fig7} we show photoionization models of the ionization 
fraction of relevant Fe species as a function of ionization parameter.
In the standard 3-zone ionization model of \ion{H}{2} regions, 
three Fe species are expected to contribute to the abundances:
Fe$^+$ in the neutral- to low-ionization zone, 
Fe$^{+2}$ in the low- to intermediate-ionization zone, and
Fe$^{+3}$ in the high-ionization zone. 
For Fe$^+$  we weakly detected the [\ion{Fe}{2}] \W4287.39, \W5158.79, \W7637.51 emission lines.
However, most of the [\ion{Fe}{2}] lines are significantly affected by fluorescence \citep{rodriguez99}. 
An exception is the [\ion{Fe}{2}] \W8617 emission line, as it is nearly insensitive to the effects 
of UV pumping \citep[][]{rodriguez03}, but this line was not detected in J104457 or J141851.
Fortunately, the Fe$^+$ ion has an ionization potential that mostly spans the neutral zone 
(7.902--16.188 eV). 
We therefore forego an Fe$^+$ abundance determination.

Fe$^{+2}$ is the species of Fe that is most commonly used for abundance determinations 
in \ion{H}{2} regions.
In the LBT/MODS spectra, we detect [\ion{Fe}{3}] \W4658.50, \W4701.53, \W4880.99, and \W5270.40.
Of these lines, both [\ion{Fe}{3}] \W4701 and \W4881 are very sensitive to density, while
[\ion{Fe}{3}] \W4659 is the strongest (by a factor of 2--4).
We determine Fe$^{+2}$/H$^+$ abundances from the \W5270 line that are larger by a 
factor of 2 than those determined by the \W4659.
Given the wide usage of the [\ion{Fe}{3}] \W4659 line in Fe abundance determinations,
and its dominance of the Fe lines in our spectra, we therefore use the [\ion{Fe}{3}] \W4659 
alone to determine Fe$^{+2}$/H$^+$ abundances.

In the proposed 4-zone model, Fe$^{+3}$ bridges the intermediate- and high-ionization zones, 
while Fe$^{+4}$ is a {\it pure} very-high-ionization ion.
Fe$^{+3}$ is often undetected owing to its relatively weak emissivities.
In fact, for the high electron temperatures of our targets, the emissivities of [\ion{Fe}{4}] \W4907 and 
\W5234 are only 6.3\% and 2.5\%, respectively, relative to the [\ion{Fe}{3}] \W4659 line. 
Thus we only weakly detect [\ion{Fe}{4}] \W4906.56 in J104457 and [\ion{Fe}{4}] \W5233.76 in J141851,
but are able to use these lines to estimate Fe$^{+3}$/H$^+$ abundances.

For Fe$^{+4}$, we detect emission from [\ion{Fe}{5}] \W4143.15 and \W4227.19  
in the optical LBT/MODS spectra of J104457 and J141851.
However, this is not terribly surprising given the very-high ionization of our EELGs and the strong 
emissivities of these lines at high electron temperatures 
($j_{\lambda4143}$/$j_{\lambda4659}$ = 0.27 and $j_{\lambda4227}$/$j_{\lambda4659}$ = 1.39).
While [\ion{Fe}{5}] \W4227 is rare, it has also been reported for other EELGs, 
such as SBS 0335-052 \citep[][]{izotov09}. 

Considering the Fe emission lines observed in the optical spectra of J104457 and J141851,
we calculate Fe/H abundances four different ways:
\begin{align}
	1.\ \frac{\mbox{Fe}}{\mbox{H}} &= {\frac{\mbox{Fe}^{+2}}{\mbox{H}^{+}}}\times{\mbox{ICF}(\mbox{Fe}^{+2})}, \nonumber \\ 
	2.\ \frac{\mbox{Fe}}{\mbox{H}} &= {\frac{\mbox{Fe}^{+2}}{\mbox{H}^{+}}}\times{\mbox{ICF}(\mbox{Fe}^{+2})} \mbox{\ (I09)}, \nonumber \\ 
	3.\ \frac{\mbox{Fe}}{\mbox{H}} &= {\frac{\mbox{Fe}^{+2}+\mbox{Fe}^{+4}}{\mbox{H}^{+}}}\times{\mbox{ICF}(\mbox{Fe}^{+2,4})}, \mbox{ and} \nonumber \\
	4.\ \frac{\mbox{Fe}}{\mbox{H}} &= {\frac{\mbox{Fe}^{+2}+\mbox{Fe}^{+3}+\mbox{Fe}^{+4}}{\mbox{H}^{+}}}\times \nonumber
	                                    {\mbox{ICF}(\mbox{Fe}^{+2,3,4})}. \nonumber
\end{align}

The first two equations follow the common method of determining Fe/H from Fe$^{+2}$,
where the ICF is the Fe$^{+2}$ ionization fraction, $X(\rm{Fe}^{+2}$), from our photoionization
models for Equation~1 and is from \citet{izotov09} for Equation~2. 
The third method incorporates our [\ion{Fe}{5}] \W4227 observations such that the ICF
must only correct for Fe$^{+}$ and Fe$^{+3}$, as in the middle panel of Figure~\ref{fig7}. 
Fourth, we used all of the observed Fe species in our optical spectra with the ICF from the bottom panel of Figure~\ref{fig7},
where ICF(Fe$^{+2,3,4}$)= ICF(Fe$^{+2}+$Fe$^{+3}+$Fe$^{+4})= X(\rm{Fe}^{+2})+X(\rm{Fe}^{+3})+X(\rm{Fe}^{+4})$.
Finally, we used the four methods of Fe/H abundance determinations to derive relative Fe/O, 
or Fe/$\alpha$, as reported in Table~\ref{tbl5}.


\section{Insights Into Physical Properties of EELGs}\label{sec:5}
\begin{figure}
\begin{center}
    \includegraphics[width=1.0\columnwidth,trim=0mm 0mm 0mm 0mm,clip]{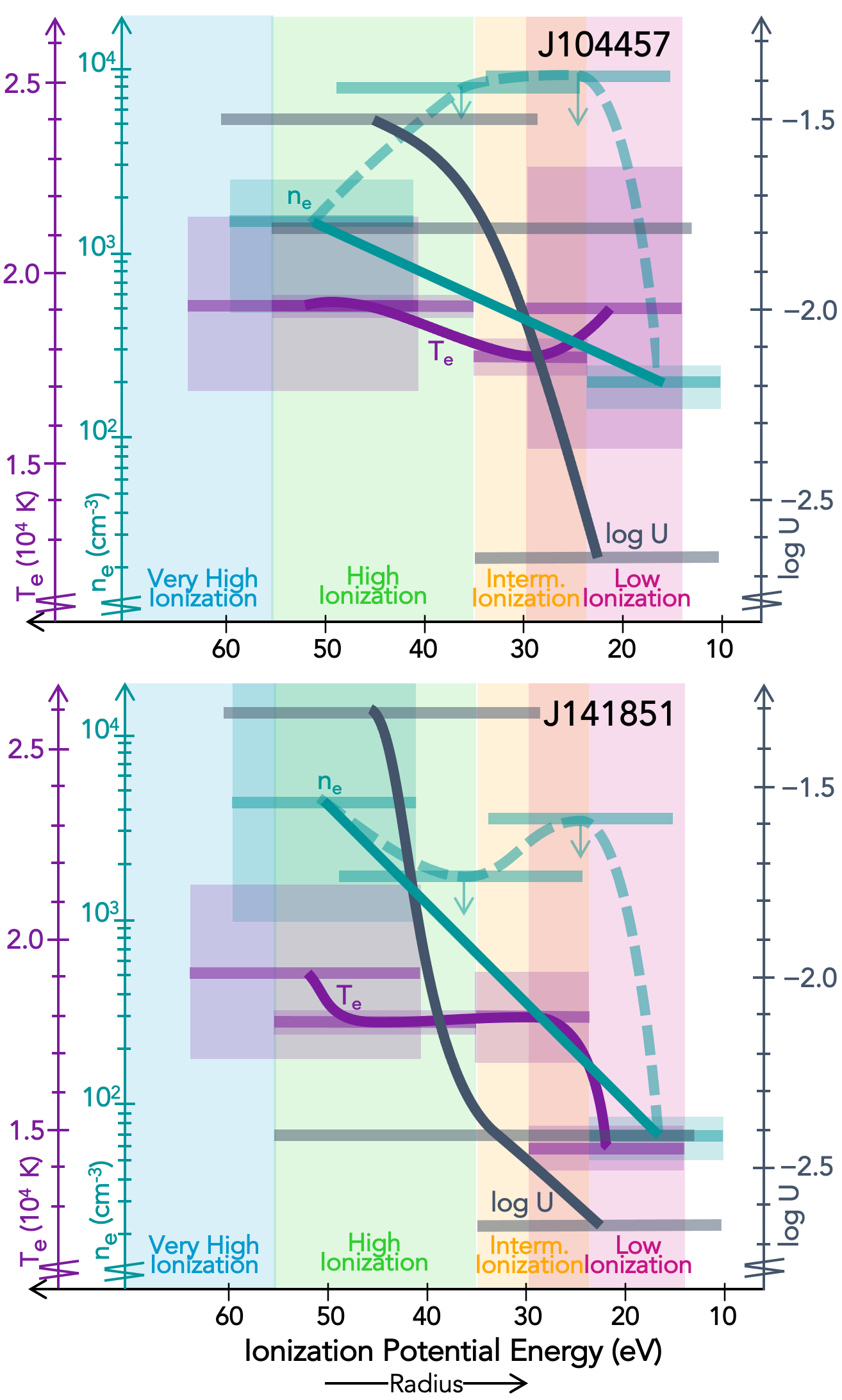} 
\caption{
A toy model view of the structure of a spherical EELG \ion{H}{2} region.
We used the electron temperature (purple), electron density (green),
and ionization parameter (gray) measurements for multiple ionization
zones to trace their changing nature as a function of radius for 
J104457 (top) and J141851 (bottom).
Each measurement is plotted as a horizontal bar extending the length
of it's corresponding ion's ionization potential energy and, for
$T_e$ and $n_e$, with a shaded vertical box representing the uncertainty.
Note that two density structures are traced for each galaxy:
(1) the darker green line connects the two density measurements 
made from the optical spectra that characterize the very-high- and
the low-ionization zones and
(2) the lighter green line connects all four density measurements,
including the upper limits for the high- and intermediate-ionization
zones derived from the UV spectra.}
\label{fig8}
\end{center}
\end{figure}

In this work we have explored the physical properties of two EELGs for both the 
classical 3-zone ionization model and the proposed 4-zone ionization model.  
In Sections~\ref{sec:3.1.1} and \ref{sec:3.2.2} we showed that examining multiple optical 
elemental line ratios allows us to probe the sub-volumes that compose a nebula
in terms of their the temperature, density, and ionization structures.
We map out these measurements in Figure~\ref{fig8}.
If we visualize our \ion{H}{2} regions with our simplified concentric shells model,
then we can over-plot the general shapes of how temperature, density, and ionization 
change as a function of radius. 


\startlongtable
\begin{deluxetable*}{lccc}
\tablecaption{3-Zone and 4-Zone Ionic and Total Abundances for EELGs}
\tablehead{
\CH{}                        & \CH{Ion. Zone}    & \CH{J104457}                        & \CH{J141851}                       \\ 
\multicolumn{1}{l}{Property} &\CH{3 Zone\M\ 4 Zone}& \CH{3 Zone\M\ 4 Zone}            & \CH{3 Zone\M\ 4 Zone}               }
\startdata
O$^0$/H$^+$ (10$^{-6})$     & N/A\M\ L          & 0.22\PM0.05                         & 0.84\PM0.12                          \\
O$^+$/H$^+$ (10$^{-6})$     & L                 & 1.14\PM0.25                         & 4.58\PM0.67                          \\
O$^{+2}$/H$^+$ (10$^{-6})$  & H                 & 27.0\PM0.67                         & 37.0\PM0.94                          \\
O$^{+3}$/H$^+$ (10$^{-6}$)  & \ N/A\M\ VH       & \hspace{7.5ex} N/A\M\ 0.011\PM0.001 & \hspace{7.5ex} N/A\M\ 0.013\PM0.001  \\
O$^{+3}$/H$^+_{UV}$ (10$^{-6}$)
                            & \ N/A\M\ VH   & \hspace{2.5ex} N/A\M\ 1.1\PM0.6     & \hspace{4.5ex} N/A\M\ 0.62\PM0.70        \\
O$^{0}$/O$_{tot.}$          &                   & 0.008\M\ 0.007                      & 0.020\M\ 0.019                       \\
O$^+$/O$_{tot.}$            &                   & 0.040\M\ 0.038                      & 0.108\M\ 0.104                       \\
O$^{+2}$/O$_{tot.}$         &                   & 0.952\M\ 0.917                      & 0.872\M\ 0.862                       \\
O$^{+3}$/O$_{tot.}$         &                   & \hspace{0.8ex}N/A\M\ 0.037          & \hspace{0.8ex}N/A\M\ 0.015           \\
12 + log(O/H)	            & All               & 7.44\PM0.01\M\ 7.44\PM0.01          & 7.62\PM0.01\M\ 7.62\PM0.02           \\
12 + log(O/H)$_{UV}$  	    & All               & \hspace{5ex} N/A\M\ 7.47\PM0.03     & \hspace{5ex} N/A\M\ 7.63\PM0.02      \\
\nicefrac{(O/H)}{(O/H)$_{\odot}$}
                            & All               & 0.056\PM0.002\M\ 0.058\PM0.003      & 0.084\PM0.003\M\ 0.087\PM0.004       \\
\\[-1.5ex] 
C$^{+2}$/O$^{+2}$		    & H                 & 0.174\PM0.037\M\ 0.186\PM0.041      & 0.147\PM0.035\M\ 0.169\PM0.037       \\ 
ICF(C$^{+2}$) 			    &                   & 1.212\PM0.201\M\ 1.281\PM0.202      & 0.960\PM0.200\M\ 1.095\PM0.201       \\ 
log(C/O)					& All               & $-0.76$\PM0.09\M\ $-0.73$\PM0.09    & $-0.83$\PM0.09\M\ $-0.77$\PM0.09     \\
\nicefrac{(C/H)}{(C/H)$_{\odot}$}
                            & All               & 0.018\PM0.004\M\ 0.020\PM0.006      & 0.022\PM0.005\M\ 0.027\PM0.006       \\
\\[-1.5ex] 
N$^{+}$/H$^+$ (10$^{-8})$	& L                 & 4.34\PM0.38                         & {14.4\PM1.4\ }                      \\
\multicolumn{2}{l}{ICF(N$^+$)}                  & 23.738\PM1.550\M\ 29.970\PM3.371    & \hspace{1.25ex}6.232\PM0.929\M\ 15.950\PM2.338 \\
\multicolumn{2}{l}{\ {\it ICF(N$^+$) (PC69)}}   & {\it 26.149\PM3.371\ }              & {\it 9.626\PM1.430}                 \\ 
\multicolumn{2}{l}{\ {\it ICF(N$^+$) (E20)}}    & {\it 31.507\PM0.100\ }              & {\it 11.845\PM0.100\ }              \\
12+log(N/H)                 & All               & 6.04\PM0.07\M\ 6.11\PM0.06          & 5.95\PM0.07\M\ 6.37\PM0.07          \\
\ {\it 12+log(N/H) (PC69)}  & All               & {\it 6.05\PM0.07}                   &{\it 6.14\PM0.07}                    \\
\ {\it 12+log(N/H) (E20)}   & All               & {\it 6.14\PM0.04}                   &{\it 6.23\PM0.04}                    \\
log(N$^+$/O$^+$)	        & L                 & $-1.41$\PM0.06\hspace{2ex}          & $-1.48$\PM0.07\hspace{2ex}          \\
log(N/O)		            & All               & $-1.43$\PM0.10\M\ $-1.35$\PM0.06    & $-1.66$\PM0.07\M\ $-1.26$\PM0.07    \\
\ {\it log(N/O) (PC69)}     & All               & {\it $-$1.41\PM0.07\hspace{2ex}}    &{\it $-$1.49\PM0.07\hspace{2ex}}     \\
\ {\it log(N/O) (E20)}      & All               & {\it $-$1.33\PM0.05\hspace{2ex}}    &{\it $-$1.39\PM0.04\hspace{2ex}}     \\
\nicefrac{(N/H)}{(N/H)$_{\odot}$}
                            & All               & 0.015\PM0.004\M\ 0.019\PM0.002      & 0.013\PM0.002\M\ 0.035\PM0.006      \\
\ {\it \nicefrac{(N/H)}{(N/H)$_{\odot}$} (PC69)}
                            & All               & {\it 0.017\PM0.002}                 & {\it 0.021\PM0.002}                 \\
\ {\it \nicefrac{(N/H)}{(N/H)$_{\odot}$} (E20)}  
                            & All               & {\it 0.020\PM0.002}                 & {\it 0.026\PM0.002}                 \\
\\[-1.5ex] 
S$^{+}$/H$^+$ (10$^{-7})$	& L                 & 0.34\PM0.06                         & 0.77\PM0.09                              \\ 
S$^{+2}$/H$^+$(10$^{-7})$	& I                 & 2.44\PM0.12                         & 3.56\PM0.39                              \\
S$^{+3}$/H$^+_{UV}$(10$^{-7})$	
                            & H                 & 3.39\PM2.04                         & 1.58\PM3.33                              \\
\multicolumn{2}{l}{ICF(S$^{+1,2}$)}         & 1.568\PM0.387\M\ 1.740\PM0.185      & 1.101\PM0.164\M\ 1.351\PM0.198           \\
\multicolumn{2}{l}{\ \it ICF(S$^{+1,2}$) (Th95)}          
                                                & {\it 5.260\PM0.526}                 & {\it 2.383\PM0.238}                      \\
12+log(S/H)                 & All               & 5.64\PM0.09\M\ 5.69\PM0.05          & 5.68\PM0.07\M\ 5.77\PM0.06               \\
12+log(S/H)$_{UV}$          & All               & 5.79\PM0.12                         & 5.77\PM0.19                              \\     
{\ \it 12+log(S/H) (Th95}   & All               &{\it 6.17\PM0.04}                    &{\it 6.01\PM0.05}                         \\
log(S/O)				    & All               & $-$1.80\PM0.09\M\ $-$1.78\PM0.06    & $-$1.94\PM0.07\M\ $-$1.86\PM0.06	     \\
log(S/O)$_{UV}$			    & All               & \hspace{6.8ex} N/A\M\ $-1.67$\PM0.13& \hspace{6.8ex} N/A\M\ $-1.86$\PM0.19	 \\
{ \it log(S/O) (Th95)}      & All               &{\it $-$1.30\PM0.05\hspace{2ex}}     &{\it $-$1.62\PM0.05\hspace{2ex}}          \\
\nicefrac{(S/H)}{(S/H)$_{\odot}$}
                            & All               & 0.033\PM0.007\M\ 0.037\PM0.004      & 0.036\PM0.006\M\ 0.045\PM0.006           \\
\nicefrac{(S/H)$_{UV}$}{(S/H)$_{\odot}$}
                            & All               & \hspace{7.3ex} N/A\M\ 0.037\PM0.004 & 0.045\PM0.021\M\ 0.045\PM0.006           \\
\ {\it \nicefrac{(S/H)}{(S/H)$_{\odot}$} (Th95)} 
                            & All               & {\it 0.111\PM0.010}                 & {\it 0.079\PM0.008}                      \\
\\[-1.5ex] 
Ar$^{+2}$/H$^+$(10$^{-8})$	& I                 & 6.50\PM0.39                         & 7.86\PM0.99                              \\ 
Ar$^{+3}$/H$^+$(10$^{-8})$	& \ \ H\M\ VH       & 6.35\PM0.23\M\ 6.09\PM0.32          & 16.6\PM0.60\M\ 11.0\PM2.90		         \\
\multicolumn{2}{l}{ICF(Ar$^{+2,3}$)}            & 1.014\PM0.250\M\ 1.010\PM0.107      & 1.062\PM0.158\M\ 1.021\PM0.150           \\
\multicolumn{2}{l}{\it \ ICF( Ar$^{+2,3}$) (Th95)}         
                                                & {\it 1.008\PM0.101}                 & {\it 1.013\PM0.101}                      \\
12+log(Ar/H)                & All               & 5.12\PM0.07\M\ 5.11\PM0.07          & 5.41\PM0.05\M\ 5.29\PM0.11               \\
\ {\it 12+log(Ar/H) (Th95}  & All               &{\it 5.11\PM0.03\M\ 5.11\PM0.07}     &{\it 5.39\PM0.04\M\ 5.28\PM0.11}          \\
log(Ar/O) 				    & All               & $-2.33$\PM0.07\M\ $-2.36$\PM0.08    & $-2.20$\PM0.05\M\ $-2.35$\PM0.12	     \\
\ {\it log(Ar/O) (Th95)}	& All               &{\it $-$2.33\PM0.04\M\ $-$2.36\PM0.08}&{\it $-$2.22\PM0.04\M\ $-$2.35\PM0.11}   \\
\nicefrac{(Ar/H)}{(Ar/H)$_{\odot}$}
                            & All               & 0.052\PM0.008\M\ 0.051\PM0.008      & 0.104\PM0.012\M\ 0.077\PM0.013           \\
\ {\it \nicefrac{(Ar/H)}{(Ar/H)$_{\odot}$} (Th95)} 
                            & All               & {\it 0.052\PM0.004\M\ 0.051\PM0.008}& {\it 0.099\PM0.008\M\ 0.076\PM0.012}     \\
\\[-1.5ex] 
Ne$^{+2}$/H$^+$(10$^{-6})$  & \ H\M\ VH         & 4.36\PM0.15\M\ 4.37\PM0.26          & 6.29\PM0.22\M\ 4.37\PM0.11		         \\ 
\multicolumn{2}{l}{ICF(Ne$^{+2}$)}              & 1.030\PM0.254\M\ 1.023\PM0.109      & 1.156\PM0.172\M\ 1.049\PM0.154           \\
\multicolumn{2}{l}{\it \ ICF(Ne$^{+2}$) (PC69)} & {\it 1.040\PM0.085}                 & {\it 1.117\PM0.049}                      \\
12+log(Ne/H)                & All               & 6.66\PM0.10\M\ 6.66\PM0.15          & 6.86\PM0.06\M\ 6.66\PM0.14               \\
\ {\it 12+log(Ne/H) (PC69)} & All               &{\it 6.66\PM0.02\M\ 6.66\PM0.15}     &{\it 6.85\PM0.03\M\ 6.39\PM0.13}          \\
log(Ne/O)   				& All               & $-0.79$\PM0.03\M\ $-0.81$\PM0.15    & $-0.75$\PM0.03\M\ $-0.97$\PM0.15         \\
\ {\it log(Ne/O) (PC69)} 	& All               &{\it $-$0.78\PM0.03\M\ $-$0.81\PM0.15}&{\it $-$0.77\PM0.03\M\ $-$0.94\PM0.14}   \\
\nicefrac{(Ne/H)}{(Ne/H)$_{\odot}$}
                            & All               & 0.053\PM0.011\M\ 0.053\PM0.018      & 0.085\PM0.012\M\ 0.054\PM0.014           \\
\ {\it \nicefrac{(Ne/H)}{(Ne/H)$_{\odot}$} (PC69)} 
                            & All               & {\it 0.054\PM0.002\M\ 0.054\PM0.019}& {\it 0.082\PM0.004\M\ 0.057\PM0.014}     \\
\\[-1.5ex] 
Fe$^{+2}$/H$^+$(10$^{-8})$  & L                 & 2.81\PM0.53                         & 7.03\PM0.36		                         \\ 
Fe$^{+3}$/H$^+$(10$^{-8})$  & H                 & 10.1\PM3.60                         & 26.1\PM4.1\hspace{1ex}                   \\ 
Fe$^{+4}$/H$^+$(10$^{-8})$  & VH                & \hspace{5ex} N/A\M\ 0.20\PM0.08     & \hspace{4.7ex} N/A\M\ 0.64\PM0.27		 \\ 
\multicolumn{2}{l}{ICF(Fe$^{+2}$)$_1$}          & 15.783\PM1.051\M\ 20.083\PM2.132    & \ 3.996\PM0.595\M\ 10.292\PM1.508        \\
\multicolumn{2}{l}{{\ \it ICF(Fe$^{+2}$)$_2$ (I09)}}
                                                &{\it 36.098\PM4.653\hspace{1ex}}     & {\it 13.199\PM1.961\hspace{1ex}}         \\
\multicolumn{2}{l}{ICF(Fe$^{+2,4}$)$_3$}        & \hspace{8.5ex} N/A\M\ 20.083\PM2.132& \hspace{8.5ex} N/A\M\ 10.292\PM1.508     \\
\multicolumn{2}{l}{ICF(Fe$^{+2,3,4}$)$_4$}      & \hspace{7.3ex} N/A\M\ 1.005\PM0.107 & \hspace{7.3ex} N/A\M\ 0.635\PM0.148      \\
12+log(Fe/H)$_1$            & All               & 5.65\PM0.12\M\ 5.77\PM0.06          & 5.42\PM0.08\M\ 5.88\PM0.08               \\
{\ \it 12+log(Fe/H)$_2$ (I09}   
                            & All               &{\it 5.96\PM0.12\M\ 6.02\PM0.07}     &{\it 5.94\PM0.08\M\ 5.99\PM0.08}          \\
12+log(Fe/H)$_3$            & All               & \hspace{5ex} N/A\M\ 5.80\PM0.06     & \hspace{5ex} N/A\M\ 5.92\PM0.08          \\
12+log(Fe/H)$_4$            & All               & \hspace{5ex} N/A\M\ 5.08\PM0.18     & \hspace{5ex} N/A\M\ 5.55\PM0.12          \\
log(Fe/O)$_1$	 			& All               & $-1.80$\PM0.11\M\ $-1.69$\PM0.07    & $-2.19$\PM0.08\M\ $-1.75$\PM0.08         \\
{\ \it log(Fe/O)$_2$ (I09)}	& All               &{\it $-$1.49\PM0.12\M\ $-$1.44\PM0.08}&{\it $-$1.68\PM0.08\M\ $-$1.64\PM0.08}   \\
log(Fe/O)$_3$	 			& All               & \hspace{7ex} N/A\M\ $-1.67$\PM0.07  & \hspace{7ex} N/A\M\ $-1.72$\PM0.08       \\
log(Fe/O)$_4$	 			& All               & \hspace{7ex} N/A\M\ $-2.38$\PM0.19  & \hspace{7ex} N/A\M\ $-2.09$\PM0.12       \\
\nicefrac{(Fe/H)$_1$}{(Fe/H)$_{\odot}$}
                            & All               & 0.015\PM0.004\M\ 0.019\PM0.002      & 0.009\PM0.002\M\ 0.024\PM0.002           \\
\ {\it \nicefrac{(Fe/H)$_2$}{(Fe/H)$_{\odot}$} (I09)} 
                            & All               & {\it 0.031\PM0.008\M\ 0.033\PM0.005}& {\it 0.029\PM0.005\M\ 0.031\PM0.005}     \\
\nicefrac{(Fe/H)$_3$}{(Fe/H)$_{\odot}$}
                            & All               & \hspace{7.5ex} N/A\M\ 0.020\PM0.003 & \hspace{7.5ex} N/A\M\ 0.026\PM0.003      \\
\nicefrac{(Fe/H)$_4$}{(Fe/H)$_{\odot}$}
                            & All               & \hspace{7.5ex} N/A\M\ 0.004\PM0.002 & \hspace{7.5ex} N/A\M\ 0.011\PM0.003           
\enddata
\tablecomments{
Ionic and total abundances for J104457 and J141851 using both the 3- and 4-zone models.
Column 2 specifies the ionization zone(s) of each property, where L = low, I =
intermediate, H = high, VH = very high, and All = all ionization zones.  
Abundances derived using an ICF from the literature are italicized.
Abundances relative to solar are given using the following notation:
[X/H] = (X/H)$/$(X/H)$_\odot$.
Specific notes for each element are provided below: \\
{\bf Oxygen}: The 4-zone O/H uses O$^{+3}$/H$^+$, which was determined in two ways:
(1) using the O$^{+3}$/O$^{+2}$ ratio predicted from photoionization models (see Fig.~\ref{fig9}); 
(2) using the UV \ion{O}{4} \W\W1401,1407 line detections relative to \ion{O}{3}] \W1666.\\
{\bf Carbon}: C/O was determined from the UV emission lines only. \\
{\bf Nitrogen}: N/H and N/O were determined using three different ICFs:
(1) this work (see Fig.~\ref{fig6}); 
(2) \citet{peimbert69}; 
(3)\citet{esteban20}.\\
{\bf Sulfur}: The corrections for missing ionization states for S/H and S/O were determined in 3 ways: 
(1) ICF from this work (see Fig.~\ref{fig6}); 
(2) using the UV \ion{S}{4} \W1406 line detection;
(3) ICF from \citet{thuan95}.\\
{\bf Argon}: Ar/H and Ar/O were determined using ICFs from:
(1) this work (see Figure~\ref{fig6}); 
(2) \citet{thuan95}. \\
{\bf Neon}: Ne/H and Ne/O were determined using ICFs from:
(1) this work (see Fig.~\ref{fig6}); 
(2) \citet{peimbert69}. \\
{\bf Iron}: Fe/H and Fe/O were determined using four different ICFs:
(1), (3), and (4) from this work (see Figure~\ref{fig7}); 
(2) \citet{izotov09}.}
\label{tbl5}
\end{deluxetable*}


All together, the multiple temperature, density, and ionization measurements used 
in this work provide a unique picture of the physical properties in EELGs.
Specifically, the ionization parameter measurements inform us of the basic shape of the ionizing radiation 
field and subsequent ionization structure\footnote{The shape of the ionizing radiation field and 
the resulting nebular emission lines are also significantly affected by dust. 
However, this is not likely a concern for the EELGs studied here, which have well-determined, very low reddening values.}.
For J104457 and J141841, the line ratios are indicative of a steep ionization gradient 
that is more highly ionized in the center and decreases with radius. 
The resulting ionization parameters measure an extreme change in the density of high-energy
ionizing photons across these nebulae, suggesting that most of the high-energy ionizing photons are
absorbed in the inner high- and very-high-ionization volumes. 

This new physical model of EELGs has important implications for the understanding and interpretation 
of both local EELGs and their high-redshift counterparts that are likely important drivers of reionization.
Perhaps most importantly, the ionization parameters in EELGs are misrepresented by the standard 3-zone model, 
indicating that harder radiation fields are present in these galaxies than previously thought. 
Here we discuss further differences between the two models and their implications
for interpreting EELGs in both the nearby and early universe. 

\subsection{Implications for Abundance Determinations}\label{sec:5.1}

\subsubsection{CNO Abundances}

The oxygen abundance of a galaxy is an important quantity, used to characterize its chemical
and evolutionary maturity.
Reassuringly, even for the high-energy ionizing radiation fields present in EELGs, the fraction of 
total O ions that are in a very-high ionization state (e.g., O$^{+3}$) is very small, and so the 
measured O/H abundance is essentially unaffected by the choice of a 3- versus 4-zone model. 
Interestingly, while the O$^{+3}$/O$_{tot.}$ fractions indicate that only 1--2\% of the oxygen 
ions are in the O$^{+3}$ state, the elevated log $U_{high}$ and $n_e$ values 
determined for the high- to very-high-ionization volume in Section~\ref{sec:3.2} suggests that a 
significant fraction of the ionizing photons are absorbed by the very-high-ionization volume. 

Arguably the next most useful abundance for characterizing galaxies is the relative N/O abundance.
The observed scaling of nitrogen with oxygen has long been understood as a combination of primary 
(metallicity independent) nitrogen production plus a linearly increasing fraction of secondary 
(metallicity dependent) nitrogen that comes to dominate the total N/O relationship at intermediate 
metallicities \citep[e.g.,][]{vce93, vanzee06a, berg12}. 
Owing to this important trend, ratios of the strength of N
emission relative to O emission and H$\alpha$ emission are popular strong-line
metallicity diagnostics,
especially the [\ion{N}{2}] \W6584/H$\alpha$ \W6563 diagnostic for studies of moderate redshift galaxies. 
However, because oxygen is primarily produced by massive stars on relatively short time scales ($\lesssim 40$ Myr),
while nitrogen is produced by both massive and intermediate mass stars on longer times scales ($\sim100$ Myr),
the N/O ratio is sensitive to the past star formation efficiency of a galaxy and serves as a 
clock since its most recent burst \citep[e.g.,][]{henry00,berg20}.
Therefore, while N/O variations may pose a challenge to simple strong line calibrations, 
they also serve as a powerful tool when considered as part of the abundance profile of a galaxy.
In this context, the very low N/O values measured for J104457 and J141851 may indicate
that the most recent burst of star formation is very young.

The total C/O abundances determined in Section~\ref{sec:4.4} rely on the emission ratio 
from C$^{+2}$ and O$^{+2}$ ions, which benefit from having similar excitation potentials 
and overlapping ionization ranges, and so shouldn't be significantly affected by non-uniform 
density and temperature distributions in the nebulae.
As expected, both J104457 and J141851 have C/O abundances of roughly 
30\% (C/O)$_\odot$ that vary by less than 10\% between the 3-zone and 4-zone ionization models.

In contrast to C/O abundances, ionic C abundances
show large differences between the 3-zone and 4-zone ionization models.
Using our detailed nebular analysis of J104457 and J141851 as constraints, we can use {\sc cloudy} 
models to predict the intrinsic \ion{C}{4} \W\W1548,1550 flux produced by photoionization.
For example, assuming reasonable values for our EELGs: 
Z$_{neb}=[0.05,0.10]$Z$_\odot$, C/O$_{neb} = 0.25$(C/O)$_\odot$, log$U=-1.5$, a stellar 
population age of [10$^{6.5},10^{6.7}$] yr, and uniform density of $n_e = [10^2,10^3]$ cm$^{-3}$,
the predicted \ion{C}{4} \W\W1548,1550/\ion{C}{3}] \W\W1907,1909 ratio is 0.40--0.72.
These flux ratios correspond to model C$^{+3}$/C$^{+2}$ ratios of 0.23--0.43.

In comparison, the observed \ion{C}{4} \W\W1548,1550/\ion{C}{3}] \W\W1907,1909 ratio is [0.91, 0.26],
corresponding to C$^{+3}$/C$^{+2}$ ratios of [0.22, 0.07] for [J104457, J141851]. 
These values suggest that J104457 produces \ion{C}{4} emission in close agreement 
of what current models predict, while only $\sim$20\%, at most, of the predicted 
\ion{C}{4} emission produced by J141851 is escaping the galaxy
\citep[c.f.,][]{berg19b}. 

As shown in this work, interpreting the physical properties of EELGs is complicated.
The relative emission from different ions is affected by many parameters,
and, thus, accurate abundance measurements require a detailed understanding 
of the physical conditions of the nebulae.
Ionic abundance determinations for ions spanning different ionization zones are 
particularly sensitive to the temperature distribution (see Appendix~\ref{sec:A1}).
Interestingly, density also plays a large role in our interpretation of \ion{C}{4}].
The true density distributions of our nebulae are likely complex, 
clumpy structures, but our simplified 4-zone model in which \ion{C}{3}] emission 
originates primarily from the intermediate-ionization zone and \ion{C}{4} emission 
from the very-high-ionization zone provides an informative upper limit.
If we employ the full range of densities that we measure such that \ion{C}{3}] emission
is associated with $n_e$ = (320, 130) cm$^{-3}$ gas and \ion{C}{4} emission is associated 
with $n_e$ = (1550, 2110) cm$^{-3}$ gas for J104457, J141851, respectively, then the 
photoionization models can reach remarkably large \ion{C}{4} \W\W1548,50/\ion{C}{3}] 
\W\W1907,09 ratios of [8.54, 5.18] for [0.05,0.10]Z$_\odot$.
While the intrinsic flux of the \ion{C}{4} \W\W1548,1550 resonant doublet can theoretically be used to
estimate the escape of \ion{C}{4} emission as a proxy for the escape of high energy photons through
high-ionization gas \citep{berg19b}, the models are currently too unconstrained to be useful. 

\subsubsection{\texorpdfstring{$\alpha$}{Alpha} Abundances}
\noindent {\it Neon:} 
For Ne/O abundances, we observe strong [\ion{Ne}{3}] emission, 
which originates partially from the very-high-ionization zone and 
partially from the dominant high-ionization zone.
Therefore, the Ne ICFs from both our models and \citet{peimbert69} are close to unity, 
implying small corrections and uncertainties.
As shown in Table~\ref{tbl5}, the Ne/O abundances for both cases are approximately 
solar for J104457 and the 3-zone model of J141851, but subsolar for the 4-zone model.
At the higher temperatures used in the 4-zone model, there must be a smaller fraction of 
Ne$^{+2}$ ions to produce the observed emission lines (see Figure~\ref{figA1} in 
Appendix~\ref{sec:A1}), resulting in lower abundances.
Interestingly, the fact that Ne/O abundances of the 3-zone model agree more closely 
with the expected solar ratio may indicate that very little of the [\ion{Ne}{3}] \W3869 
emission originates from the very-high-ionization zone.
Because the high-ionization zone dominates our nebulae (O$^{+2}$) and Ne${+2}$ is a 
bridge ion, this is not terribly surprising. 

\noindent {\it Argon:} 
For Ar/O abundances, we observe strong [\ion{Ar}{3}] and [\ion{Ar}{4}] 
emission, spanning, in part, the dominant high-ionization zone and beyond.
Therefore, similar to Ne, the Ar ICFs are close to unity for both our models and 
those of \citet{thuan95}, who also utilized [\ion{Ar}{4}] emission when present.
The Ar/O abundances we derive are approximately equal to or greater than solar for the 3-zone 
model, but are slightly subsolar for the 4-zone model, regardless of our choice of ICF. 
    
\noindent {\it Sulfur:} 
For S/O, we measure strong emission lines from [\ion{S}{2}] and [\ion{S}{3}], 
spanning the low- and intermediate-ionization zones, but no strong features 
probing the dominate high-ionization zone.
Therefore, to determine S/O ICFs, we created models specifically tailored 
to the conditions of EELGs. 
Adopting these new ICFs (see Figure~\ref{fig6} and the 4-zone ionization model, 
we find S/O abundances that are roughly [62\%, 51\%](S/O)$_\odot$.

As a test of our ICFs, we can use our weakly observed \ion{S}{4} \W1406 
fluxes to estimate the S$^{+3}$/S$^{+2}$ ratio.
Using a combination of line ratios from both our UV and optical spectra, we determine
S$^{+3}$/S$^{+2}$ = (S$^{+3}$/O$^{+2}$)$_{UV}$/(S$^{+2}$/O$^{+2}$)$_{opt.}$ 
values of [1.36,0.44] for [J104457, J141851] using the 4-zone model. 
In comparison, these S$^{+3}$/S$^{+2}$ fractions are significantly higher than
those predicted by our photoionization models: 
[0.60,0.77] for the [3-zone,4-zone] model in J104457 and 
[0.11,0.38] for the [3-zone,4-zone] model in J141851. 

Correcting for the estimated S$^{+3}$ contributions would require additional ICFs of 
[1.77, 1.16] and result in higher S/O abundances of [110\%, 60\%](S/O)$_\odot$.
We see large variations between the 3-zone and 4-zone models and variations 
between the model ICFs and using the UV line fluxes of up to 80\%. 
These large uncertainties associated with sulfur abundances in our EELGs are due to the 
lack of secure measurements of ions from the dominant high-ionization zones. 
However, the S/O abundances of our EELGs seem to be subsolar regardless of the model used. 
    

\subsection{The High-Energy Ionizing Photon Production Problem}\label{sec:5.2}

\begin{figure}
\begin{center}
    \includegraphics[width=0.8\columnwidth,trim=0mm 0mm 0mm 0mm,clip]{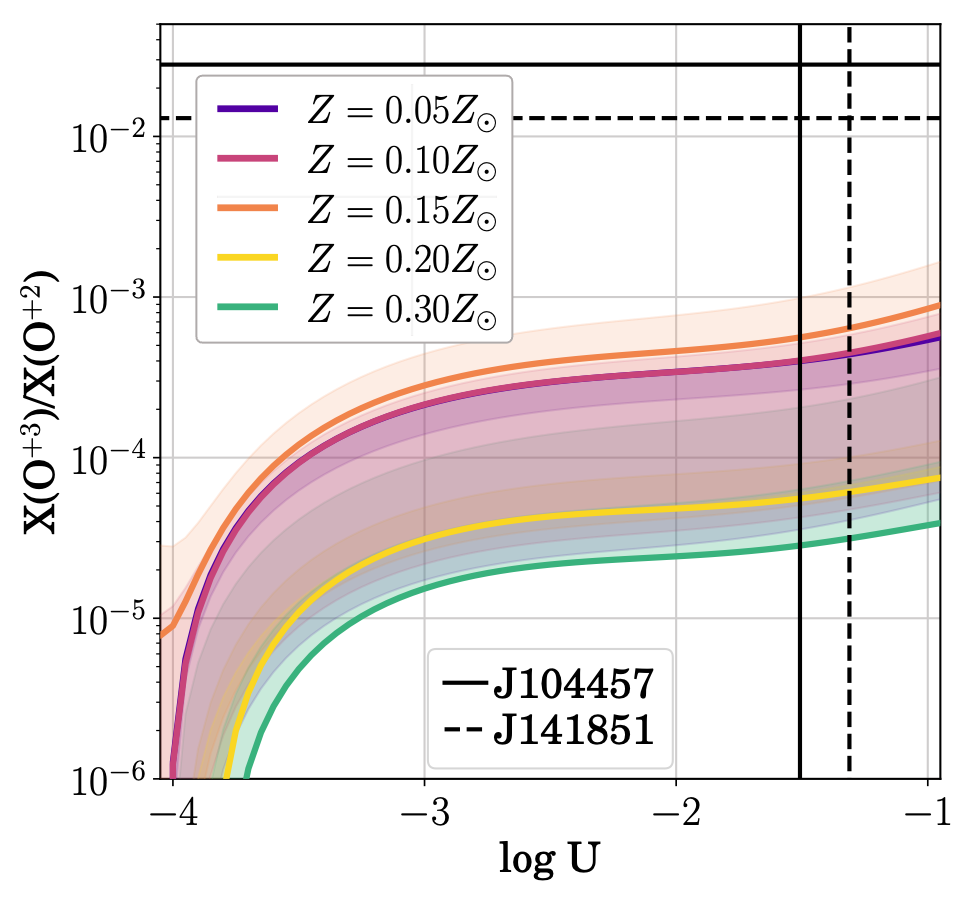}
\caption{
Photoionization models of the ionization fraction of O$^{+3}$ relative to O$^{+2}$
versus ionization parameter.
Models of a given metallicity (indicated by color) are shown for a burst age of $10^{6.7}$ yrs
(solid line) and spanning $10^6 - 10^7$ yrs (shaded regions).
Even for the low metallicities and high log$U_{high}$ values (vertical lines) measured
for J104457 and J141851, the models fail to reproduce the O$^{+3}$ fraction determined from 
the UV \ion{O}{4} emission lines (horizontal lines).}
\label{fig9}
\end{center}
\end{figure}

In the very-high ionization nebulae of EELGs, proper ICFs are especially important to account
for the potentially significant unseen ionization states.
We can test the robustness of our ICFs by comparing our observations of {\it pure} very-high-ionization
species, such as O$^{+3}$ and Fe$^{+4}$, to their model predictions.
In Figure~\ref{fig9} we consider the O$^{+3}$/O$^{+2}$ ratio, comparing our photoionization models 
(colored lines) with the measured ratios from the UV spectra of J104457 and J141851. 
Similar to the long-standing problem of nebular \ion{He}{2} production in blue compact 
dwarf galaxies \citep[e.g.,][]{kehrig15,kehrig18,berg18,senchyna19,stanway19}, we are clearly 
unable to reproduce O$^{+3}$ ionizing flux with stellar populations alone. 
Even for the extreme conditions in J104457 and J141851 and the small O$^{+3}$ contribution 
fractions measured for them, the photoionization models under-predict the O$^{+3}$/O$^{+2}$ 
ratio by more than an order of magnitude.
Additionally, we detect emission from [\ion{Fe}{5}] \W4143.15 and \W4227.19 in the optical 
LBT/MODS spectra of J104457 and J141851, and yet, the model Fe$^{+4}$ ion fraction in Figure~\ref{fig7} 
is negligible even for very-high ionization parameters.\footnote{
Note that the ICF(Fe$^{+2}$) in the middle panel of Figure~\ref{fig7} is equivalent to 
ICF(Fe$^{+2}+$ Fe$^{+4})= X(\rm{Fe}^{+2})+X(\rm{Fe}^{+4})$ because the $X(\rm{Fe}^{+4}$) 
contribution is erroneously negligible in the models.
Thus, ICF$_1=$ ICF$_3$ in Table~\ref{tbl5}. }
This discrepancy is indicative of the failure of photoionization models to accurately represent 
the conditions producing the very-high ionization-zone in EELGs.
Specifically, there seems to be a
{\it high-energy ionizing photon production problem} (HEIP$^3$)
that current stellar population synthesis models alone cannot solve. 

The HEIP$^3$ has little effect on our resulting oxygen abundance measurements (see Table~\ref{tbl5}) 
because we observed emission from the dominant ion in the nebula, O$^{+2}$. 
Similarly, for Ne and Ar we measure the species covering the dominant high-ionization zone,
and so our modeled ICFs, and subsequent abundance determinations, have small uncertainties.
On the other hand, the HEIP$^3$ may significantly affect our interpretation of elements with only
low-ionization species observed, such as S and N.
For S we observe S$^+$ and S$^{+2}$, but both become trace ions in very-high-ionization nebulae 
such that they require large, robust ICFs to determine accurate S/H abundances.  
The case for N/H abundances is similar.
Therefore, our N/H, S/H, and S/O abundances are likely biased low.
Relative N/O abundances may be an exception, as the N$^+$ and O$^+$ ions used in the calculation
trace each other fairly closely, and so should require similar ICFs.
However, the values in Table~\ref{tbl5} suggest that N/O determinations are in fact rather sensitive 
to different ICFs in the extreme environments of EELGs.

In general, for our EELGs, the HEIP$^3$ results in underestimated ionization parameters and 
ICFs that underestimate total abundances for observations of
low-ionization species, but has negligible effect when bridge and pure very-high-ionization species are observed. 
These biases may be even more extreme at high redshifts, where conditions are expected to be more extreme.
Iron, however, presents an interesting exception to this rule due to the so-called {\it [\ion{Fe}{4}] discrepancy}:
ICFs derived from photoionization models {\it over-predict} Fe$^{+3}$ abundances by more than a factor of four
compared to observations \citep[][and references therein]{rodriguez05}.
Similarly, we find our ICFs to predict a larger contribution from Fe$^{+3}$ than we measure 
directly from [\ion{Fe}{4}], by a factor of $\sim2-5$. 
Given these Fe$^{+3}$ and Fe$^{+4}$ discrepancies, further high S/N observations are needed to 
empirically constrain the Fe$^{+3}$ and Fe$^{+4}$ contributions to the Fe ICF.


\begin{figure}
\begin{center}
    \includegraphics[width=0.9\columnwidth,trim=0mm 0mm 15mm 0mm,clip]{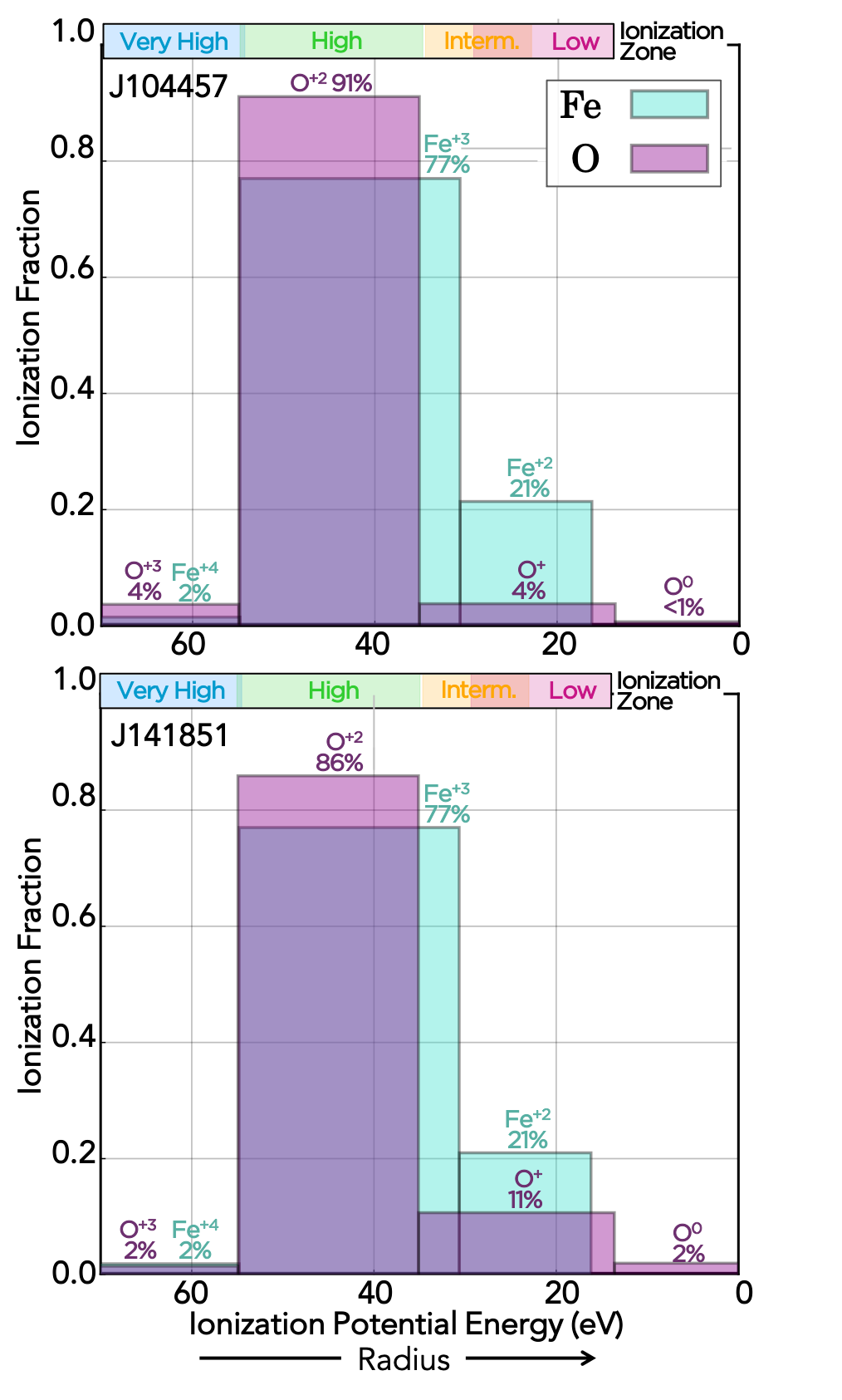} 
\caption{Ionization structure of two EELGs based on the O and Fe ionic abundance fractions 
reported in Table~\ref{tbl5}.
The O and Fe ions trace slightly different ionization potential ranges,
yet for both EELGs, the nebular ionization structure is similar.
While the high-ionization zone clearly dominates the ionic abundances for both EELGs,
the very-high-ionization ions are still important for interpreting
the volume-averaged ionization parameter.}
\label{fig10}
\end{center}
\end{figure}


\setlength{\tabcolsep}{3pt}
\begin{deluxetable}{lcc}
\tablecaption{Abundance Profiles of Two EELGs}
\tablehead{
{}                                   & {J104457}                             & {J141851} \\
{Solar Fraction}                     & \CH{3 Zone\M\ 4 Zone}                 & {3 Zone\M\ 4 Zone}}
\startdata
\multicolumn{3}{c}{\bf $\alpha$-Elements:} \\
\nicefrac{(O/H)}{(O/H)$_{\odot}$}   & 0.056\PM0.002\M\ 0.058\PM0.003       & 0.084\PM0.003\M\ 0.087\PM0.003   \\
\nicefrac{(Ar/H)}{(Ar/H)$_{\odot}$} & 0.052\PM0.008\M\ 0.051\PM0.008       & 0.104\PM0.012\M\ 0.077\PM0.013   \\
\nicefrac{(Ne/H)}{(Ne/H)$_{\odot}$} & 0.053\PM0.011\M\ 0.053\PM0.018       & 0.085\PM0.012\M\ 0.054\PM0.014   \\ 
\vspace{-0.5ex} \\
\multicolumn{3}{c}{\bf Non $\alpha$-Elements:} \\
\nicefrac{(N/H)}{(N/H)$_{\odot}$}   & 0.015\PM0.004\M\ 0.019\PM0.002       & 0.013\PM0.002\M\ 0.035\PM0.006   \\
\nicefrac{(C/H)}{(C/H)$_{\odot}$}   & 0.018\PM0.004\M\ 0.020\PM0.006       & 0.022\PM0.006\M\ 0.027\PM0.006   \\
\nicefrac{(Fe/H)}{(Fe/H)$_{\odot}$} & 0.015\PM0.004\M\ 0.020\PM0.003       & 0.009\PM0.002\M\ 0.026\PM0.003
\enddata
\tablecomments{
Total abundances of J104457 and J141851 relative to solar.
The abundances are split into two groups of similar chemical composition.
For N, Ar, Ne, and Fe, where we reported we reported multiple abundances 
derived from different ICFs, we have adopted the abundances using ICFs
derived in this work. 
For Fe, we use ICF(Fe$^{+2}$)$^2$ for the 3-zone model and ICF(Fe$^{+2}+$Fe$^{+4}$)$^3$ for the 4-zone model.}
\label{tbl6}
\end{deluxetable}
 


\subsection{The Abundance Profile of EELGs:  \\
Evidence of Hard Radiation Fields?}\label{sec:5.3}

We examined ionic abundances of O, N, C, Ne, Ar, S, and Fe in the previous sections
to better understand the ionic structure of EELGs.
Figure~\ref{fig10} summarizes our resulting ionization model
using the ionic abundances measured for O and Fe, which allow us to map 
each of the zones in the 4-zone ionization model.
For both J104457 and J141851, the high-ionization ions clearly dominate their 
respective elements such that the very-high-ionization zone has little effect on 
the total abundance.
However, the very-high-ionization zones do play an important role in our
interpretation of the volume-averaged ionization parameter of a galaxy.

In general, we found that
(1) photoionization models with stellar-population SEDs generally fail to reproduce
the high-ionization species observed in EELGs (the HEIP$^3$),
(2) the higher ionization parameters determined for the 4-zone model indicate a 
larger fraction of metals are in high-ionization states than predicted in the 3-zone model,
and 
(3) the higher temperatures assumed in the 4-zone model can reduce abundances of 
high-ionization species.
The latter point means that if high temperatures play a critical role in EELGs, 
then nebular abundances determined primarily from their high-ionization lines may 
be somewhat overestimated by the standard 3-zone model.
On the other hand, point (2) results in elemental abundances determined primarily 
from their lower-ionization species (i.e., N, S, and Fe) being underestimated
by the 3-zone model.

Regardless of the ionization model, we measured all elements to be 
significantly subsolar for J104457 and J141851, as expected for low-mass galaxies.
However, the various elements span a range of abundances relative to solar from
1.9\%--5.8\%\ for J104457 and 2.6\%--8.7\%\ for J141851, using the 4-zone model.
To better compare the abundance profiles for J104457 and J141851, as determined by the 3- and 4-zone
models, we report the adopted elemental abundances relative to solar in Table~\ref{tbl6}.
In general, we have adopted the abundances determined with the ICFs of this work. 
Given the sensitivity of the S/H calculations to the assumed ICF and, subsequently,
the large uncertainties, we do not analyze the S/H abundances further.
For Fe, we consider the opposing challenges of the photoionization models for different Fe ions, 
and thus adopt the log(Fe/O) abundances based solely on the Fe$^+$ measurements for the 
3-zone model and the log(Fe/O) based on Fe$^+$ + Fe$^{+3}$ abundances for the 4-zone model. 

We find that the abundances in Table~\ref{tbl6} naturally split into two populations:
(1) $\alpha$-elements: Ar/H, and Ne/H, which have abundances relative to solar consistent with O/H, and
(2) non-$\alpha$-elements: N/H, C/H, and Fe/H, which have solar-scaled abundances that are deficient 
relative to O/H and the other $\alpha$-elements.
The $\alpha$-element trend aligns with typical nucelosynthetic arguments, where O, S, Ar, and Ne 
are all produced predominantly on short timescales ($\lesssim$ 40 Myr) by core-collapse supernovae (SNe), 
and so should all follow a consistent abundance profile.
However, the recent chemical evolution models of \citet{kobayashi20} indicate that 
while oxygen is produced mostly by CC SNe and a bit by AGB stars, S and Ar also have a 
significant contribution from Type Ia SNe (29\% and 34\% respectively), resulting in 
delayed enrichment relative to the CC SNe production. 
In this case, the variations in abundances ranging roughly 5--6\% Z$_\odot$ and 5--9\% Z$_\odot$ 
in the $\alpha$-elements for J104457 and J141851, respectively, may be explained by the 
individual star formation histories of these galaxies.

In contrast, the solar scaled non-$\alpha$-element abundances are roughly 2--3$\times$ lower than those of the $\alpha$-elements.
One concern is that Fe strongly depletes onto dust and so the Fe abundances may be strongly 
affected such that Fe/H is biased to even lower abundances. 
\citet{izotov06} measured subsolar Fe/O abundance ratios from metal-poor galaxies in the 
Sloan Digital Sky Survey DR3, suggesting that iron depleting onto dust grains was responsible, 
but that this effect decreases with decreasing metallicity. 
Therefore, the effect should be small for the low metallicities and very small reddening values of our EELGs. 
Further, Table~\ref{tbl6} shows that Fe and N have similar abundance levels relative to solar, 
but N doesn't deplete onto dust, so the level of Fe dust depletion is likely to be small.
If high dust destruction rates are occurring in these metal-poor galaxies due to 
their very hard radiation fields, then Fe may not be significantly locked up in dust.
But even if we are missing a fraction of the Fe abundance, the N/H and C/H abundances 
suggest that the $\alpha$-elements are truly under-abundant, perhaps due to the 
longer timescales of their production via Type Ia SNe and AGB stars.

\subsubsection{\texorpdfstring{$\alpha$}{Alpha}/Fe Enrichment}
This result of enhanced $\alpha$/Fe abundances has been suggested as the source
of the extremely hard radiation fields inferred from the stellar continua and
emission line ratios in chemically-young, high-redshift galaxies 
\citep[e.g.,][]{steidel18, shapley19, topping20}.
These authors argue that the highly super-solar O/Fe abundances ($\sim4-5\times$(O/Fe)$_\odot$)
are expected for the brief star formation histories of $z\sim2-3$ galaxies,
whose enrichment is dominated by the 
core-collapse SNe products of their recent burst of star formation. 
In this case, O dominates the cooling and emission lines of the ionized gas,
whereas Fe determines FUV opacity and mass-loss rate from massive stars 
(although this effect is less certain in very metal poor stars),
resulting in a harder extreme-UV radiation field than would be generated by
stars and gas with a solar O/Fe composition.


\begin{figure}
\begin{center}
    \includegraphics[width=0.85\columnwidth]{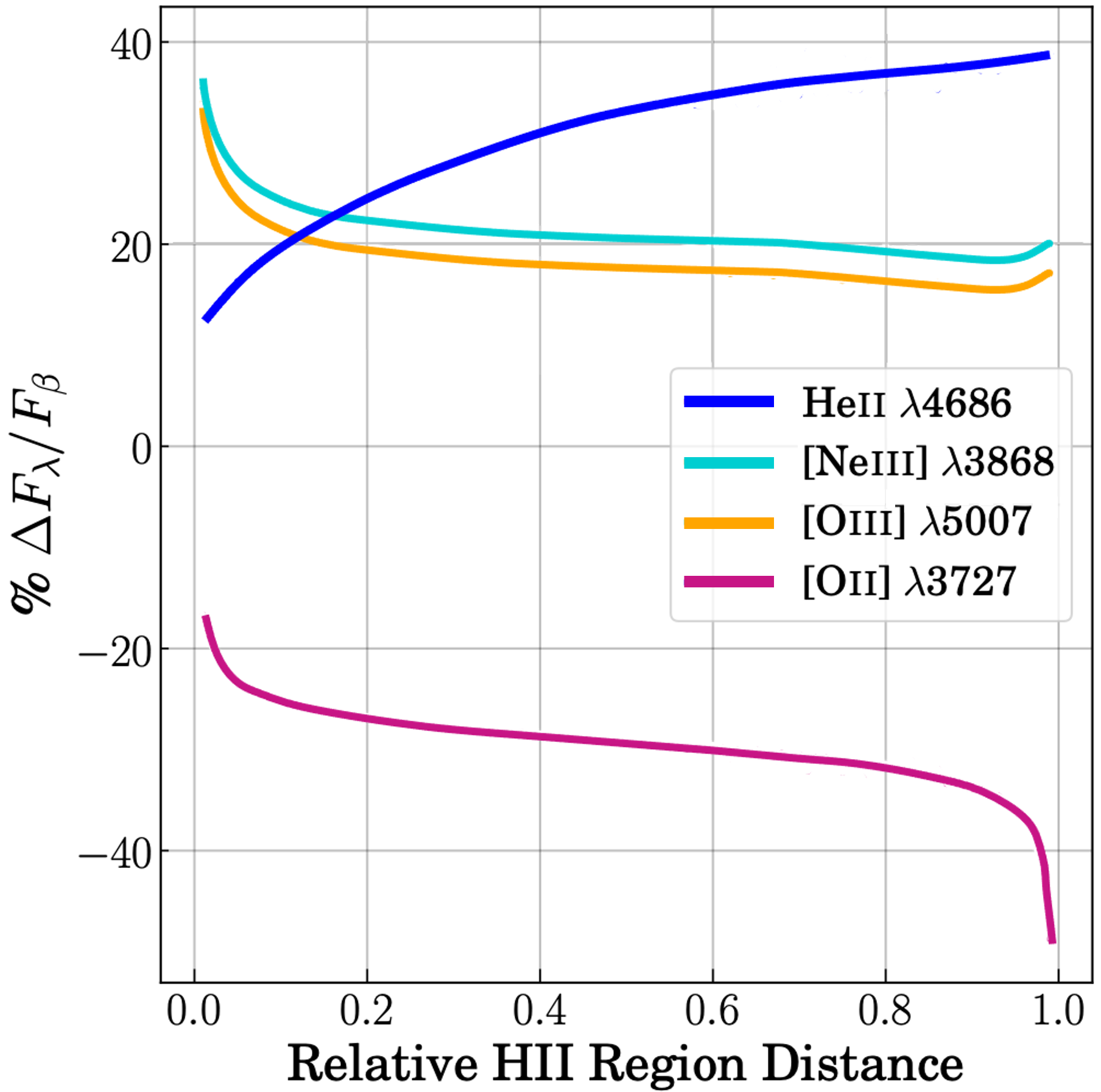}
\caption{
Fractional difference in emission line fluxes for an $\alpha$/Fe-enriched 
photoionization model (i.e., gas abundance $>$ stellar abundance) relative
to a classic model (i.e., gas abundance $=$ stellar abundance) for metal-poor
($Z=0.05Z_\odot$) EELGs with log$U=-1$ and a BPASS ionizing SED with a
burst age of $t=10^{6.5}$ yrs.
In general, $\alpha$/Fe enhancement results in low-ionization emission such as 
[\ion{O}{2}] \W3727 decreasing and high-ionization emission like [\ion{O}{3}] \W5007
and [\ion{Ne}{3}] \W3868 increasing. 
Interestingly, very-high-ionization emission like \ion{He}{2} \W4686 becomes
more radially extended as a result of $\alpha$/Fe-enrichment.}
\label{fig11}
\end{center}
\end{figure}


BCDs have also been argued to be cosmologically young systems that formed most of their 
stellar mass in the past $\sim2$ Gyr \citep[e.g.,][]{guseva01,pustilnik04,papaderos08}.
However, \citet{papaderos08} found the SDSS imaging morphologies of a sample of BCDs, 
including J104457, to reveal a redder stellar host implying these galaxies are unlikely
to be forming their first generation of stars. 
Further, \citet{janowiecki17} found that the spectral energy distributions of BCDs in the 
Local Volume Legacy survey \citep[LVL;][]{dale09} were better fit with a two-burst model 
than a single-burst, with old stellar populations ranging from $\sim3-5$ Gyr in age. 
Therefore, Fe-poor massive stars are not expected to be prevalent in
local BCD galaxies due to their longer, complicated star formation histories that 
result in Fe enrichment from older stars. 

For J104457 and J141851, we measure O/Fe abundances that are similar to the values 
inferred from $z\sim2-3$ galaxies ($\gtrsim3\times$(O/Fe)$_\odot$), but seem to  
have a different origin.
Due to the different timescales of CC SNe and Type Ia SNe,
young bursts of star-formation in EELGs may result in increased O/Fe ratios due to 
a recent injection of O, where the associated Fe has not yet been released by Type Ia SNe.
This idea is supported by the chemical evolution models of \citet{weinberg17}, where a 
sudden burst of star formation with an initial abundance of $Z=0.3 Z_\odot$ can 
temporarily boost O/Fe by up to $\sim0.45$ dex if a significant fraction of the 
gas is consumed and the core-collapse SNe yields are retained. 
The effect is expected to be even stronger for the very low metallicities of our EELGs,
and could then account for the $\alpha$/Fe enhancement observed. 
However, this model is complicated by the expectations for low-mass galaxies 
to have high metal-loading factors \citep{peeples11,chisholm18b} and a range of effective 
oxygen yields \citep{yin11,berg19a}.

We explore the potential impact of $\alpha$/Fe enrichment on the observed 
emission-line ratios of EELGs in Figure~\ref{fig11}.
To do so, we compared two {\sc cloudy} photoionization models with gas-phase parameters 
matched to our EELGs and an input {\sc BPASS} ionizing SED with a metallicity either 
matched to the gas-phase ($Z = 0.05 Z_\odot$) or 10 times deficient relative to the 
gas-phase ($Z = 0.005 Z_\odot$).
The latter case mimics extreme $\alpha$/Fe enrichment.
The resulting percent differences in the emission-line fluxes are plotted
in Figure~\ref{fig11} as a function of radius. 
The $\alpha$/Fe enriched model produces larger flux ratios for high-ionization
emission lines, by up to $\sim$40\% for lines such as \ion{He}{2} \W4686,
[\ion{O}{3}] \W5007, and [\ion{Ne}{3}] \W3868, and smaller flux ratios for
low-ionization lines such as [\ion{O}{2}] \W3727.
However, even the boosted \ion{He}{2} fluxes of this extreme $\alpha$/Fe enriched
model fail to reproduce the observed \ion{He}{2}/H$\beta$ ratios of EELGs (0.01--0.02) 
by a factor of 5--10.
Therefore, $\alpha$/Fe enrichment may be responsible for the observed properties 
of typical $z\sim2-3$ galaxies, such as the BPT offset to higher [\ion{O}{3}]/H$\beta$ values
\citep[e.g.,][]{strom17}, but it catastrophically fails to solve the HEIP$^3$ in EELGs.


\begin{figure}
\begin{center}
    \includegraphics[width=1.0\columnwidth,trim=0mm 0mm 0mm 0mm,clip]{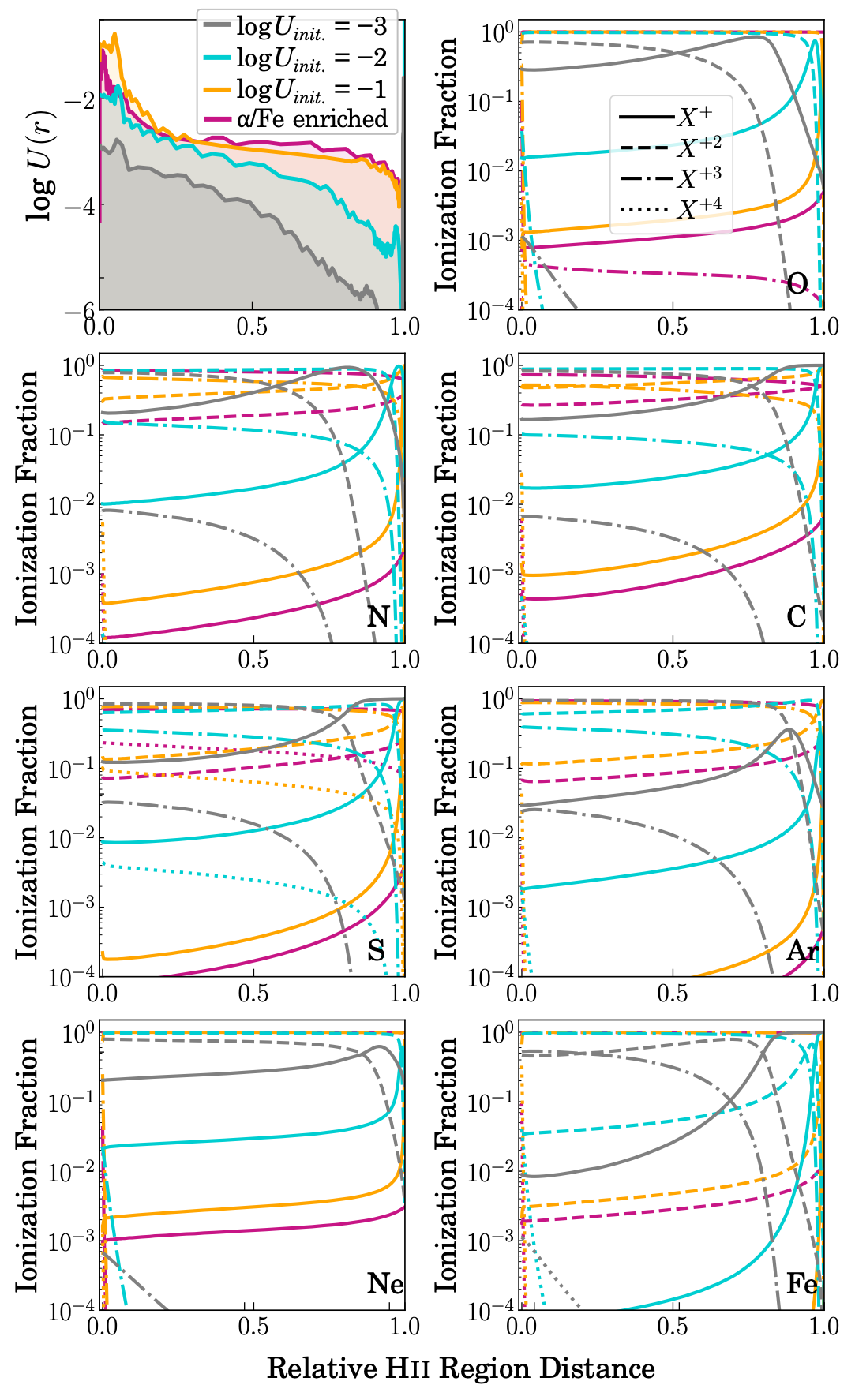}
\caption{Photoionization models showing how the ionization structure of a simple sphere
changes as a function of radius within an \ion{H}{2} region for different ionization parameters, 
as defined at the initial face of the cloud.
Specifically, the top left panel shows the ionization parameter gradient and the other panels
compare how the ionization fraction profiles of a given element compare.}
\label{fig12}
\end{center}
\end{figure}


\begin{figure*}[t]
\begin{center}
    \includegraphics[width=0.9\textwidth,trim=5mm 3mm 0mm 0mm,clip]{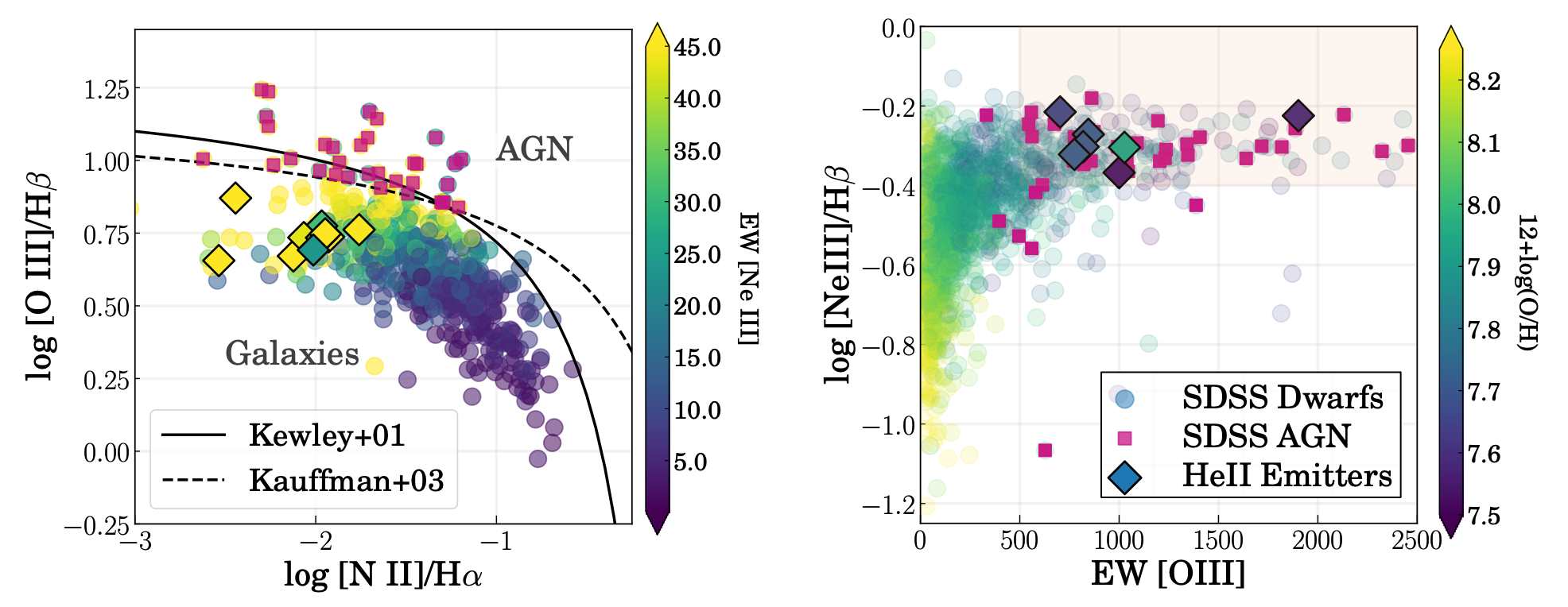}
\caption{
Emission-line ratios characterizing EELG \ion{He}{2}-emitters.
{\it Left:} The [\ion{O}{3}]\W5007/H$\beta$ versus [\ion{N}{2}]\W6584/H$\alpha$ BPT 
diagram is plotted for a low-mass ($M_\star < 10^{9} M_\odot$) subset of the SDSS 
DR14 (circles) and color-coded by the [\ion{Ne}{3}] \W3868 EW as a proxy for the 
strength of very-high-ionization lines. 
The solid lines are the theoretical starburst limits from \citet{kewley01} and
\citet{kauffmann03}, designating outliers as AGN (magenta squares).
In comparison, EELG \ion{He}{2}-emitters from \citet{berg16} and \citet{berg19a}
occupy upper-left tail where excitation is high and metallicity is low.
{\it Right:} For [\ion{Ne}{3}]\W3868/H$\beta$ versus the EW of [\ion{O}{3}] \W5007,
the EELG \ion{He}{2}-emitters occupy the same upper-right, extreme parameter space as AGN.
These plots indicate that 4-zone EELGs are likely produced by the combination of 
very high excitation and ionization with powerful emission relative to the stellar continuum.}
\label{fig13}
\end{center}
\end{figure*}

\subsection{The Ionization Structure of EELG Nebulae}\label{sec:5.4}
In Section~\ref{sec:3.2.2}, we calculated three different volume-averaged
ionization parameters to characterize different ionization zones: 
log $U_{low}$([\ion{S}{3}]/[\ion{S}{2}]), 
log $U_{int}$([\ion{O}{3}]/[\ion{O}{2}]), and
log $U_{high}$([\ion{Ar}{4}]/[\ion{Ar}{3}]). 
Together, these three log $U$ estimates describe an ionization structure 
gradient that is more highly ionized in the center and decreases with radius.

We demonstrate the theoretical ionization structures of photoionized 
nebulae in the top left panel of Figure~\ref{fig12}.
Here we compare three {\sc cloudy} photoionization models with parameters matched 
to our EELGs, but different ionizing radiation field strengths, as defined by the 
initial ionization parameter, log $U_{init}$ at the illuminated face of the cloud.
Keeping all other parameters identical, the only differing variable between the gray, turquoise, 
and gold models is the surface flux of ionizing photons as a function of radius.
The pink model is the same log$U_{init}=-1$ $\alpha$/Fe enriched model as plotted 
in Figure~\ref{fig11}.
We use the log $U_{init} = -1$ and log $U_{init} = -2$ models as a good approximation
for the large log $U_{high}$ values determined for the very-high-ionization zones
of our EELGs and use the log $U_{init} = -3$ model to represent typical \ion{H}{2} 
regions \citep[e.g.,][]{dopita00,moustakas10}.
In comparison, the inner ionization structure of EELGs is steeper than is 
expected for typical \ion{H}{2} regions by a factor of 10.
This extreme ionization structure is also demonstrated by the large $\Delta$log $U$ 
values ($-1.14, -1.35$) observed across the different ionization volumes of our EELGs, 
suggesting that most of the high-energy ionizing photons 
are absorbed in the inner high- and very-high-ionization volumes.

While we have shown that this type of detailed analysis is very useful
for interpreting the physical properties of nearby galaxies, 
it does not directly translate to practical applications for high redshift galaxies,
where only a few emission lines are typically observed. 
Therefore, we also quantified an ionization parameter typical of the 
{\it entire} 4-zone nebulae of EELGs.
To do so, we consider the ionization structure of our EELGs in Section~\ref{sec:4.1},
where the ionization fractions of oxygen suggested that the high-ionization zone is 
dominant in EELG nebulae, with very-high-ionization zones that can be comparable 
to the low- to intermediate-ionization zones.

For J104457 and J141851, we found that the volume-averaged ionization parameters 
for the 4-zone model (log $U_{ave} = -1.70, -1.91$, respectively) are noticeably 
higher than those of the 3-zone model (log $U_{ave} = -1.77, -2.42$, respectively).
This exercise has an important implication:
ionization parameters in EELGs determined by the standard 3-zone model misrepresent 
(underestimate) the steepness of the ionization structure and the volume-average
ionization parameter. 
Given the known luminosity of observed EELGs, this may imply that the 3-zone model 
underestimates the hardness of the underlying radiation field. 
Our forthcoming work (G. Olivier et al. 2021, in preparation)
will further examine any implications the 4-zone
model has on the shape of the ionizing spectrum by simultaneously modeling 
the ionizing stellar population and suite of observed emission lines.

\subsection{Recommendations for EELG Studies}\label{sec:5.5}
Although our detailed abundance analyses are only for two EELGs,
they provide an important benchmark with which to compare other EELGs, 
both near and far, and suggest diagnostic corrections. 
To determine when corrections are appropriate and most important, we examine 
how the properties of EELGs compare to a more general population of dwarf galaxies.
In Figure~\ref{fig13} we plot extreme UV \ion{He}{2} emitters (EW(\ion{He}{2}\W1640)$>1$ \AA)
from \citet{berg19a} compared to dwarf galaxies ($M_\star < 10^9 M_\odot$) 
from the SDSS Data Release 14 \citep{blanton17}.

Figure~\ref{fig13} shows that \ion{He}{2} EELGs occupy the limits of the 
high-ionization, low-metallicity parameter space in both the BPT (left panel)
and the [\ion{Ne}{3}]\W3868/H$\beta$ vs EW([\ion{O}{3}]\W5007) (right panel) diagnostic diagrams. 
These properties align with our expectations for EELGs and observed early universe galaxies:
very-high-ionization emission lines, such as \ion{He}{2}, are produced in
metal-poor galaxies with high star formation rates and hard radiation fields, 
as indicated by their large equivalent widths of high-ionization emission lines. 
These conditions, along with the observed {\it pure} very-high-ionization 
emission lines, are indicative of the presence of a very-high-ionization zone
and require caution in interpreting their spectra.

By considering the parameter space occupied by nebular \ion{He}{2} emitters, 
Figure~\ref{fig13} can be used to predict which galaxies are likely to have a very-high-ionization zone.
Specifically, good candidates can be identified with a combination of large [\ion{Ne}{3}]/H$\beta$ 
line ratios ($>-0.4$) indicating hard radiation fields and large [\ion{O}{3}] EWs ($> 500$ \AA) 
indicating large specific star formation rates.

We have shown that EELGs have enhanced high-ionization
zones and smaller low-ionization zones.
In general, at higher ionization parameters, low-ionization lines, such as [\ion{O}{2}] \W3727, 
[\ion{N}{2}] \W\W6548,84, and [\ion{S}{2}] \W\W6717,31, have smaller fluxes and high-ionization 
lines, such as [\ion{Ne}{3}] \W3868 and [\ion{O}{3}] \W5007, have larger fluxes. 
This results in additional errors in strong-line metallicity measurements beyond their 
standard biases, and is especially important for high redshift galaxies, 
where direct abundances are rarely accessible. 
Many of the strong-line calibrations involving low-ionization lines will 
{\it underestimate} both the calibrator and the true metallicity.
These include:
\begin{itemize}
    \setlength\itemsep{-0.35em}
    \item N2 = log([\ion{N}{2}]\W6584/H$\alpha$),
    \item S2 = log([\ion{S}{2}]\W\W6717,31/H$\alpha$),
    \item O32 = log([\ion{O}{3}]\W5007/[\ion{O}{2}]\W3727),
    \item Ne3O2 = log([\ion{Ne}{3}]\W3868/ [\ion{O}{2}]\W3727), and
    \item O3N2 = log(([\ion{O}{3}]\W5007/H$\beta$)/ ([\ion{N}{2}]\W6584/H$\alpha$)),
\end{itemize}
\noindent Other oxygen-based calibrations are double valued, such that 
R3 = log([\ion{O}{3}]\W5007/H$\beta$) will {\it overestimate} and 
R23 = log(([\ion{O}{2}]\W3727+[\ion{O}{3}]\W5007)/H$\beta$) will {\it underestimate}
the true metallicity for 12+log(O/H) $< 8.0$ and $>8.0$, respectively. 
For the \citet{berg19a} \ion{He}{2} emitters, we measure log $U_{high}$ to be greater than
log $U_{int}$ by 0.30$\pm$0.15 dex.

We, therefore, recommend the following guidelines for very-high-ionization galaxy candidates:
(1) Ionization parameters inferred from [\ion{O}{3}]/[\ion{O}{2}] should be
considered lower limits for very-high-ionization galaxy candidates.
(2) Strong-line oxygen abundances should be considered lower limits for most calibrations,
while the R3 and R23 calibrators should be avoided all together. 


\section{Summary and Conclusions} \label{sec:6}

We investigated the high-quality {\it HST}/COS UV and LBT/MODS optical spectra 
of two nearby, metal-poor {\it extreme emission line galaxies} (EELGs), 
J104457 ($Z=0.058Z_\odot$) and J141851 ($Z=0.087Z_\odot$).
These galaxies have very strong high-ionization nebular emission-line features,
including \ion{He}{2} \W1640, \ion{C}{4} \W\W1548,1550,
[\ion{Fe}{5}] \W4227, \ion{He}{2} \W4686, and [\ion{Ar}{4}] \W\W4711,4740
(see Figures~\ref{fig2} and \ref{fig3}), that liken them to reionization-era systems.

Typical stellar population models produce radiation fields that can only significantly 
ionize oxygen up to the O$^{+2}$ species.
As a result, a fully ionized \ion{H}{2} region is typically characterized by a 3-zone model, 
with a low-ionization zone defined by N$^+$ (14.5--20.6 eV),
an intermediate-ionization zone defined by S$^{+2}$ (23.3--34.8 eV), and
a high-ionization zone defined by O$^{+2}$ (35.1--54.9 eV).
We showed this structure in Figure~\ref{fig4}, where the outer edge of an \ion{H}{2} region 
is defined by the lower H-ionizing edge ($> 13.6$ eV) and extends inward to the upper limit of 
O$^{+2}$ (< 54.9 eV). 
However, the ionization potentials of the very-high-ionization lines observed here
extend to energies higher than the upper limit of the standard 3-zone ionization model 
and are indicative of very hard radiation fields.
To better characterize the extreme, extended ionization parameter space of the nebular 
environments of EELGs, we define a new 4-zone ionization model that includes the addition
of a very-high-ionization zone, characterized by the He$^{+2}$ ion 
($>54.4$ eV; see Figure~\ref{fig4}).

Using the 4-zone model, we measured the nebular properties and ionic abundances in
each ionization zone of the two EELGs studied here and then used the results
to re-evaluate their average properties and structure. 
In general, we find that the addition of a very-high-ionization zone 
has little to no effect on the integrated nebular abundances, but does
change the interpretation of several physical properties of EELGs.
Specifically, the main results of this work are:

\begin{enumerate}[leftmargin=*]
\item Our 4-zone determination of the nebular properties showed
that temperature, density and ionization parameter peak in the
very-high-ionization zone of EELGs (Figure~\ref{fig8}).
To measure this, we adopted the commonly-used temperature and density diagnostics
for each zone of the standard 3-zone model and added 
$T-e$([\ion{Ne}{3}] \W3342/\W3869) ($41.0 - 63.5$ eV) and 
$n_e$([\ion{Ar}{4}] \W4711/\W4740) ($40.7 - 59.8$ eV) diagnostics
for the very-high-ionization zone.
Using these diagnostics, in the very-high ionization zones, 
we measured temperatures of $T_e = 19,200\pm2,300$ K and $23,600\pm3,200$ K and
densities of $n_e = 1,550\pm1,100$ cm$^{-3}$ and $2,100\pm1,300$ cm$^{-3}$ for 
J104457 and J141851, respectively.
We also considered, for the first time, multiple ionization parameters to 
characterize different ionization zones, adopting:
log$U_{low}$([\ion{S}{3}]/[\ion{S}{2}]),
log$U_{int}$([\ion{O}{3}]/[\ion{O}{2}]), and
log$U_{high}$([\ion{Ar}{4}]/[\ion{Ar}{3}]).

\item Our 4-zone determinations of total abundances in EELGs are consistent
with the 3-zone model when all relevant ions are observed.
Specifically, we measured ionic abundances for all the relevant O ions 
(O$^0$, O$^+$, O$^{+2}$, and O$^{+3}$) and Fe ions 
(Fe$^+$, Fe$^+2$, Fe$^{+3}$, and Fe$^{+4}$) spanning the 4-zone model.
This result is also true for elements with ions observed from the dominant 
(high-)ionization zone, such as Ne and Ar. 
On the other hand, elements that only have observations of trace ions, 
such as N and S, likely have underestimated abundances. 

\item We found a model-independent dichotomy in the abundance patterns, 
where the abundances for J104457 and J141851 fall into two groups:
(1) $\alpha$-element ratios that are consistent with measured oxygen abundances
and a solar-abundance pattern (O/H, Ar/H, Ne/H) and 
(2) relatively deficient element ratios (N/H, C/H, Fe/H).

\item The two abundance groups suggest that these EELGs are 
$\alpha$/Fe-enriched by a factor of 3 or more, but this 
result alone cannot account for the properties of EELGs:
\begin{itemize}[noitemsep,topsep=-4pt,partopsep=0pt,leftmargin=*]
    \item We used photoionization models to show that $\alpha$/Fe-enriched conditions in EELGs 
    can produce high-ionization flux ratios that are augmented by up to 40\%\ relative to
    solar-$\alpha$/Fe EELGs (Figure~\ref{fig11}.
    However, these models still fail to reproduce the large \ion{He}{2}/H$\beta$ ratios 
    observed for EELGs by a factor of 5--10.
    \item While $\alpha$/Fe enrichment may be responsible for the observed properties of typical
    $z\sim2-3$ galaxies \citep[e.g.,][]{strom17}, we conclude there is an unsolved 
    {\it high-energy ionizing photon production problem}, or HEIP$^3$, in EELGs.
\end{itemize}

\item Regardless of the source, the very hard radiation fields in EELGs 
seem to produce higher central nebular temperatures, densities, 
and ionization parameters than previously thought.
Using the measured O ion fractions as weights, we determined average ionization 
parameters of the 4-zone model to be log$U=-1.66$ and $-1.93$ for J104457 and J141851, 
respectively, that are notably higher than the 3-zone model average ionization
parameters (log$U = -1.77$ and $-2.42$, respectively). 
Importantly, we showed in Figure~\ref{fig11} that these conditions support the 
model of a steeper central ionization structure than seen in more typical
\ion{H}{2} regions, which must be accounted for when determining properties of EELGs. 
\end{enumerate}

In summary, we found that the 4-zone model is a more accurate representation of 
EELGs than the classical 3-zone model, and the adoption of the 4-zone model
has a few important implications for the interpretation of these galaxies.
Specifically, using the 4-zone model reveals:
(1) the presence of a central, compact very-high-ionization zone,
(2) higher central gas-phase temperatures and densities and ionization parameters, 
(3) higher volume-averaged ionization parameters (log$U$) indicative of harder radiation fields,
(4) increased abundances of N/H, S/H, and Fe/H, 
(5) negligible to small reductions in relative abundances of C/O, Ar/O and Ne/O, and
(6) negligible changes in the overall O/H abundance, and
(7) an unsolved {it high-energy ionizing photon production problem} (HEIP$^3$).
This work suggests that EELGs in both the local and distant universe
have more extreme properties than previously thought. 
Therefore, future work with JWST and ELTs will likely require the 4-zone model
to diagnose accurate conditions within reionization-era galaxies.

\begin{acknowledgements}

DAB is grateful for the support for this program, HST-GO-15465, that was provided by 
NASA through a grant from the Space Telescope Science Institute, 
which is operated by the Associations of Universities for Research in Astronomy, 
Incorporated, under NASA contract NAS5- 26555. 
DKE is supported by the US National Science Foundation (NSF) through the 
Astronomy \& Astrophysics grant AST-1909198.
We are also grateful to the referee for thoughtful feedback that greatly 
improved the clarity of this paper. 

This work also uses observations obtained with the Large Binocular Telescope (LBT).
The LBT is an international collaboration among institutions in the
United States, Italy and Germany. LBT Corporation partners are: The
University of Arizona on behalf of the Arizona Board of Regents;
Istituto Nazionale di Astrofisica, Italy; LBT Beteiligungsgesellschaft,
Germany, representing the Max-Planck Society, The Leibniz Institute for
Astrophysics Potsdam, and Heidelberg University; The Ohio State
University, University of Notre Dame, University of
Minnesota, and University of Virginia.

This paper made use of the modsIDL spectral data reduction reduction pipeline
developed in part with funds provided by NSF Grant AST-1108693 and a generous
gift from OSU Astronomy alumnus David G. Price through the Price Fellowship in
Astronomical Instrumentation. 
Photoionization models were run with {\sc cloudy}, which is currently supported 
by grants from NSF (AST 1816537 and AST 1910687 National Aeronautics and Space 
Administration 19-ATP19-0188) and Space Telescope Science Institute (HST-AR-15018). 
These models also made use of v2.1 of the Binary Population and Spectral Synthesis 
(BPASS) models, as last described in \citet{eldridge17}.

\end{acknowledgements}

\facilities{LBT (MODS), HST (COS)}

\software{
{\sc cloudy} 17.00 \citep{ferland13}
DUSTMAPS \citep{green18}, 
modsIDL pipeline \citep{croxall19}, 
STARLIGHT \citep{fernandes05}, 
XIDL (\url{http://www.ucolick.org/~xavier/IDL/}), 
{\sc PyNeb} \citep{luridiana12, luridiana15}}

\clearpage
\newpage


\appendix

\section{Structure of Ionic Species}\label{sec:A1}
\begin{figure}[H]
\begin{center}
    \includegraphics[width=0.925\textwidth,trim=5mm 0mm 5mm 0mm,clip]{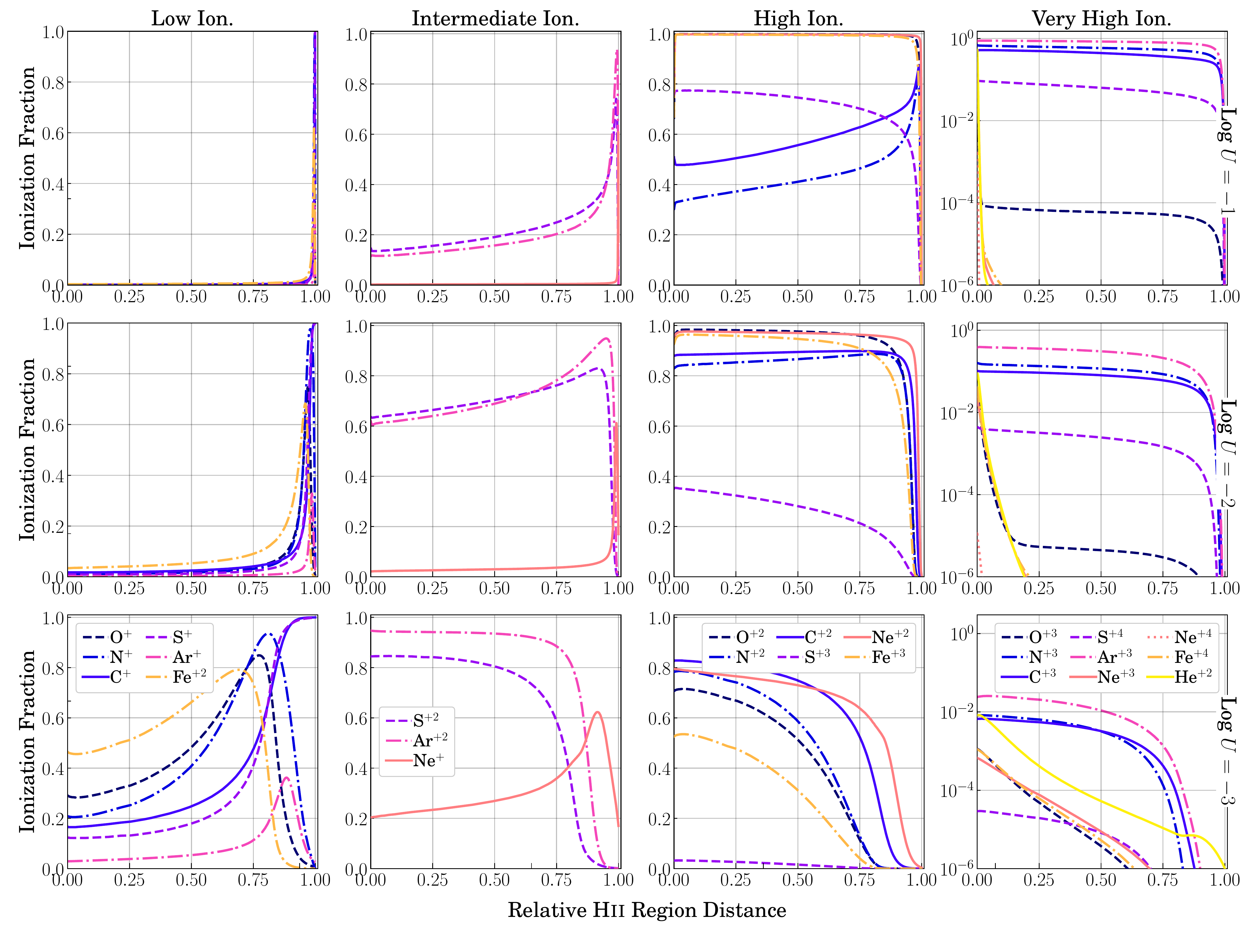}
    \vspace{-3ex}
\caption{
Photoionization models showing how ionization fractions of different species change as 
a function of relative \ion{H}{2} region radius.}
\label{figA1}
\end{center}
\end{figure}

To better understand the structure of EELGs, we explore the ionization structure
of different elements using the photoionization model grid described in Section~\ref{sec:3.1.1}.
In Figure~\ref{figA1} we plot how the ionization fractions of different species change as 
a function of relative \ion{H}{2} region radius from the central ionizing source and
for different input ionization parameters.
We show plots for three different ionization parameters, with
EELGs represented by the log$U=-1$ model in the top row, compared to
the log$U=-2$ model in the middle row, and the log$U=-3$ model that is characteristic
of average \ion{H}{2} regions in the bottom row.
Figure~\ref{figA1} is further separated into columns categorized by the ionization zones
of the 4-zone model.


As a whole, the log$U=-3$ model in Figure~\ref{figA1} shows some ions contributing
significantly from each of the low-, intermediate-, and high-ionization zones,
where the high-ionization ions dominate in the inner 50--75\%\ of the
nebula and the low-ionization ions dominate in the complimentary outer regions. 
In this model, no very-high-ionization ion makes a significant contribution, and 
therefore confirms the 3-zone model as the appropriate model for typical \ion{H}{2} regions.
In contrast, the log$U=-1$ model in Figure~\ref{figA1} shows 
that the high- and very-high-ionization ions dominate the ionization fractions 
over the majority of the nebula.
Very little contribution is seen from low- or intermediate-ionization ions on average,
as their small contributions only take effect at the very outer edges of the nebula. 
This structure highlights the importance of using the 4-zone model to interpret EELGs.

When comparing the different log$U$ models for the very-high-ionization zone (last column),
we not only see that ionization fractions of these elements significantly increase 
with higher log$U$, but also see their shapes drastically change. 
For example, in the log$U=-3$ model, the He$^{+2}$ ion (solid yellow line)
peaks in the center of the nebula at a fraction of $\sim1\%$, and then somewhat
gradually falls off with radius, reaching $\sim0.0001\%$ at the outer edge of the nebula.
In contrast, in the log$U=-1$ model, the He$^{+2}$ ion reaches a much higher peak
of $\sim50\%$, but quickly falls off to the same $\sim0.0001\%$ at only $\sim5\%$
of the relative radius of the nebula.
This supports the idea of a central, very compact very-high-ionization zone,
where the ionization structure is much more steeply declining than that 
of a typical nebula. 


\bibliography{mybib}

\clearpage

\end{document}